\documentclass[12pt,english]{article}
\usepackage{libertine}
\usepackage[T1]{fontenc}
\usepackage[latin9]{inputenc}
\usepackage{geometry}
\usepackage{color}
\usepackage{babel}
\usepackage{float}
\usepackage{amsmath}
\usepackage{amsthm}
\usepackage{amssymb}
\usepackage{graphicx}
\usepackage{setspace}
\usepackage[authoryear]{natbib}
\usepackage[unicode=true,pdfusetitle,
 bookmarks=true,bookmarksnumbered=false,bookmarksopen=false,
 breaklinks=false,pdfborder={0 0 1},backref=page,colorlinks=false]
 {hyperref}
\hypersetup{
 colorlinks,citecolor=darkblue,linkcolor=maroon,urlcolor=maroon,bookmarks=true,backref=page}

\makeatletter

\theoremstyle{definition}
\newtheorem{defn}{\protect\definitionname}
\theoremstyle{plain}
\newtheorem{assumption}{\protect\assumptionname}
\theoremstyle{plain}
\newtheorem{lem}{\protect\lemmaname}
\theoremstyle{plain}
\newtheorem{thm}{\protect\theoremname}
\theoremstyle{plain}
\newtheorem{prop}{\protect\propositionname}
\theoremstyle{plain}
\newtheorem{cor}{\protect\corollaryname}
\theoremstyle{definition}
 \newtheorem{example}{\protect\examplename}

\@ifundefined{date}{}{\date{}}
\usepackage{tikz}
\usetikzlibrary{positioning}
\usetikzlibrary{calc}
\usetikzlibrary{arrows} 
\usetikzlibrary{patterns}

\usepackage{color}
\definecolor{darkblue}{rgb}{0.0,0,.6}
\definecolor{maroon}{rgb}{0.68,0,0}
\definecolor{darkgreen}{rgb}{0,0.369,0.086}
\definecolor{gray}{rgb}{.5,.5,.5}

\definecolor{ao}{rgb}{0.0,0.5,0.0}

\definecolor{darkred}{rgb}{.6,0,0}
\definecolor{darkgreen}{rgb}{0,.6,0}
\definecolor{aliceblue}{rgb}{0.94, 0.97, 1.0}
\definecolor{bluegray}{rgb}{0.4, 0.6, 0.8}
\definecolor{shadecolor}{rgb}{.94,.97,1}

\usepackage{titletoc}
\usepackage[page,header]{appendix}

\@ifundefined{showcaptionsetup}{}{%
 \PassOptionsToPackage{caption=false}{subfig}}
\usepackage{subfig}
\makeatother

\providecommand{\assumptionname}{Assumption}
\providecommand{\corollaryname}{Corollary}
\providecommand{\definitionname}{Definition}
\providecommand{\examplename}{Example}
\providecommand{\lemmaname}{Lemma}
\providecommand{\propositionname}{Proposition}
\providecommand{\theoremname}{Theorem}

\begin{document}
\title{Getting the Agent to Wait\thanks{We owe the title of this paper to Orlov, Skrzypacz and Zryumov's work!
We thank several seminar and conference participants as well as Nageeb
Ali, Arjadha Bardhi, James Best, Lukas Bolte, Joyee Deb, Alexey Kushnir,
George Loewenstein, and Vasiliki Skreta.}}
\author{Maryam Saeedi\\
Carnegie Mellon University\\
\textcolor{blue}{msaeedi@andrew.cmu.edu}\and Yikang Shen\\
Carnegie Mellon University\\
\textcolor{blue}{yikangs@andrew.cmu.edu}\and Ali Shourideh\\
Carnegie Mellon University\\
\textcolor{blue}{ashourid@andrew.cmu.edu}}
\date{July 11, 2024}
\maketitle
\begin{abstract}
We examine the strategic interaction between an expert (principal)
maximizing engagement and an agent seeking swift information. Our
analysis reveals: When priors align, relative patience determines
optimal disclosure---impatient agents induce gradual revelation,
while impatient principals cause delayed, abrupt revelation. When
priors disagree, \emph{catering to the bias} often emerges, with the
principal initially providing signals aligned with the agent's bias.
With private agent beliefs, we observe two phases: one engaging both
agents, followed by catering to one type. Comparing personalized and
non-personalized strategies, we find faster information revelation
in the non-personalized case, but higher quality information in the
personalized case.
\end{abstract}

\section{Introduction}

Maximizing engagement is a central objective across many economic
settings, from traditional expert services such as management consulting
and legal advice to modern digital platforms with content provision
and recommender systems. In these contexts, an expert (principal)
aims to prolong user engagement to maximize value extraction, while
the user (agent) incurs costs from extended information acquisition
periods. In this paper, we develop a general framework for analyzing
such interactions.

To do so, we consider a model in which a principal and an agent interact
through engagement. The agent engages to collect information about
a payoff-relevant decision, while the principal aims solely to maximize
the duration of this engagement. This engagement is irreversible and
represents a trade-off: it is costly for the agent but valuable for
the principal. A key feature of our model is allowing for the agent
to have a different prior from the principal which could be publicly
observable to the principal (in the spirit of \citet{aumann1976agreeing}'s
agree-to-disagree framework) or privately known to the agent. This
approach enables us to explore how differences in beliefs and information
asymmetry impact optimal engagement strategies.

Our analysis reveals three key patterns that determine optimal information
provision in this environment. First, the relative patience of the
agent and principal determines whether information is revealed abruptly
or gradually. Second, when the agent and the principal agree to disagree
on their prior, it is often the case that optimal disclosure features
\textit{catering to the bias --} the principal first reveals the
state towards which the agent is more biased towards. Third, when
the agent's belief is private information, there is a trade-off between
speed and quality as information is revealed faster but engagements
end with more uncertain beliefs.

When the principal and the agent share the same prior, the main determinant
of information disclosure is how the cost of engagement for the agent
evolves over time relative to its benefit for the principal. More
specifically, suppose that both the principal and the agent use exponential
discounting. The principal then values each moment of engagement at
an exponentially declining rate, i.e., he values early engagement
more than later ones. When the agent is more \emph{impatient} than
the principal, exponential discounting implies that the relative cost
of engagement decreases over time. This in turn means that by some
gradual revelation early on, the principal is able to postpone revelation
for a time when the cost of engagement is lower for the agent. Technically,
when the discount rate of the principal is lower than that of the
agent, the payoff of the agent is convex in the payoff of the principal,
and thus the agent and principal both benefit from random or gradual
disclosure. This disclosure takes the form of a Poisson arrival of
the news. In contrast, when the agent is more \emph{patient}, the
cost of being engaged increases for the agent relative to the benefit
for the principal, which will lead to abrupt information revelation.

The equilibrium dynamics shift significantly when the principal and
agent have different prior beliefs. This divergence introduces a crucial
asymmetry in how each party values engagement across different states.
Consider a scenario where the agent is more optimistic about state
$\omega=0$ than the principal. In this case, revealing information
about $\omega=0$ carries higher value for the agent but represents
a lower cost for the principal. This asymmetry arises because the
principal, believing $\omega=0$ to be less probable, perceives the
promise of revelation in this state as less likely to materialize
than the agent anticipates. Consequently, promising to disclose information
about $\omega=0$ becomes a cost-effective strategy for maintaining
agent engagement. This is what we refer to as \textit{catering to
the bias.}

When the agent is more \emph{patient}, catering to the bias always
occurs. There is a phase in which no information is revealed; mirroring
the optimal disclosure pattern under shared priors. This is followed
by a second phase where the state towards which the agent is more
biased is gradually revealed according to a time-varying Poisson rate.
Finally, at the end of the second phase, the other state is revealed
instantaneously.

With a more \emph{patient} principal, catering to the bias would occur
when the difference between the prior of the agent and the principal
is sufficiently large. In this case, a frontier steady state pair
of beliefs exists such that the beliefs are going to converge to it.
When the priors agree, this will be the point of maximum uncertainty
and maximum engagement. Along this frontier, the optimal disclosure
involves symmetric, constant-rate Poisson disclosure of both states.
As such, beliefs remain stable on the frontier over time until a disclosure
event occurs. 

The evolution of beliefs (probability that $\omega=1$ for the agent
(x-axis) and the principal (y-axis)) away from this steady state is
depicted in Figure \ref{fig:Catering to the Bias-1}. Below the stationary
engagement curve, the state $\omega=1$ is revealed according to a
time-varying Poisson process, and above it the state $\omega=0$ is
going to be revealed until we land on the steady state curve. As shown
in Figure \ref{fig:Catering to the Bias-1}, the steady state curve
crosses the $45^{\circ}$ line exactly at the midpoint when $\mu_{P}=\mu_{A}=\frac{1}{2}$;
which is the special case of starting with the same priors. When the
priors do not agree, the point of stationary engagement is no longer
at $\mu_{A}=\frac{1}{2}$. This shift occurs because at $\mu_{A}=1/2$,
the principal believes one state is less likely than the other, revealing
information about that state becomes less costly for the principal
while maintaining the same value for the agent. Consequently, the
optimal strategy gravitates towards revealing the state the principal
deems less likely. The point of steady state is the one that equates
costs and benefits of revealing the state given the differences in
beliefs.
\begin{flushleft}
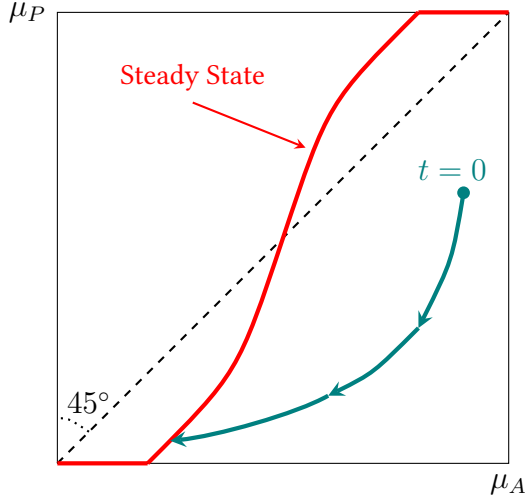
\begin{figure}[t]
\begin{centering}
\begin{tikzpicture}[scale=0.6]
\centering
\draw (0,0) -- (10,0) node[anchor=north] {$\mu_A$} -- (10,10) -- (0,10) node[anchor=east] {$\mu_P$} -- (0,0);
\draw[thick,dashed] (0,0) -- (10,10);
\draw[ultra thick, red] (2,0) .. controls (4,2) .. (5,5) .. controls (6,8) .. (8,10);
\draw[ultra thick, red] (0,0) -- (2,0);
\draw[ultra thick, red] (8,10) -- (10,10);
\draw[thick,dotted] (0.7,0.7) .. controls (.25,1) .. (0,1);
\draw (.7,1.35) node{{$45^\circ$}};
\draw[ultra thick, -stealth, teal] (9,6.) .. controls (8.75,4.5) .. (8,3);
\draw[ultra thick, -stealth, teal] (8,3) .. controls (7.,2.) .. (6,1.5);
\draw[ultra thick, -stealth, teal] (6,1.5) .. controls (5.3,1.1) and (3.5,.6) .. (2.5,.5);
\fill[teal,radius=4pt] (9,6) circle;
\draw[teal] (8.75,6.5) node{$t=0$};
\draw[thick, -stealth, red] (3,8) node[above]{\small Steady State} -- (5.5,7);
\end{tikzpicture}
\par\end{centering}
\caption{Evolution of beliefs under catering to the bias with a patient principal
\label{fig:Catering to the Bias-1}}
\end{figure}
\par\end{flushleft}

Building on these insights, we extend our model to examine cases where
agents' beliefs are private information. This extension enables us
to compare optimal information revelation strategies in a non-personalized
setting, where agents with different beliefs are exposed to the same
source of information (analogous to mass media in the context of news)
with the personalized strategies (similar to personalized social media
news feeds) discussed earlier. Our focus is on how these different
approaches impact belief evolution, speed of information revelation,
and quality of information.

In this extended model, we consider an economy with two agent types,
each identified by a different prior belief. To reflect non-personalized
communication, all information is public and non-targeted, i.e., types
are privately known. This non-personalized environment shares some
key features with its personalized counterpart. In both cases, we
identify a steady state, in the non-personalized case determined through
via a simple constrained concavification.

However, the non-personalized model diverges in a few important aspects.
First, the non-personalized setting may see one agent type departing
with incomplete information as the principal may wish to keep only
one type around while letting the other type leave. This is in contrast
with the personalized setting, where agents exit only upon full state
revelation. In other words, the principal strategically balances retaining
one type while allowing the other to leave. If only one type of agents
remains, we revert to personalized strategy. Moreover, in order to
make information acquisition incentive compatible, the principal should
increase the speed of exit. In other words, there is a trade-off between
speed and quality of information arrival.

Finally, we conclude our paper by incorporating more general forms
of discounting and adding random exogenous exit for the agents. The
model with exit allows us to investigate and compare numerically personalized
and non-personalized delivery of news. We observe two main differences
between these two modes of communication: their speed of delivery
and the quality of delivery.

When news is targeted and personalized, the principal can keep the
agent longer and maintain a lower speed of news delivery. In contrast,
the principal has to increase the speed of news delivery in the non-personalized
case. In the non-personalized case, two forces lead to this increase
in speed. First, the principal cannot necessarily keep both types
as uncertain as she wishes, given that they start with different priors.
Second, upon delivery of news, the principal may wish to keep one
type of agent while letting the other go, so the signal will not reveal
the state perfectly and therefore the increase in the value for agents
is going to be smaller. Both of these forces, higher steady state
value for agents and smaller increase in this value upon arrival of
news, lead to higher speed of news delivery to keep them engaged.

The quality of news is consistently higher in the case of personalized
news as the principal will never wish to have an agent leave without
full information. We also compare the impact of mode of delivery on
polarization. We see evidence of polarization in both cases and no
clear increase in the personalized case. This suggests that personalization
of news delivery may not necessarily exacerbate polarization more
than non-personalized approaches.

Ultimately, we view our paper as providing a framework to analyze
incentives for attention and its manipulation via engagement. This
is especially important as the internet and social media become an
integral part of daily life, and engagement and advertising still
remain the main source of monetization on the internet. In fact, there
is some evidence that our model does capture some behavior of advertisers
and web designers, see the discussion in section \ref{subsec:Examples-of-the}.
Specifically, various academics and policy makers, concerned by the
spread of misinformation on social media, have advocated taxes on
digital advertising.\footnote{In 2021, the state of Maryland put into effect a tax on digital advertising
revenues for those above \$100 millions revenue. See also the following
articles by Paul Romer in the New York Times, (see \href{https://www.nytimes.com/2019/05/06/opinion/tax-facebook-google.html}{https://www.nytimes.com/...},
accessed on July 11, 2024), and \citet{AcemogluJohnsonTax} for advocating
a flat 50\% tax on advertising revenues.} Through the lens of our model, a flat proportional tax on advertising
does not necessarily change the engagement strategy of the principal
as long as it does not remove incentives for this activity altogether.
Moreover, our model also suggests that belief polarization is not
necessarily worse with personalization of news (due to the trade-off
between speed and quality).

\paragraph*{Related Literature}

Our paper relates to several strands of the literature on contracting,
mechanism design, and information design. The most relevant is the
literature on dynamic Bayesian persuasion as in \citet{10.1257/aer.20150218},
\citet{RENAULT2017329}, \citet{doi:10.1086/704387}, \citet{doi:10.1086/706687},
\citet{10.1257/mic.20170405}, and \citet{doi:10.1086/722985}, among
many others, which build on the static persuasion model of \citet{10.1257/aer.101.6.2590}.
The key difference in our setting is that the agent's final decision
does not affect the principal's payoff, i.e., the principal only cares
about the duration of the game. We assume that both players are long-lived,
which is different from the myopic agent settings in \citet{10.1257/aer.20150218}
and \citet{RENAULT2017329}. The myopic setting suggests that gradual
revelation of information is optimal. In contrast, \citet{doi:10.1086/706687}
suggest that having a long-lived agent changes the greedy nature of
the principal's equilibrium strategy, as future information disclosure
can be used as an incentive device. Depending on the commitment power
of the principal, the agent could be persuaded to wait for future
information. In contrast with this literature, we are able to establish
results even in settings in which the belief of the agent is private
information.

A very recent set of papers have studied problems similar to ours,
namely that of communication when the objective of the principal is
to lengthen engagement by the agent. These papers include \citet{knoepfle2020dynamic},
\citet{hebert2022engagement}, \citet{koh2022attention}, and \citet{koh2024persuasion}.
\citet{knoepfle2020dynamic} studies a dynamic model of information
transmission where multiple senders compete for the attention of a
decision maker by strategically revealing information over time. In
their set up the decision maker has a fixed marginal cost of staying
each period. They find that in equilibrium senders use simple ``all-or-nothing''
strategies to reveal information, leading to full information transmission
in minimal time. \citet{hebert2022engagement} study this problem
assuming that the agent's time cost is separable and linear in time
and that the speed of learning is bounded above -- which is modeled
as a bound on the change in a generalized entropy function. They show
that despite the existence of constraints on learning, the principal
is able to reduce the value of the agent to her outside option.

Perhaps the closest paper to ours is that of \citet{koh2022attention}.
In their paper, they assume that the cost of waiting for the agent
is linear and separable over time of engagement while the payoff of
the principal is a general function of the time-engagement. They provide
general principles that determine the optimal mechanisms. There are
two key distinctions between their environment and that of ours: 1.
The use of linear time cost for the agent which implies that at each
point in time, only the expected value of engagement matters to the
agent; 2. We allow for mis-specified beliefs and agree-to-disagree
which then allows us to talk about catering to the bias as well as
communication in the presence of private information. As we show,
an important implication of geometric or multiplicative discounting
by both parties is that the commitment assumption of the principal
is often binding. \citet{koh2024persuasion} also study the general
version of this problem (under belief agreements) and extend to the
case when commitment is not binding. 

Note that our result on the gradual information revelation is purely
driven by the convexity of the agent's time preferences. Similar forces
occur in other settings where the precise shape of time preferences
and their relationship between the parties involved in contracting
matters. These include models of inspection (\citet{ball2023should}),
and choice under uncertainty about the timing of rewards (\citet{dejarnette2020time}),
among others. In our setup, the relative curvature of the payoffs,
what we refer to as the marginal cost of engagement, and its evolution
over time that determines the form of dynamics over time.

Finally, our paper is also related to the extensive literature on
experimentation and contracting for discovery. Specifically, in our
model, a time-varying Poisson bandit model as in \citet{keller2005strategic}
and \citet{keller2010strategic} arises endogenously.\footnote{Much of this literature focuses on strategic interactions among several
experimenters -- see also \citet{strulovici2010learning}. While
this would be relevant to our question, it is beyond the scope of
current paper.} Several papers have studied contracting in models of experimentation
(\citet{guo2016dynamic}, \citet{halac2016optimal}, \citet{henry2019research},
among others). This literature often takes as given the process for
experimentation and studies ways in which the conflict of interest
between the parties can be handled either via delegation or persuasion
or by the use of monetary incentives. In our setup, on the other hand,
the principal is choosing the process of discovery guided by his interest
in maximizing engagement.

The rest of the paper is organized as follows: Section \ref{sec:The-General-Model}
presents our general model, Section \ref{sec:A-Simple-Example} provides
a simple example, Sections \ref{sec:Optimal-Communication-under}
analyzes the case of common priors, Section \ref{sec:Catering-to-the}
explore scenarios with different priors, Section \ref{sec:personalized}
examines the case of private beliefs, and Section \ref{sec:Extensions-and-Applications}
discusses various extensions of our framework.

\section{The General Model\label{sec:The-General-Model}}

In this section, we present our general model. The model consists
of a principal (referred to as \textquotedbl he\textquotedbl ) and
an agent (referred to as \textquotedbl she\textquotedbl ). The principal
provides information to the agent about a payoff-relevant state over
time, and the agent collects this information in order to take a final
action. Time, denoted by $t$, is a continuous variable belonging
to $\left[0,\infty\right)$. The payoff of the agent is given by
\[
u\left(T,\omega,a\right)=\delta_{A}e^{-\delta_{A}T}\hat{u}\left(\omega,a\right)
\]
where $\omega\in\Omega=\left\{ 0,1\right\} $ represents the underlying
state, $a\in A$ is the action taken by the agent, and $T$ is the
time spent acquiring information. Throughout our analysis, we assume
that the agent prefers to take the action sooner, therefore, $\delta_{A}$
and $\hat{u}\left(\omega,a\right)$ are both positive. Additionally,
we assume that the payoff of the principal if the agent quits at $T$
is $u_{P}\left(T\right)=\int_{0}^{T}e^{-\delta_{P}t}dt=\frac{1-e^{-\delta_{P}T}}{\delta_{P}}$.
In other words, the principal seeks to maximize the engagement time
$T$. The fundamental disagreement between the principal and the agent
is on the value of engagement: while the agent desires to acquire
information as quickly as possible, the principal prefers longer engagements.\textit{\emph{
As we will show, the key determinant of the principal's strategy is
the relative patience of the principal to that of the agent or whether
$\delta_{A}<\delta_{P}$ or $\delta_{A}\geq\delta_{P}$.}}

\begin{figure}[H]
\begin{centering}
\textbf{}\begin{tikzpicture} 
\sffamily{
\draw (-1,1) node[right]{$\cdots$};
\draw (0.07,1) -- (2,1) ;
\filldraw[radius=2pt,black] (4.15,1) circle node[above right=.1cm and .02cm]{\small A};
\draw (4.15,1) -- (6.15,0)  node[midway, below,sloped] {\small Quit}; 
\draw (4.15,1) -- (6.15,2.)   node[midway, above,sloped] {\small Stay};
\draw[radius=2pt,black] (0,1) circle node[above=.1cm]{{\small $t$}};
\filldraw[radius=2pt,black] (2,1) circle node[above=.1cm]{{\small P}};
\draw (2,1) -- (4,2);
\draw (2,1) -- (4,0) node[midway,below]{\small $s_t$};
\draw [thin] (4,2) ..  controls (4.2,1) ..  (4,0);
\filldraw[radius=2pt,black] (4,2) circle;
\filldraw[radius=2pt,black] (4,0) circle;
\filldraw[radius=2pt,black,fill=white] (6.15,2) circle node[above=.1cm]{\small $t+dt$};
\filldraw[radius=2pt,black] (6.15,0) circle node[right=.2cm]{\small $a_{t+dt}\in A$} node[above=.05cm]{\small A};
\draw (6.4,2) node[right]{$\cdots$};
}
\end{tikzpicture} 
\par\end{centering}
\caption{The timing of the actions within a time interval $\left[t,t+dt\right]$\label{fig:The-timing-of}}
\end{figure}
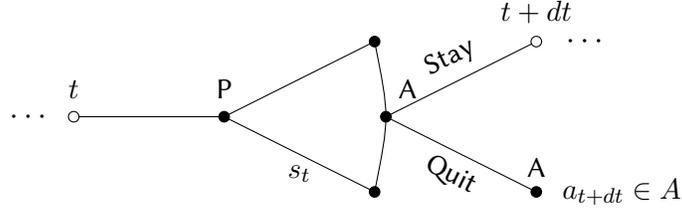

Figure \ref{fig:The-timing-of} depicts the timing of the communication
game in a short time period between $t$ and $t+dt$. Conditional
on the agent staying engaged up until $t$, at $t$ the principal
sends a signal $s_{t}\in S_{t}$ to the agent whose distribution depends
on the state $\omega$ and the history of signal realizations in the
past. This history is represented by the function $s^{t}=\left\{ s_{\tau}\right\} _{\tau\in\left[0,t\right]}$.
The agent, having observed $s_{t}$ -- and the history of public
signals -- decides whether to stay engaged after $t+dt$ or quit
at $t+dt$ and choose $a_{t+dt}\in A$ to maximize her expected payoff
$\mathbb{E}u\left(t+dt,\omega,a\right)$ where the expectation is
taken with respect to her belief at $t+dt$ about $\omega$.

\textbf{}

The strategies and learning process of the players can be explained
as follows. The principal chooses an information structure: a mapping
from the space of history realizations to probability distributions
over signals at $t$. More formally, the principal's strategy is a
quadruple $\left(S_{\infty}\times\Omega,\mathcal{F},\mathbb{P}_{P},\left\{ \mathcal{F}_{t,P}\right\} _{t\in\mathbb{R}_{+}}\right)$
where $S_{\infty}$ is the set of history of signal realizations,
i.e., each member is of the form $s^{\infty}$, $\Sigma$ is a $\sigma$-algebra
over $S_{\infty}\times\Omega$, $\mathbb{P}_{P}$ is its associated
probability measure from the principal's perspective, and finally,
$\mathcal{F}_{t,P}\subset\mathcal{F}$ is a filtration, i.e., a family
of increasing $\sigma$-algebras representing the information at time
$t$. Here, the probability measure over the signal realizations is
specified from the principal's perspective, which helps to identify
the optimal strategies when the players have different prior beliefs
over $\omega$. We can think of the filtration $\mathcal{F}_{t,P}\left(\omega\right)$
as the $\sigma$-algebra induced by the function $\hat{s}^{t}\left(s^{\infty},\omega\right)$.
In words, this is the function that translates a complete history
of the game $\left(s^{\infty},\omega\right)$ into history up to $t$.
Let $\mathcal{F}_{t,A}$ be the filtration associated with the agent's
information at time $t$. That is,
\begin{equation}
\mathcal{F}_{t,A}=\left\{ S\subset S_{\infty}|\exists S'\in\mathcal{F}_{t},s^{\infty}\in S\Leftrightarrow\left(s^{\infty},1\right)\in S'\text{ or }\left(s^{\infty},0\right)\in S'\right\} \label{eq: FtA}
\end{equation}
The agent's strategy is a stopping time $\tau$, associated with quitting,
i.e., $T=\tau$, with respect to the filtration $\left\{ \mathcal{F}_{t,A}\right\} _{t\in\mathbb{R}_{+}}$
together with a decision rule $a_{\tau}:S_{\infty}\rightarrow A$
which is progressively measurable with respect to the filtration $\mathcal{F}_{t,A}$.

For the learning process, we assume that the principal's and the agent's
prior beliefs about $\omega$ are given by $\mu_{P}=\Pr_{P}\left(\omega=1\right)=\mathbb{P}_{P}\left(S_{\infty}\times\left\{ 1\right\} \right)$
and $\mu_{A}=\Pr_{A}\left(\omega=1\right)$, respectively. We thus
allow the priors to disagree, but this disagreement is common knowledge.\footnote{In Section \ref{sec:personalized}, we study an extension of this
model where the principal is not informed about the type of the agent
and cannot send personalized signals.} Given the priors, the agent uses Bayes' rule to update her belief.
Hence, her belief $\mu_{A}\left(t\right)$ is a progressively measurable
stochastic process with respect to filtration $\mathcal{F}_{t,A}$
and follows:
\[
\mu_{A}\left(t\right)=\mathbb{E}_{A}\left[\mathbf{1}\left[\omega=1\right]|\mathcal{F}_{t,A}\right].
\]
This conditional expectation operator maps members of $\mathcal{F}_{t,A}$
to $\left[0,1\right]$. Note that since the agent and the principal
may disagree on their prior belief about $\omega$, the above expectation
is taken with respect to the agent's probability measure over $S_{\infty}\times\Omega$,
$\mathbb{P}_{A}$. This probability measure can be constructed from
those of the principal according to
\[
\mathbb{P}_{A}\left(S\right)=\mu_{A}\mathbb{P}_{P}\left(S|S_{\infty}\times\left\{ 1\right\} \right)+\left(1-\mu_{A}\right)\mathbb{P}_{P}\left(S|S_{\infty}\times\left\{ 0\right\} \right)
\]
This means that even if the agent and the principal disagree on their
prior beliefs, they agree on the probability distribution chosen by
the principal over signals conditional on the state $\omega$.

Throughout our analysis, we assume that the principal is committed
to his strategy, while the agent is not. Hence, a Perfect Bayesian
Equilibrium is defined as follows:
\begin{defn}
A PBE of the game consists of a strategy profile for the principal
$\left(S_{\infty},\mathcal{F},\mathbb{P}_{P},\left\{ \mathcal{F}_{t,P}\right\} _{t\geq0}\right)$
and a strategy profile for the agent $\left(\tau,a_{\tau}\right)$
such that:
\begin{enumerate}
\item Given $\left(\tau,a_{\tau}\right)$, $\left(S_{\infty},\mathcal{F},\mathbb{P}_{P},\left\{ \mathcal{F}_{t,P}\right\} _{t\geq0}\right)$
maximizes principal's payoff
\[
\mathbb{E}_{P}\left[1-e^{-\delta_{P}T}\right]
\]
\item Given $\left(S_{\infty},\mathcal{F},\mathbb{P}_{P},\left\{ \mathcal{F}_{t,P}\right\} _{t\geq0}\right)$,
at any point in time $t$ and conditional on not quitting, the agent
chooses $\left(\tau,a_{\tau}\right)_{\tau\geq t}$ to maximize her
payoff 
\[
\mathbb{E}_{A}\left[\delta_{A}e^{-\delta_{A}\tau}\hat{u}\left(\omega,a\right)|\mathcal{F}_{t,A}\right]
\]
where $\mathcal{F}_{t,A}$ is derived from $\mathcal{F}_{t}$ using
(\ref{eq: FtA}).
\end{enumerate}
\end{defn}

\subsection{Examples of the Environment\label{subsec:Examples-of-the}}

Our model can be applied to several settings where an expert provides
information while maximizing engagement. Here, we discuss a few examples.

\textbf{Consulting} \textbf{and Legal Services}: An important application
of our model is in the context of consulting and legal services, where
compensation is often a function of the so-called ``billable hours,''
i.e., the time spent on the project by the expert. This structure
inherently creates large information asymmetries between the service
provider and the customer. In the legal services domain, \citet{hadfield2007legal}
and \citet{hadfield2022legal} argue that the American Bar Association's
industry regulations on organizational structure, contract specification,
and other areas create perverse incentives. A key inefficiency they
highlight is the complexity and opaqueness of contracts, which often
do not specify total costs, thereby incentivizing longer engagements.
Our results can be interpreted as a theory linking the cost and benefit
of engagement to the intertemporal preferences and beliefs of the
parties involved in expert services contracts. Given that marginal
benefit and cost of engagement are key determinants of contract structure,
and are often influenced by factors such as cost of capital and organizational
structures, our model provides testable predictions. Specifically,
it suggests examining the effectiveness of expert services in relation
to these variables, offering a framework for empirical investigation
of contract design and service provision in consulting and legal industries.

\textbf{Market for News}: For individual users, news consumption represents
another common scenario of information acquisition with potential
conflicts of interest, mirroring the dynamics in our model. News providers,
ranging from traditional outlets like television and newspapers to
online sources such as Google News, often rely on user engagement
and advertisement as primary revenue sources. An important question
in the literature on the economics of the media is that of the effect
of private incentives on the quality of news and ultimately political
competition -- see, for example, \citet{stromberg2004mass}. This
issue becomes particularly salient in the context of personalized
news delivery, where there is a possibility of news being \textit{catered
to the biases} of consumers. Our model can thus be used to shed light
on the effect of these new media sources, personalized and catered
media as opposed to mass media, on political outcomes. We specifically
illustrate this application in Section \ref{sec:personalized} and
\ref{subsec:Personalization-and-Belief}.

\textbf{Social media and the Internet:} Social media has become an
increasingly dominant platform for news dissemination. For instance,
more than half of US adults now rely on the Internet as their primary
source of health information (\citet{doi:10.1080/10410236.2020.1748829}).
The revenue model of social media platforms, heavily dependent on
advertising and user engagement, aligns closely with the dynamics
explored in our model, offering insights into the impact of this business
approach on information distribution. Empirical evidence from social
media algorithm designs appears to corroborate our theoretical predictions.
\Citet{10.1257/aer.20191777} conducted a large field experiment on
Facebook by offering participants to subscribe to outlets with random
political attitudes. Their study shows that \emph{`` ... Facebook's
algorithm is less likely to supply individuals with posts from counter-attitudinal
outlets, conditional on individuals subscribing to them,'' }suggesting
that\emph{ ``social media algorithms may limit exposure to counter-attitudinal
news.''} This observation aligns with our model's prediction of \emph{catering
the news to the bias}, where, in the presence of biased beliefs and
personalized news, the information provider (principal) tends to prioritize
information about states that the user (agent) considers more likely.
Additionally, along the same lines of argument, \citet{10.1257/aer.20190658}
conduct an experiment on Facebook and show that deactivating Facebook
significantly reduces the polarization of views on policy issues.

One can also view our model as one in which an advertiser or web designer
decides how to deliver content to a user. Designers have a choice
of how to deliver content that an interested user is wishing to learn:
they can bombard a page with advertising and require a lot of scrolling
until the end or randomly choose instances during a video to advertise,
etc. According to IAB UK, an industry body for digital advertising,
21\% of impressions on the web were via the so-called ``made for
advertising'' websites.\footnote{See \href{https://www.iabuk.com/news-article/guide-identifying-made-advertising-websites}{https://www.iabuk.com/news-article/...}
(accessed on July 11, 2024) for more guidelines on how to design a
website for effective digital marketing. According to this article,
``\textit{Made for Advertising sites will often have unusual navigation
and user journeys in order to maximize ad exposure. For example, they
may tempt users to view \textquoteleft 20 actors from the nineties,
you won\textquoteright t believe what number 17 looks like now!\textquoteright{}
The content for an article like this will be laid out with one actor
per page, so the user must click through 17 pages, being exposed to
multiple ads on each page of the journey} ...''.}While design experts often advise against requiring users to watch
or scroll through large segments of advertising, providers still use
such strategies. According to our model, this behavior can be explained
by the relative patience of the provider to that of the user. This
is in contrast to providing \textit{``... salient information within
a page's initially viewable area ..''} together with \textit{``...
while placing the most important stuff on top, don't forget to put
a nice morsel at the very bottom..}''.\footnote{Quoted from \href{https://www.nngroup.com/articles/scrolling-and-attention-original-research/}{https://www.nngroup.com/...},
a User Experience (UX) Research company (accessed on July 11, 2024).} Through the lens of our model, such behavior, interpreted as gradual
revelation of information with the most likely information arriving
first, is associated with a more patient principal.

\subsection{Solving the Model\label{subsec:Solving-the-Model}}

In this section, we describe the technical results required to characterize
the solution of the principal's optimal choice of communication.

First notice that similar to the formulation of the standard Bayesian
Persuasion model of \citet{10.1257/aer.101.6.2590}, it is sufficient
to describe the evolution of beliefs of the agent from the perspective
of the principal, given that the agent and the principal could disagree
on their priors. More specifically, an application of Bayes' rule
implies that
\[
\frac{\mu_{t,A}\left(s^{t}\right)}{1-\mu_{t,A}\left(s^{t}\right)}\frac{1-\mu_{t,P}\left(s^{t}\right)}{\mu_{t,P}\left(s^{t}\right)}=\frac{\mu_{A}}{1-\mu_{A}}\frac{1-\mu_{P}}{\mu_{P}}
\]
In words, since both the principal and the agent use the same signal
structure to update their beliefs, the relative likelihood ratio of
the agent's belief to that of the principal remains constant over
time. Let us define $\ell$ as this ratio
\[
\ell=\frac{\mu_{A}}{1-\mu_{A}}\frac{1-\mu_{P}}{\mu_{P}}
\]
Given $\ell$ and the belief of the principal, the belief of the agent
can be calculated using the above and is given by
\begin{equation}
\mu_{t,A}\left(s^{t}\right)=\frac{\ell\mu_{t,P}\left(s^{t}\right)}{\ell\mu_{t,P}\left(s^{t}\right)+1-\mu_{t,P}\left(s^{t}\right)}\label{eq: mpAmuP}
\end{equation}
 An implication of this stationarity is that it is sufficient to describe
the strategy of the principal via a stochastic process over his beliefs
$\sigma_{t,P}\left(\mu_{t,P}|\mu_{P}^{t-}\right)$ and rewrite the
payoff of the agent in terms of that of the principal. Under this
reformulation, we can refer to the history of signals as $\mu_{P}^{t}$
or history of beliefs for the principal. The strategy of the principal
then is simply choosing a distribution over such histories given by
$\sigma_{P}\left(\mu_{P}^{t}\right)$.

Subsequently, we can define the value of the agent upon exiting as
a function of her belief, $v\left(\mu_{A}\right)$, as
\[
v\left(\mu_{A}\right)=\max_{a\in A}\mathbb{E}_{\mu_{A}}\left[\hat{u}\left(a,\omega\right)\right]
\]
It is evident that $v\left(\mu_{A}\right)$ is a convex function of
$\mu_{A}$. For convenience, we make the following assumption about
$v\left(\mu_{A}\right)$:
\begin{assumption}
\label{assu:The-payoff-function}The payoff function $v\left(\mu_{A}\right)$
is strictly convex, differentiable, and symmetric around $\mu_{A}=1/2$.
\end{assumption}
We should note that this assumption is rather innocuous and allows
us to conveniently characterize optimal strategies of the principal
via first-order conditions. One can always normalize the payoffs of
the principal and the agent to make $v$ symmetric. Moreover, any
convex symmetric function can be approximated by a sequence of strictly
convex and differentiable functions. Assumption \ref{assu:The-payoff-function}
allows us to take derivatives, which streamlines the analysis.

Our first result concerns the beliefs of the agent upon exit.
\begin{lem}
\label{lem:Suppose-that-Assumption}Suppose that Assumption \ref{assu:The-payoff-function}
holds. If the agent exits after the belief history $\mu_{P}^{t-}$,
then $\mu_{P}\left(t\right)=\mathbb{E}_{P}\left[\omega|\mu_{P}^{t-}\right]\in\left\{ 0,1\right\} $
almost surely.
\end{lem}
The intuition behind this result is straightforward. Consider a scenario
where $\mu_{P}\left(t\right)\notin\left\{ 0,1\right\} $ for a positive
measure of histories in which the agent exits. In this case, the principal
can improve the information revelation strategy by splitting the signal
into two fully revealing signals, $\left\{ 0,1\right\} $, with probabilities
$1-\mu_{P}\left(t\right)$ and $\mu_{P}\left(t\right)$ respectively.
This modification is a mean preserving spread of the beliefs. Given
that $v\left(\cdot\right)$ is strictly convex (Assumption\textbf{
}\ref{assu:The-payoff-function}), this spread increases the agent's
expected payoff without inducing earlier exits. Hence, such a strategy
can be combined with an initial period of no information revelation
at the beginning of the game, thus inducing a profitable deviation
for the principal.

Given our reformulation of the problem, we can apply the Carathéodory
theorem and show that three signals are enough for each period. This
together with Lemma \ref{lem:Suppose-that-Assumption} implies that
these signals are given by $\Omega\cup\left\{ \text{No News}\right\} $.
Therefore, in each period, the agent either quits with full information
or updates her belief based on the fact that the state is not revealed.\footnote{This feature of optimal communication is similar to models of Poisson
experimentation a la \citet{keller2010strategic} where no news leads
to a gradual change in beliefs.} As a result, we can summarize the strategy of the principal by the
use of two distribution functions:
\begin{align*}
G_{P,1}\left(t\right)= & \mathbb{P}_{P}\left(\text{exit}\geq t,\omega=1\right)\\
G_{P,0}\left(t\right)= & \mathbb{P}_{P}\left(\text{exit}\geq t,\omega=0\right)\\
\mu_{P}\left(t\right)= & \mathbb{P}_{P}\left(\left\{ \omega=1\right\} |\text{stay until }t\right)\\
= & \frac{G_{P,1}\left(t\right)}{G_{P,1}\left(t\right)+G_{P,0}\left(t\right)}=\frac{G_{P,1}\left(t\right)}{G_{P}\left(t\right)}
\end{align*}
where $G_{P,1}\left(0\right)=\mu_{P}=1-G_{P,0}\left(0\right)$, and
$G_{P}\left(t\right)=G_{P,1}\left(t\right)+G_{P,0}\left(t\right)$.
Note that $G_{P,\omega}\left(t\right)$'s are decumulative distribution
functions and are thus decreasing over time. We will refer to $G_{P,\omega}\left(t\right)$'s
as \textit{engagement probability functions}.

This simplification of the problem allows us to rewrite the payoffs
of the agent and the principal in simpler forms:
\begin{align*}
u_{P} & =-\int_{0}^{\infty}\frac{1-e^{-\delta_{P}t}}{\delta_{P}}dG_{P}\left(t\right)\\
u_{A,t} & =-\int_{t}^{\infty}e^{-\delta_{A}\left(s-t\right)}v\left(1\right)\frac{d\left(\ell G_{P,1}\left(s\right)+G_{P,0}\left(s\right)\right)}{\ell G_{P,1}\left(t\right)+G_{P,0}\left(t\right)}
\end{align*}
In the above, the negative signs represent the fact that $G_{p,\omega}\left(t\right)$
is a decumulative probability function. Moreover, the probability
used for the agent is adjusted to account for their difference in
their priors while we have used the fact that $v\left(0\right)=v\left(1\right)$.
Additionally, $u_{A,t}$ is the continuation payoff of the agent in
the event that she does not exit until $t$. Using integration by
parts, we can write these payoffs as
\begin{align*}
u_{P} & =\int_{0}^{\infty}e^{-\delta_{P}t}G_{P}\left(t\right)dt\\
u_{A,t} & =v\left(1\right)-\frac{\delta_{A}v\left(1\right)\int_{t}^{\infty}e^{-\delta_{A}\left(s-t\right)}\left(\ell G_{P,1}\left(s\right)+G_{P,0}\left(s\right)\right)ds}{\ell G_{P,1}\left(t\right)+G_{P,0}\left(t\right)}
\end{align*}

The above calculations imply that the principal's optimal information
provision problem, which yields $V(\mu_{P},\ell)$, is given by:
\begin{equation}
V\left(\mu_{P},\ell\right)=\max_{G_{P,0},G_{P,1}}\int_{0}^{\infty}e^{-\delta_{P}t}\left[G_{P,1}\left(t\right)+G_{P,0}\left(t\right)\right]dt\tag{P}\label{eq:P}
\end{equation}
subject to
\begin{align}
v\left(1\right)G_{A}\left(t\right)-v\left(1\right)\delta_{A}\int_{t}^{\infty}G_{A}\left(s\right)e^{-\delta_{A}\left(s-t\right)}ds & \geq G_{A}\left(t\right)v\left(\frac{\ell G_{P,1}\left(t\right)}{\ell G_{P,1}\left(t\right)+G_{P,0}\left(t\right)}\right),\forall t\label{eq: Obedience}\\
G_{P,\omega}\left(t\right) & \text{ : non-increasing},\omega\in\left\{ 0,1\right\} \nonumber \\
G_{P,1}\left(0\right)= & 1-G_{P,0}\left(0\right)=\mu_{P}\nonumber 
\end{align}
where in the above, $G_{A}\left(t\right)=\ell G_{P,1}\left(t\right)+G_{P,0}\left(t\right)$
is the probability of engagement until $t$ from the agent's perspective
and is used for brevity.

In the above problem, when the agent and the principal agree on their
prior, i.e., $\ell=1$, the gain from information revelation in each
state is given by $v\left(1\right)-v\left(\mu_{A}\right)$. Given
our symmetry assumption, this is maximized at $\mu_{A}=1/2$, i.e.,
when agent is maximally uncertain. Thus, depending on the evolution
of Marginal Cost of Engagement (MCE), the principal wishes to maximize
the amount of time spent at this belief or as close as possible to
this level.

In contrast, when $\ell<1$, the agent is more optimistic about the
state $\omega=0$ relative to the principal. In this case, while the
gain from revelation from the agent's perspective is still maximized
at $\mu_{A}=1/2$, since the principal disagrees with the agent on
her beliefs, this maximum gain does not necessarily lead to longer
engagement from the principal's perspective.

Finally, we should note that given Assumption \ref{assu:The-payoff-function},
the function $\left(\ell G_{P,1}+G_{P,0}\right)v\left(\frac{\ell G_{P,1}}{\ell G_{P,1}+G_{P,0}}\right)$
is strictly convex. As a result, the constraint set in (\ref{eq:P})
is convex. Together with the fact that the objective is linear in
$G_{P,1}\left(t\right)$ and $G_{P,0}\left(t\right)$, the standard
results from convex optimization (See, for example, \citet{luenberger1997optimization},
sections 8 and 9) apply.\footnote{The Luenberger's results cannot readily apply because the constraint
set does not have a non-empty interior. In the appendix, we first
show that when time is finite, the constraint set has a non-empty
interior. We then send the time horizon to infinity, show that multipliers
converge and use Berge's maximum Theorem.} Therefore, we can set $\mathcal{G}$ to be the set of decumulative
distribution functions $\left(G_{P,1},G_{P,0}\right)$ with $G_{P,1}\left(0\right)=1-G_{P,0}\left(0\right)=\mu_{P}$.
Then, we can use integration by parts to write the Lagrangian associated
with \ref{eq:P} as:
\begin{align*}
\mathcal{L} & =\int_{0}^{\infty}e^{-\delta_{P}t}\left[G_{P,1}\left(t\right)+G_{P,0}\left(t\right)\right]dt\\
 & +\int_{0}^{\infty}e^{-\delta_{P}t}G_{A}\left(t\right)\left[v\left(1\right)-v\left(\frac{\ell G_{P,1}\left(t\right)}{\ell G_{P,1}\left(t\right)+G_{P,0}\left(t\right)}\right)\right]d\Lambda\left(t\right)\\
 & -\delta_{A}v\left(1\right)\int_{0}^{\infty}G_{A}\left(t\right)e^{-\delta_{A}t}\left[\Lambda\left(t\right)-\Lambda\left(0\right)\right]dt
\end{align*}
where $\Lambda\left(t\right)$ is a weakly increasing function which
is constant at $t$ whenever the incentive constraint is slack. 

With this reformulation, $d\Lambda\left(t\right)$ is the value of
gains from information revelation, and the last term captures the
benefits of revelation at $t$ on all previous periods' incentives.
If $\partial\mathcal{L}\left(G_{1},G_{0};h\right)$ is the Gateau
derivative of $\mathcal{L}$ along the direction $h\in\mathcal{G}$,
we can use the result on convex programming in \citet{luenberger1997optimization}
and state the following.
\begin{thm}
\label{thm: thm1}If $\left(G_{P,1}^{*},G_{P,0}^{*}\right)$ is the
solution to \ref{eq:P}, then there exists an increasing function
$\Lambda\left(t\right)$ such that 
\begin{align*}
\partial\mathcal{L}\left(G_{P,1}^{*},G_{P,0}^{*};G_{P,1}^{*},G_{P,0}^{*}\right) & =0\\
\partial\mathcal{L}\left(G_{P,1}^{*},G_{P,0}^{*};h\right) & \leq0,\forall h\in\mathcal{G}.
\end{align*}
Conversely, if for some increasing $\Lambda\left(t\right)$, $\left(G_{P,1}^{*},G_{P,0}^{*}\right)$
satisfies the above conditions, then it is the solution to \ref{eq:P}.
\end{thm}
We should note that the above does not imply that the monotonicity
constraints on $G_{\omega}$'s are not binding. Rather, it implies
that sometimes ironing is needed. Our solution technique is similar
to a strand of the literature on Bayesian Persuasion that uses linear
and convex programming techniques in infinite-dimensional spaces.
See \citet{dworczak2019simple}, \citet{https://doi.org/10.3982/ECTA18312},
and \citet{rating} among many others. Note that since our constraint
set is strictly convex, it is often the case that the solution to
the optimization problem \ref{eq:P} is unique.

\section{A Simplified Example\label{sec:A-Simple-Example}}

In order to build intuition for our first set of results, we illustrate
some of the basic ideas in a simple version of the model discussed
above, where the principal can only choose to reveal the state or
not.

Suppose that the value to the agent of engagement until $T$ is given
by
\[
\frac{1}{2}e^{-\delta_{A}T}\left(1+\mathbf{1}\left[\text{informed by }T\right]\right),
\]
where $\delta_{A}$ is the agent's discount rate. The value to the
principal is given by $T$.

The principal can commit to a revelation strategy, while the agent
cannot commit to an engagement strategy. In this example, we have
abstracted from the details of the agent's value of information. Additionally,
we have only allowed the principal to reveal the state or not.

Given that the principal is able to commit to any, possibly random,
revelation strategy, we can represent his strategy as $F\left(T\right)$,
an increasing function of time that represents the cumulative probability
of revelation until time $T$. Hence, the principal's problem can
be written as
\begin{align}
\max_{F\in\mathcal{F}}\, & \mathbb{E}_{F}T\nonumber \\
\text{s.t. } & \mathbb{E}_{F}\left[e^{-\delta_{A}T}|T>t\right]\ge\frac{1}{2}e^{-\delta_{A}t},\,\forall t\ge0\label{eq: ICex1}
\end{align}
where $\mathcal{F}$ is the set of increasing functions over $\mathbb{R}_{+}\cup\left\{ 0\right\} $
with $F\left(0\right)=0,\lim_{T\rightarrow\infty}F\left(T\right)=1$.
Moreover, the constraint is the incentive constraint of the agent
that remains uninformed at $t$ and needs to be incentivized to stay
engaged.

To understand the forces at play in solving the principal's problem,
consider the naive strategy of delivering the information at a predetermined
period $\hat{T}$. For this to be incentive compatible for the agent,
we must have
\[
e^{-\delta_{A}\hat{T}}\geq\frac{1}{2}\Rightarrow\hat{T}\leq\frac{\log2}{\delta_{A}}.
\]

\begin{figure}
\begin{centering}
\begin{tikzpicture}[scale=.6]
\centering
\draw[->] (0,0) -- (8,0) node[anchor=north] {$T$};     \draw[->] (0,0) -- (0,8) node[anchor=east] {$u(T)$};
    \draw[thick,color=blue] (0,6) .. controls (2,2) and (4,1) .. (8,0.5) node[anchor=south,scale=1]{$e^{-\delta_A T}$};          
\draw[thick,dashed,color=darkgreen] (0.75, 4.61) -- (5.75,0.88);     
\filldraw[radius=3pt,color=darkgreen] (0.75, 4.61) circle;     \filldraw[radius=3pt,color=darkgreen] (5.75,0.88) circle;     \filldraw[radius=3pt,color=red] (2.75,2.28) circle;
\draw[thick,dashed,color=darkgreen] (3.93,0)  -- (3.93,2.28);
\filldraw[radius=3pt,color=darkgreen] (3.93,2.28) circle;
\draw[thick,dashed,color=red] (0,2.28) node[anchor=east]{$\frac{1}{2}$} -- (2.75,2.28);
\draw[thick,dashed,color=darkgreen]  (2.75,2.28) -- (3.93,2.28);
\draw[thick,dashed,color=red] (2.75,0)  -- (2.75,2.28);
    \draw (-0.3,6) node{{$1$}};     \draw (0,-0.3) node{{$0$}};     \draw[color=red] (2.75,-0.6) node{{$\frac{\log2}{\delta_A}$}};
\end{tikzpicture}
\par\end{centering}
\caption{A mean preserving spread of revelation time\label{fig:A-mean-preserving}}
\end{figure}
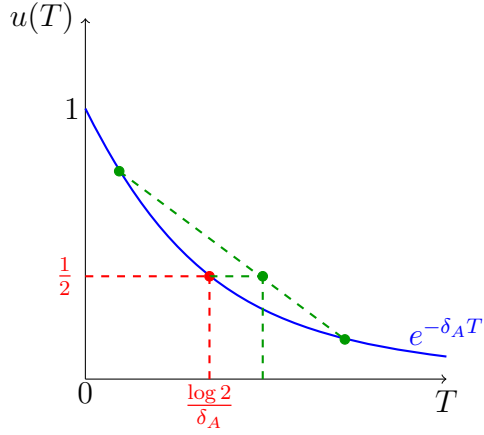

Therefore, the optimal time is $\hat{T}=\frac{\log2}{\delta_{A}}$.
The convexity of the function $e^{-\delta_{A}T}$ in $T$ implies
that a small mean-preserving spread of $\hat{T}$ increases the expected
value for the agent at all times while keeping $\mathbb{E}\left(T\right)$
unchanged. Moreover, if the spread is small enough, the agent is willing
to remain engaged. This implies that it can be accompanied by an increase
in all values of $T$, consequently increasing the principal's value.
This is depicted in Figure \ref{fig:A-mean-preserving}.

Such a perturbation is always feasible when the incentive constraint
is slack for any interval of time $t$, as it can be implemented within
that interval. Consequently, in the optimal solution, the incentive
constraint must bind for all values of $t$. A multiplication of \ref{eq: ICex1}
by $1-F\left(t\right)$ and differentiation with respect to $t$ yields:
\[
-e^{-\delta_{A}t}f\left(t\right)=-\frac{1}{2}e^{-\delta_{A}t}f\left(t\right)-\delta e^{-\delta_{A}t}\left(1-F\left(t\right)\right)
\]
which in turn implies that $F\left(t\right)=1-e^{-\delta_{A}t}$.
In other words, the information is revealed according to a Poisson
process with arrival rate $\delta_{A}$.

The key property driving this result is the convexity of the agent's
payoff function with respect to the principal's payoff. To see this,
consider the case in which the principal discounts future engagements
at rate $\delta_{P},$ where $0\leq\delta_{A}<\delta_{P}$, i.e.,
the principal's payoff is given by $u_{P}=\int_{0}^{T}e^{-\delta_{P}t}dt$,
while the agent's payoff remains as before. We can write the agent's
payoff as a function of the payoff of the principal:
\begin{align*}
u_{P} & =\frac{1-e^{-\delta_{P}T}}{\delta_{P}}\rightarrow T=\frac{-\log\left(1-\delta_{P}u_{P}\right)}{\delta_{P}}\rightarrow u_{A}=\frac{1}{2}\left(1-\delta_{P}u_{P}\right)^{\frac{\delta_{A}}{\delta_{P}}}\left(1+\mathbf{1}\left[\text{informed by }T\right]\right).
\end{align*}

Since $u_{P}$ is an increasing function $T$, choosing a distribution
of $T$, $F\left(T\right)$, is equivalent to choosing a distribution
of $u_{P}$, $G\left(u_{P}\right)$. Thus, the principal's problem
of delivering the information is given by:
\begin{align*}
 & \max_{G\in\mathcal{G}}\mathbb{E}_{G}\left[u_{P}\right],\\
 & \text{s.t. }\mathbb{E}_{G}\left[\left(1-\delta_{P}u_{P}\right)^{\frac{\delta_{A}}{\delta_{P}}}|u_{P}>v\right]\ge\frac{1}{2}\left(1-\delta_{P}v\right)^{\frac{\delta_{A}}{\delta_{P}}},\,\forall v\in\left[0,1\right].
\end{align*}
The key distinction between this case and the previous one lies in
the relationship between $\delta_{P}$ and $\delta_{A}$. Since $\delta_{P}>\delta_{A}$,
the function $\left(1-\delta_{P}u_{P}\right)^{\frac{\delta_{A}}{\delta_{P}}}$
is decreasing and concave. As a result, starting from any distribution
of $u_{P}$, a change to revelation at $\hat{T}$ given by
\[
\frac{1-e^{-\delta_{P}\hat{T}}}{\delta_{P}}=\mathbb{E}_{G}\left[u_{P}\right],
\]
 does not change the payoff of the principal while it increases the
payoff of the agent at all times. This implies that the best strategy
of the principal is to reveal the information with certainty at $\hat{T}=-\log2/\delta_{A}$.
This perturbation is depicted in Figure \ref{fig:A-mean-preserving-contraction}
for a two point distribution of $u_{P}$.
\begin{center}
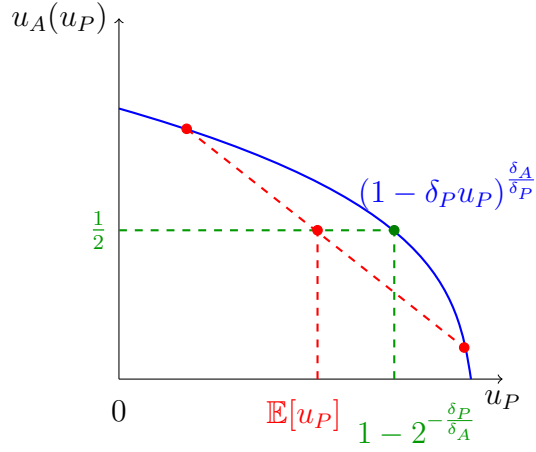
\begin{figure}
\begin{centering}
\begin{tikzpicture}[scale=0.6]
\centering
\draw[->] (0,0) -- (8.5,0) node[anchor=north] {$u_P$};     \draw[->] (0,0) -- (0,8) node[anchor=east] {$u_A(u_P)$};
\draw[thick,color=blue] (0,6) .. controls (7,4) and (7.5,2) .. (7.8,0);
\draw (7.3,3.5) node[anchor=south,scale=1,color=blue]{$(1-\delta_P u_P)^{\frac{\delta_A}{\delta_P}}$};    
\draw[color=darkgreen] (6.6,-1) node{{$1-2^{-\frac{\delta_P}{\delta_A}}$}};     \draw (0,-0.7) node{{$0$}};
\draw[thick,dashed,color=darkgreen] (0,3.3) node[anchor=east]{$\frac{1}{2}$} -- (6.1,3.3);
\draw[thick,dashed,color=darkgreen] (6.1,0)  -- (6.1,3.3);
\filldraw[radius=3pt,color=ao] (6.1,3.3) circle;
\draw[thick,dashed,color=red] (1.5, 5.55) -- (7.65,0.7);     
\filldraw[radius=3pt,color=red] (1.5, 5.55) circle;     \filldraw[radius=3pt,color=red] (7.65,0.7) circle;
\draw[color=red] (4.1,-.8) node{$\mathbb{E}[u_P]$};
\draw[thick, dashed, color=red] (4.4,0) -- (4.4,3.3);
\filldraw[radius=3pt,color=red] (4.4,3.3) circle;
\end{tikzpicture}
\par\end{centering}
\caption{A mean-preserving contraction of Principal's payoff \label{fig:A-mean-preserving-contraction}}
\end{figure}
\par\end{center}

The key distinction between these the two scenarios lies in relative
changes in the cost of engagement for the agent relative to its marginal
benefit for the principal. The value of extending engagement from
$T$ to $T+dT$ for the principal is $e^{-\delta_{P}T}dT$, while
the cost of postponing revelation for the agent is $\delta_{A}e^{-\delta_{A}T}dT$.
We define the ratio of these costs to benefits as the \emph{Marginal
Cost of Engagement} (MCE):
\[
\text{MCE}=\delta_{A}e^{-\left(\delta_{A}-\delta_{P}\right)T}.
\]

In the first case, when $\delta_{P}=0<\delta_{A}$, the MCE decreases
over time. Loosely speaking, the agent becomes relatively more patient
as engagement duration increases. Thus, the principal benefits from
divulging some information early on to incentivize the agent to wait
until later periods when maintaining engagement becomes easier.

In contrast, in the second case, when $0<\delta_{A}<\delta_{P}$,
the MCE increases over time, suggesting that the agent becomes relatively
less patient. Hence, early partial revelation of information is not
beneficial for the principal since to keep the agent engaged later,
information must be released faster. In the rest of the paper, we
illustrate how these two strategies can be applied to the more complex
information and utility structure.

\section{Communication under Belief Agreement\label{sec:Optimal-Communication-under}}

In this section, we analyze optimal communication strategies when
the agent and the principal share the same prior. This implies that
$\ell=1$ and that the agent does not have any bias towards any state
relative to the principal. This shared prior setting serves as a baseline
for our subsequent extensions.

\subsubsection*{More Patient Agent}

When the agent is more patient than the principal ($\delta_{P}>\delta_{A}$)
and they share the same prior, the optimal disclosure strategy takes
a simple form. Similar to the example in section \ref{sec:A-Simple-Example},
in this case, optimal information revelation occurs abruptly. This
result is formalized in the following proposition:
\begin{prop}
\label{prop:When--is}When $\delta_{P}>\delta_{A}$ and $\mu_{P}=\mu_{A}$,
the optimal solution of the principal's problem is 
\begin{align*}
G_{A,1}\left(t\right) & =\mu_{A}\mathbf{1}\left[t<t^{*}\right]\\
G_{A,0}\left(t\right) & =\left(1-\mu_{A}\right)\mathbf{1}\left[t<t^{*}\right]\\
e^{-\delta_{A}t^{*}}v\left(1\right) & =v\left(\mu_{A}\right)
\end{align*}
\end{prop}
In words, Proposition \ref{prop:When--is} states that optimal information
provision, in the case of concave discounting or when the Marginal
Cost of Engagement (MCE) increases over time, leads to abrupt revelation
of information at $t^{*}$. At $t=0$, the agent is indifferent between
this revelation at $t^{*}$, and no revelation at all or not participating.
The intuition behind this result is that when MCE increases with time,
any gradual information revelation before $t^{*}$ must be accompanied
by even faster revelation later to maintain the agent's engagement.
Hence, optimal revelation is abrupt.

\subsubsection*{More Patient Principal}

When the principal is more patient than the agent ($\delta_{P}<\delta_{A}$),
the endogeneity of beliefs leads to more complex dynamics in the optimal
provision of information compared to the simple example. As mentioned
above, without belief disagreement, the gain from information revelation
is maximized at $\mu_{A}=1/2$. However, the initial belief may differ
from half, i.e., $\mu_{A}(0)\neq1/2$. In such cases, the principal
aims to shift beliefs towards $1/2$. In this case, the principal
wants to shift the beliefs toward half. Additionally, this transition
should be smooth due to the strict convexity of the agent's payoff
in $\mu_{t,A}$ and the agent's preference for gradual belief changes. 

We show that it is optimal to first reveal the state towards which
initial beliefs are closer. For instance, if $\mu_{A}(0)>1/2$, $\omega=1$
should be revealed first. This revelation continues until beliefs
reach $\mu_{t,A}=1/2$. Once the agent reaches this point of maximum
uncertainty (and thus maximum engagement), the principal optimally
reveals both states at a rate that keeps the agent indifferent between
engaging and quitting. This strategy ensures that when the state is
not revealed, beliefs remain at the point of maximum engagement where
the gain from revelation is highest. We formalize these findings in
the following proposition:
\begin{prop}
\label{prop:When--is-1}When $\delta_{A}>\delta_{P}$ and the priors
of the agent and principal agree, the optimal solution to (\ref{eq:P})
has two phases:
\begin{enumerate}
\item A phase, $0\leq t<t^{*}$, where only state one (zero) is revealed
when $\mu_{P}>1/2$ ($\mu_{P}<1/2)$ and beliefs upon staying satisfy
\begin{align*}
\mu_{P}\left(0\right)>1/2:-\delta_{A} & =\frac{\mu_{P}'\left(t\right)}{v\left(\mu_{P}\left(t\right)\right)}\left[\frac{v\left(1\right)-v\left(\mu_{P}\left(t\right)\right)}{1-\mu_{P}\left(t\right)}-v'\left(\mu_{P}\left(t\right)\right)\right]\\
\mu_{P}\left(0\right)<1/2:\delta_{A} & =\frac{\mu_{P}'\left(t\right)}{v\left(\mu_{P}\left(t\right)\right)}\left[\frac{v\left(1\right)-v\left(\mu_{P}\left(t\right)\right)}{\mu_{P}\left(t\right)}+v'\left(\mu_{P}\left(t\right)\right)\right]
\end{align*}
\item A phase, $t\geq t^{*}$, where both states are revealed at rate $\lambda^{*}$
with
\[
\lambda^{*}=\delta_{A}\frac{v\left(1/2\right)}{v\left(1\right)-v\left(1/2\right)}
\]
 and beliefs are at the point of maximum uncertainty, $\mu_{P}=1/2$.
\end{enumerate}
\end{prop}
In this case, phase $1$ consists of revealing the state towards which
beliefs are more optimistic, at a time-varying poisson rate. For example,
if $\mu_{P}>1/2$ or state $\omega=1$ is more likely, during phase
1, there is only a possibility of receiving a signal if the true state
is $1$. If no signal was observed, both the agent and the principal
update their beliefs downwards. This process continues until beliefs
reach $1/2$, the point of maximum uncertainty. At this point, both
states are revealed at a rate $\lambda^{*}$. Figure \ref{fig:Evolution-of-beliefs}
illustrates this process.
\begin{flushleft}
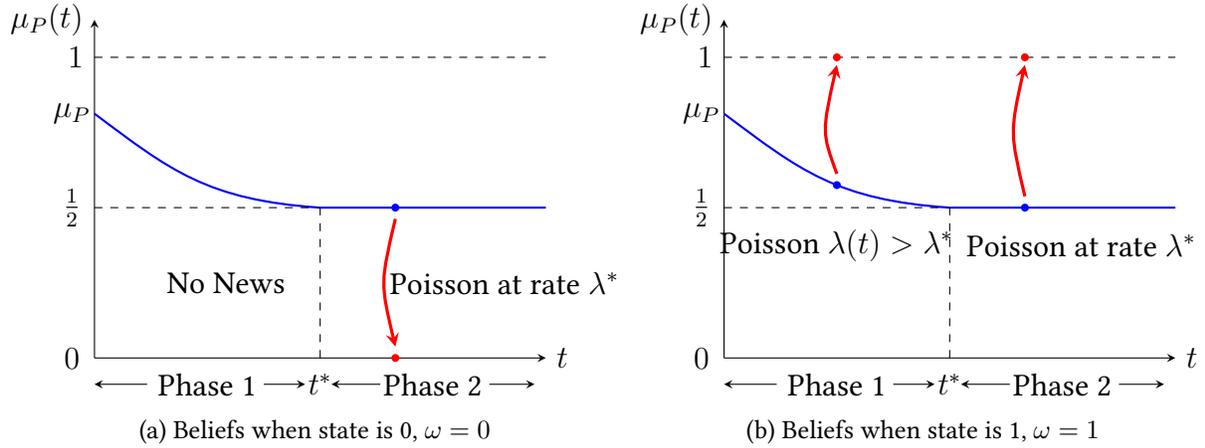
\begin{figure}
\subfloat[Beliefs when state is 0, $\omega=0$]{\begin{tikzpicture}[scale=.5]
\draw[-stealth] (0,0) -- (12,0) node[right] {$t$};
\draw[-stealth] (0,0) -- (0,9) node[left] {$\mu_P(t)$};
\draw[thick,color=blue] (0,6.5) .. controls (2,5) and (3,4.2)  .. (6,4);     \draw[thick,color=blue] (6,4) -- (12,4);
\draw[dashed] (0, 8) -- (12,8);
\draw[dashed] (0, 4) -- (6,4);     
\draw[dashed] (6, 0) -- (6,4);
\draw[-stealth] (1.4, -0.7) -- (0.,-0.7);
\draw[-stealth] (4.5, -0.7) -- (5.6,-0.7);
\draw[-stealth] (7.6, -0.7) -- (6.3,-0.7);
\draw[-stealth] (10.5, -0.7) -- (11.7,-0.7);
\draw (3,-0.7) node{{Phase 1}};
\draw (9,-0.7) node{{Phase 2}};
\draw (6,-0.7) node{{$t^*$}};
\draw (-0.5,8) node{{$1$}};
\draw (-0.6,0) node{{$0$}};
\draw (-0.6,4) node{{$\frac{1}{2}$}};
\draw (-0.6,6.5) node{{$\mu_P$}};
\draw (3.5,2) node{No News};
\fill[blue,radius = 3pt] (8,4) circle;
\fill[red,radius = 3pt] (8,0) circle;
\draw[very thick, -stealth,color=red] (8, 3.7) .. controls (7.6,2) .. (8,.3);
\draw (10.9,2.) node{Poisson at rate $\lambda^*$};
\end{tikzpicture}}\subfloat[Beliefs when state is 1, $\omega=1$]{\begin{tikzpicture}[scale=.5]
\draw[-stealth] (0,0) -- (12,0) node[right] {$t$};
\draw[-stealth] (0,0) -- (0,9) node[left] {$\mu_P(t)$};
\draw[thick,color=blue] (0,6.5) .. controls (2,5) and (3,4.2)  .. (6,4);     \draw[thick,color=blue] (6,4) -- (12,4);
\draw[dashed] (0, 8) -- (12,8);
\draw[dashed] (0, 4) -- (6,4);     
\draw[dashed] (6, 0) -- (6,4);
\draw[-stealth] (1.4, -0.7) -- (0.,-0.7);
\draw[-stealth] (4.5, -0.7) -- (5.6,-0.7);
\draw[-stealth] (7.6, -0.7) -- (6.3,-0.7);
\draw[-stealth] (10.5, -0.7) -- (11.7,-0.7);
\draw (3,-0.7) node{{Phase 1}};
\draw (9,-0.7) node{{Phase 2}};
\draw (6,-0.7) node{{$t^*$}};
\draw (-0.5,8) node{{$1$}};
\draw (-0.6,0) node{{$0$}};
\draw (-0.6,4) node{{$\frac{1}{2}$}};
\draw (-0.6,6.5) node{{$\mu_P$}};
\fill[blue,radius = 3pt] (3,4.6) circle;
\fill[red,radius = 3pt] (3,8) circle;
\draw[very thick, -stealth,color=red] (3, 4.9) .. controls (2.6,6) .. (3,7.8);
\draw (3.,3) node{Poisson  $ \lambda(t) > \lambda^*$ };
\fill[blue,radius = 3pt] (8,4) circle;
\fill[red,radius = 3pt] (8,8) circle;
\draw[very thick, -stealth,color=red] (8, 4.3) .. controls (7.6,6) .. (8,7.8);
\draw (9.5,3) node{Poisson at rate $\lambda^*$};
\end{tikzpicture}

}

\caption{Evolution of beliefs in each state when the agent is more impatient
and $\mu_{P}>1/2$\label{fig:Evolution-of-beliefs}}
\end{figure}
\par\end{flushleft}

The rate of information revelation and the length of phase 1 depends
on the curvature of the value function $v\left(\cdot\right)$. The
higher the curvature, the longer phase 1 becomes. This is due to the
strong preferences for smoothing of beliefs which implies that in
order to keep the agent engaged, it is optimal for beliefs to decline
more gradually. Consequently, this lengthens phase 1, albeit at the
cost of reducing the principal's value.

Proposition \ref{prop:When--is-1} highlights one of the key mechanisms
in our model: the principal wishes to reach the state of \textit{maximum
engagement} as fast as possible by revealing information. Under belief
agreement, this state is given by the state of maximum uncertainty
or $\mu_{A}=1/2$.

It's worth emphasizing that the value of $\delta_{P}$ does not affect
how the revelation happens beyond being gradual or abrupt. This invariance
is partly due to the principal's ability to capture all surplus in
both scenarios.

\section{Catering to the Bias\label{sec:Catering-to-the}}

In this section, we illustrate the ``catering to the bias'' force
that emerges when the principal and the agent disagree on their prior
beliefs. As we discussed in Section \ref{subsec:Solving-the-Model},
the problem of optimal communication under belief disagreement can
be formulated in terms of the distribution of beliefs from the principal's
perspective. While our primary focus is on interpreting the model
with differing priors, this framework can be equivalently interpreted
as a scenario where the principal and agent share the same prior,
but the agent exhibits a preference for information acquisition in
one state over the other--a form of confirmation bias.

\subsubsection*{More Patient Agent\label{subsec:Impatient-Principal}}

We begin our analysis of optimal communication under belief disagreement
by examining the case where the agent is more patient than the principal
$(\delta_{A}<\delta_{P})$. Recall that with shared priors, an impatient
principal would opt to reveal all information abruptly. However, we
demonstrate that this strategy changes when priors differ.

Consider the scenario where $\ell<1$, indicating that the agent believes
$\omega=0$ is more likely than the principal does.\footnote{The case of $\ell>1$ is the mirror of this case.}
In this case, the agent puts a higher weight on $G_{P,0}\left(t\right)$
than principal, as shown in Equation (\ref{eq: Obedience}). This
implies that the agent attributes greater value to information about
$\omega=0$ than the cost perceived by the principal. The principal,
believing $\omega=0$ to be relatively less likely, perceives this
cost as low and considers it unlikely to result in a high probability
of exit. This disparity creates an incentive for the principal to
first reveal the state towards which the agent is biased.

However, despite this logic, the principal remains less patient than
the agent, and forces favoring postponed revelation persist in this
case. This tension results in an optimal strategy comprising an initial
phase of no information revelation, followed by a phase of catering
to the bias. The following proposition characterizes the optimal information
revelation strategy in this case:
\begin{prop}
\label{thm2: Bias}Suppose that $\delta_{P}>\delta_{A}$ and $\ell<1$.
Then the solution to (\ref{eq:P}) consists of two phases and two
instantaneous revelations times:
\begin{enumerate}
\item In phase 1, $t\in\left[0,t_{1}^{*}\right)$, no information is revealed.
\item At $t_{1}^{*}$, $\omega=0$ is revealed with a positive probability.
\item In phase 2, $t\in\left[t_{1}^{*},t_{2}^{*}\right]$, $\omega=0$ is
revealed gradually according to a Poisson process at a rate such that
the agent's beliefs satisfy the following ODE:
\[
\delta_{A}=\frac{\mu_{A}'\left(t\right)}{\mu_{A}\left(t\right)}\frac{v\left(1\right)-v\left(\mu_{A}\left(t\right)\right)+v'\left(\mu_{A}\left(t\right)\right)\mu_{A}\left(t\right)}{v\left(\mu_{A}\left(t\right)\right)}.
\]
\item At $t_{2}^{*}$, $\omega=1$ is revealed such that $\mu_{A}\left(t_{2}^{*}\right)=1$.
\end{enumerate}
Moreover, the length of phase 2 is always positive, i.e., $t_{1}^{*}<t_{2}^{*}$.
\end{prop}
The dynamics implied by Proposition \ref{thm2: Bias} are depicted
in Figure \ref{fig:Bias}. The optimal communication strategy consists
of two distinct phases: 1. An initial phase where no information is
provided, and thus beliefs stay unchanged. 2. A subsequent phase that
begins with an abrupt, but not certain, revelation of $\omega=0,$
followed by a gradual revelation of $\omega=0$ according to a Poisson
process at a rate determined by the ODE specified in the proposition.
This phase concludes when the belief reaches 1, at which point the
true state is fully revealed. Furthermore, in the second phase, news
arrive increasingly faster over time as the poisson revelation rate
increases over time.

Figure \ref{fig:Bias} illustrates these dynamics under two scenarios:
Sub-figure (a) depicts the case where the true state is $\omega=0$.
In this scenario, information is revealed either abruptly at the end
of phase 1 or gradually revealed during phase 2. Sub-figure (b) shows
the case where the true state is $\omega=1$. Here, all information
is abruptly revealed at $t_{2}^{*}$.
\begin{flushleft}
\begin{figure}
\begin{raggedright}
\subfloat[Beliefs when $\omega=0$]{\begin{tikzpicture}[scale=.5]
\draw[-stealth] (0,0) -- (12,0) node[right] {$t$};
\draw[-stealth] (0,0) -- (0,9) node[left] {$\mu_A(t)$};
\draw[thick,color=blue] (0,2.5) -- (5,2.5);
\draw[thick,color=blue] (5,2.5) -- (5,4.5);
\draw[thick,color=blue] (5,4.5) .. controls (8,6) and (9,7.5)  .. (10,8);    
\draw[thick,color=blue] (10,8) -- (12,8);  
\draw[dashed] (0, 8) -- (12,8);
\draw[dashed] (5,0) -- (5,8);
\draw[-stealth] (1., -0.7) -- (0.,-0.7);
\draw[-stealth] (4., -0.7) -- (4.6,-0.7);
\draw[-stealth] (6.2, -0.7) -- (5.3,-0.7);
\draw[-stealth] (9.3, -0.7) -- (10.,-0.7);
\draw (2.5,-0.7) node{{Phase 1}};
\draw (7.7,-0.7) node{{Phase 2}};
\draw (5,-0.7) node{{$t_1^*$}};
\draw (-0.5,8) node{{$1$}};
\draw (-0.7,0) node{{$0$}};
\draw (-0.7,2.5) node{{$\mu_A$}};
\draw (2.3,1.8) node{No News};
\fill[blue,radius = 3pt] (7,5.7) circle;
\fill[red,radius = 3pt] (7,0) circle;
\draw[very thick, -stealth,color=red] (7, 5.4) .. controls (6.6,2) .. (7,.3);
\draw (10.3,2.) node{Poisson revelation};
\end{tikzpicture}}\subfloat[Beliefs when $\omega=1$]{\begin{tikzpicture}[scale=.5]
\draw[-stealth] (0,0) -- (12,0) node[right] {$t$};
\draw[-stealth] (0,0) -- (0,9) node[left] {$\mu_A(t)$};
\draw[thick,color=blue] (0,2.5) -- (5,2.5);
\draw[thick,color=blue] (5,2.5) -- (5,4.5);
\draw[thick,color=blue] (5,4.5) .. controls (8,6) and (9,7.5)  .. (10,8);    
\draw[thick,color=blue] (10,8) -- (12,8);  
\draw[dashed] (0, 8) -- (12,8);
\draw[dashed] (5,0) -- (5,8);
\draw[dashed] (10,0) -- (10,8);
\draw[-stealth] (1., -0.7) -- (0.,-0.7);
\draw[-stealth] (4., -0.7) -- (4.6,-0.7);
\draw[-stealth] (6.2, -0.7) -- (5.3,-0.7);
\draw[-stealth] (9., -0.7) -- (9.7,-0.7);
\draw (2.5,-0.7) node{{Phase 1}};
\draw (7.5,-0.7) node{{Phase 2}};
\draw (5,-0.7) node{{$t_1^*$}};
\draw (10,-0.7) node{{$t_2^*$}};
\draw (-0.5,8) node{{$1$}};
\draw (-0.7,0) node{{$0$}};
\draw (-0.7,2.5) node{{$\mu_A$}};
\draw (2.3,1.8) node{No News};
\draw (7.3,1.8) node{No News};
\fill[blue,radius = 3pt] (10,8) circle;
\end{tikzpicture}

}
\par\end{raggedright}
\caption{Catering to the bias with an Impatient Principal\label{fig:Bias}}
\end{figure}
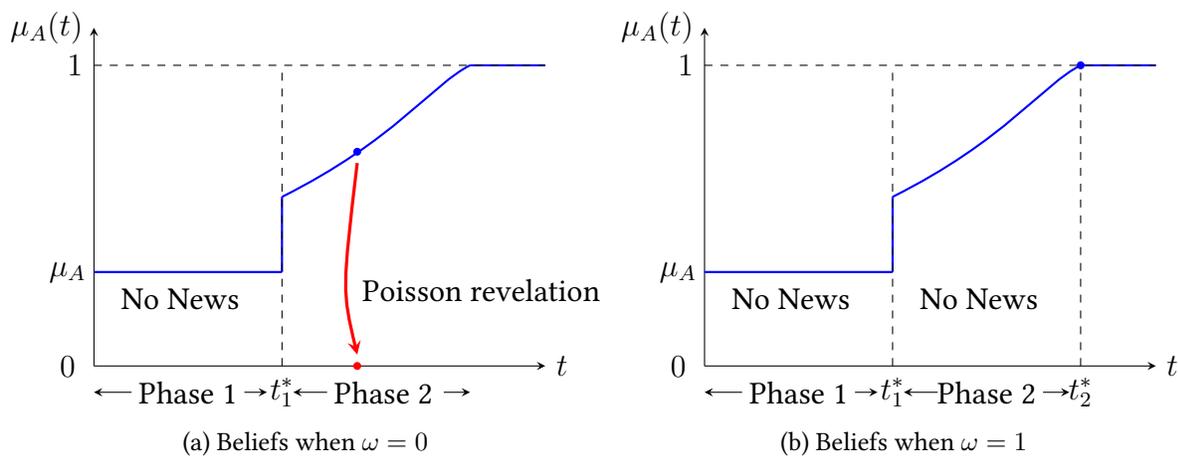
\par\end{flushleft}

Proposition \ref{thm2: Bias} illustrates the ``\textit{catering-to-the-bias}''
force present in our model. Due to belief disagreement, the optimal
information provision problem becomes equivalent to one in which the
principal prefers the agent to stay engaged in state $\omega=1$ more
than in state $\omega=0$. With bias in beliefs, the solution is to
reveal the state $\omega=0$ gradually. This gradual revelation of
$\omega=0$ serves as a low cost method of keeping the agent engaged
when the true state is $\omega=1$, exploiting the agent's biased
beliefs towards $\omega=0$. This allows the principal to postpone
the revelation of $\omega=1$ as late as possible.

While the statement that $t_{1}^{*}<t_{2}^{*}$ implies that catering
to the bias always occurs under the optimal engagement policy, it
does not rule out the possibility of phase 1 having zero length (i.e.,
$t_{1}^{*}=0$). The following corollary provides conditions under
which catering to the bias becomes so extreme that it eliminates the
no-news phase entirely. Notably, it indicates that phase 1 is more
likely to occur when belief disagreement is small.
\begin{cor}
\label{cor: 1} Given the parameters of the model, there exists $\underline{\mu}_{P}$
and $\overline{\mu}_{P}$ such that when $\mu_{A}\notin\left[\underline{\mu}_{P},\overline{\mu}_{P}\right]$,
phase 1 has zero length.
\end{cor}

\subsubsection*{More Patient Principal\label{subsec:Patient-Principal}}

Suppose now that the principal is more patient than the agent ($\delta_{A}>\delta_{P}$).
Recall that under belief agreement, the principal reveals the state
gradually, with the more likely state being revealed first, to maximize
the time the agent spends at a belief with maximum uncertainty. With
belief disagreement, the spirit of the strategy of driving the agent
to a stationary belief stays the same, yet it takes a form of ``catering
to the bias'' as discussed earlier and this stationary belief does
not maximize engagement.

More specifically, we start by showing that the point of maximum uncertainty
$(\mu_{A}=1/2)$ is not going to be steady state anymore due to the
disagreement in beliefs. To see this, suppose that $\mu_{A}=1/2$
and $\mu_{P}<1/2$, for this point to be steady state, both states
should be revealed at a same rate $\lambda$, for the agent's belief
to remain at $\mu_{A}=1/2$. The variable $\lambda$ is chosen to
keep the agent indifferent between staying engaged and quitting at
any point in time. Next, we show a profitable deviation for the principal.

Consider an alternative strategy where the principal reveals only
$\omega=1$ for a short interval of length $dt$ at a rate $2\lambda$,
before reverting to symmetric Poisson revelation. This brief asymmetric
revelation lowers the agent's belief slightly. While the symmetric
Poisson rate should increase to maintain indifference, the change
in the agent's payoff is only second-order in $dt$ due to $v'(1/2)=0$.
Thus, to first order, $\lambda$ in the symmetric revelation case
remains unchanged. However, since the principal believes the probability
of $\omega=1$ is $\mu_{P}<1/2$, revealing only disclosing $\omega=1$
for the period between 0 and $dt$ has first order gain for the principal.
This implies that keeping the agent at $\mu_{A}=1/2$ is not optimal
for the principal.

We can continue with the above heuristic argument, to find a steady
state pair of $\left(\mu_{A}^{*},\mu_{P}^{*}\right)$. To keep the
agent indifferent between quitting and staying, her payoff from quitting
at any point in time must equal $v\left(\mu_{A}\right)$. Under this
strategy, the agent's payoff from staying engaged is:
\begin{equation}
\frac{\lambda}{\lambda+\delta_{A}}v\left(1\right)=v\left(\mu_{A}\right).\label{eq: Pa}
\end{equation}

The principal's payoff from this strategy is $1/\left(\lambda+\delta_{P}\right)$.
Now, consider a deviation where the principal reveals state $\omega=1$
at rate $q$ over the time period $\left[0,\Delta\right]$ and reverts
to gradual revelation of both states at rate $\lambda+d\lambda$ after
$t=\Delta$. The agent's belief at the end of period $\Delta$ is:
\[
\mu_{A}+d\mu_{A}=\mu_{A}-\mu_{A}\left(1-\mu_{A}\right)q\Delta+o\left(\Delta^{2}\right).
\]
Thus, the principal can choose $\lambda+d\lambda$ such that:
\[
\frac{d\lambda}{\left(\lambda+\delta_{A}\right)^{2}}v\left(1\right)=v'\left(\mu_{A}\right)d\mu_{A}.
\]
The agent's payoff from this strategy, up to a second-order approximation,
is: 
\[
q\mu_{A}\left[v\left(1\right)-v\left(\mu_{A}\right)\right]\Delta-\delta_{A}v\left(\mu_{A}\right)\Delta+\frac{d\lambda}{\left(\lambda+\delta_{A}\right)^{2}}v\left(1\right).
\]
Here, the first term represents the gain from revealing $\omega=1$
with probability $q\Delta\mu_{A}$ over the time interval $\left[0,\Delta\right]$.
The second term is the change in value from discounting, while the
last term is the change in the agent's value due to the belief change.
Equating this to $v\left(\mu_{A}\right)$, implies a lower bound on
$q$:

\[
q=\frac{\delta_{A}}{\mu_{A}}\frac{v\left(\mu_{A}\right)}{v\left(1\right)-v\left(\mu_{A}\right)-\left(1-\mu_{A}\right)v'\left(\mu_{A}\right)}.
\]
The change in the principal's payoff from this deviation, to first-order
approximation, is: 
\[
-\frac{1}{\lambda+\delta_{P}}\mu_{P}q\Delta-\frac{d\lambda}{\left(\lambda+\delta_{P}\right)^{2}}+\frac{\lambda}{\lambda+\delta_{P}}\Delta.
\]
The first term represents losses from revealing $\omega=1$ over $\left[0,\Delta\right]$.
The second term is the change in payoff from postponing symmetric
revelation at rate $\lambda+d\lambda$, and the last term is the gain
from abandoning symmetric revelation at rate $\lambda$ over the interval
$\left[0,\Delta\right]$. While revelation of $\omega=1$ at rate
$q$ benefits the agent, it is costly for the principal, with this
cost decreasing for lower $\mu_{P}$. Thus, for sufficiently high
$\mu_{P}$, this deviation is not profitable, defining a threshold
$\mu_{P}^{*}$:
\[
\mu_{P}^{*}\left(\mu_{A}\right)=\mu_{A}+\frac{\mu_{A}\left(1-\mu_{A}\right)v'\left(\mu_{A}\right)}{v\left(\mu_{A}\right)+\frac{\delta_{P}}{\delta_{A}-\delta_{P}}v\left(1\right)}.
\]
For $\mu_{P}<\mu_{P}^{*}\left(\mu_{A}\right)$, an initial phase of
revealing $\omega=1$ is optimal with both $\mu_{A}\left(t\right)$
and $\mu_{P}\left(t\right)$ decreasing until the above threshold
is reached. This is depicted in Figure \ref{fig:Catering to the Bias}.
As time progresses, the likelihood ratios, $\ell=\mu_{A}/\left(1-\mu_{A}\right)\times\left(1-\mu_{P}\right)/\mu_{P}$,
remains constant. Eventually, the above holds, at which point both
states are revealed symmetrically and the beliefs conditional on not
quitting remain constant. In contrast, when $\mu_{P}>\mu_{P}^{*}\left(\mu_{A}\right)$,
the opposite happens where $\omega=0$ is revealed first until the
steady state levels are reached.
\begin{flushleft}
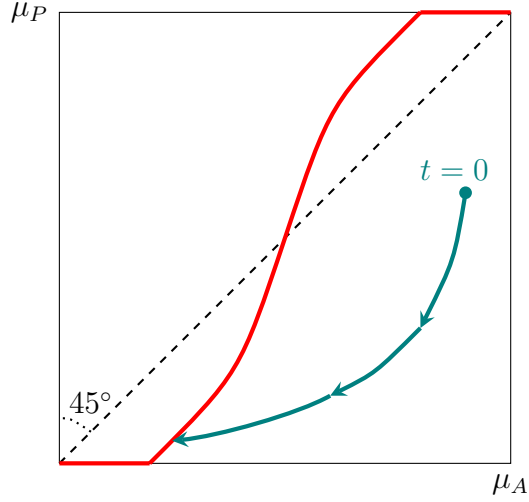
\begin{figure}
\begin{centering}
\begin{tikzpicture}[scale=0.6]
\centering
\draw (0,0) -- (10,0) node[anchor=north] {$\mu_A$} -- (10,10) -- (0,10) node[anchor=east] {$\mu_P$} -- (0,0);
\draw[thick,dashed] (0,0) -- (10,10);
\draw[ultra thick, red] (2,0) .. controls (4,2) .. (5,5) .. controls (6,8) .. (8,10);
\draw[ultra thick, red] (0,0) -- (2,0);
\draw[ultra thick, red] (8,10) -- (10,10);
\draw[thick,dotted] (0.7,0.7) .. controls (.25,1) .. (0,1);
\draw (.7,1.35) node{{$45^\circ$}};
\draw[ultra thick, -stealth, teal] (9,6.) .. controls (8.75,4.5) .. (8,3);
\draw[ultra thick, -stealth, teal] (8,3) .. controls (7.,2.) .. (6,1.5);
\draw[ultra thick, -stealth, teal] (6,1.5) .. controls (5.3,1.1) and (3.5,.6) .. (2.5,.5);
\fill[teal,radius=4pt] (9,6) circle;
\draw[teal] (8.75,6.5) node{$t=0$};
\end{tikzpicture}
\par\end{centering}
\caption{Catering to the Bias with a Patient Principal \label{fig:Catering to the Bias}}
\end{figure}
The threshold defined by the above is an increasing function of $\mu_{A}$
provided that its value is between 0 and 1. For high values of $\mu_{A}$
the threshold can exceed 1, and for low values, it can be less than
0. Thus, we can define the threshold as follows:
\par\end{flushleft}

\begin{equation}
\mu_{P}^{*}\left(\mu_{A}\right)=\max\left\{ \min\left\{ \mu_{A}+\frac{\mu_{A}\left(1-\mu_{A}\right)v'\left(\mu_{A}\right)}{v\left(\mu_{A}\right)+\frac{\delta_{P}}{\delta_{A}-\delta_{P}}v\left(1\right)},1\right\} ,0\right\} \label{eq: mups}
\end{equation}

Given this discussion, we have the following result:
\begin{prop}
\label{prop:Suppose-that-}Suppose that $\delta_{A}>\delta_{P}$.
Then there exists a threshold $\mu_{P}^{*}\left(\mu_{A}\right)$ given
by (\ref{eq: mups}) such that the solution to (\ref{eq:P}) consists
of two phases:
\begin{enumerate}
\item If $\mu_{P}>\mu_{P}^{*}\left(\mu_{A}\right)$, in phase 1 only the
state $\omega=0$ is gradually revealed such that the agent's beliefs
satisfy
\begin{equation}
\delta_{A}=\frac{\mu'_{A}\left(t\right)}{\mu_{A}\left(t\right)}\frac{v\left(1\right)-v\left(\mu_{A}\left(t\right)\right)+v'\left(\mu_{A}\left(t\right)\right)\mu_{A}\left(t\right)}{v\left(\mu_{A}\left(t\right)\right)}\label{eq: ODEmu0}
\end{equation}
\item If $\mu_{P}^{*}\left(\mu_{A}\right)>\mu_{P}$, in phase 1 only the
state $\omega=1$ is gradually revealed such that the agent's beliefs
satisfy
\begin{align}
-\delta_{A} & =\frac{\mu'_{A}\left(t\right)}{1-\mu_{A}\left(t\right)}\frac{v\left(1\right)-v\left(\mu_{A}\left(t\right)\right)-v'\left(\mu_{A}\left(t\right)\right)\left(1-\mu_{A}\left(t\right)\right)}{v\left(\mu_{A}\left(t\right)\right)}\label{eq:ODEmu1}
\end{align}
\item In phase 2, when $\mu_{P}^{*}\left(\mu_{A}\right)=\mu_{P}$, both
states are gradually revealed according to a Poisson process with
intensity $\lambda$ which satisfies $\frac{\lambda}{\lambda+\delta_{A}}v\left(1\right)=v\left(\mu_{A}\right)$.
\end{enumerate}
\end{prop}
Proposition \ref{prop:Suppose-that-} and Figure \ref{fig:Catering to the Bias}
highlight the key property of the catering to the bias effect. Specifically,
for very high and very low values of $\mu_{A}$ where $\mu_{P}^{*}$
is 1 or 0, respectively, our result implies that irrespective of the
principal's initial belief, $\mu_{P}$, it is always optimal to first
reveal the state towards which the agent is biased.

We should note that catering to the bias can become extreme, such
that no information about one of the states ever arrives before the
end of communication. This occurs when the function $\mu_{P}^{*}\left(\mu_{A}\right)$
at $\mu_{A}=0,1$ is not flat. It implies that for high or low enough
values of $\ell$, the constant $\ell$ curves -- loci of points
with the same relative likelihood ratio -- only intersect $\mu_{P}^{*}\left(\mu_{A}\right)$
at $\mu_{A}=0,1$. This implies that catering to the bias occurs during
the entire communication process. We thus have the following proposition:
\begin{prop}
Suppose that $\delta_{A}v\left(0\right)+\left(\delta_{A}-\delta_{P}\right)v'\left(0\right)>0$.
Then for all values of initial beliefs $\mu_{A},\mu_{P}$ such that
$\ell=\frac{\mu_{A}}{1-\mu_{A}}\frac{1-\mu_{P}}{\mu_{P}}\leq1+\frac{\left(\delta_{A}-\delta_{P}\right)v'\left(0\right)}{\delta_{A}v\left(0\right)}$,
only $\omega=1$ is revealed during the course of communication. Accordingly,
when $\frac{1}{\ell}\geq1+\frac{\left(\delta_{A}-\delta_{P}\right)v'\left(0\right)}{\delta_{A}v\left(0\right)}$,
only $\omega=0$ is revealed during the course of communication.
\end{prop}

\section{\label{sec:personalized}Non-Personalized vs. Personalized News}

In this section, we extend our general model to investigate the impact
of news personalization on information transmission. Specifically,
we consider a scenario where the agent's belief is privately known
and unobserved by the principal.

We analyze a variant of our model where the agent's belief is either
$\mu_{A}^{L}$ or $\mu_{A}^{H}$, with $\mu_{A}^{L}<\mu_{A}^{H}$,
while the principal's belief is $\mu_{P}$. Let $\alpha^{j}=\Pr\left(\mu_{A}^{j}\right),$
$j\in{\{L,H\}}$ denote the probability of the agent being of type
$j$. The agent discounts future payoffs at rate $\delta_{A}>0$,
while the principal's discount rate is $\delta_{P}<\delta_{A}$, thus
the principal is more patient than the agent.

Our goal in this section is to compare the personalized benchmark
of previous sections with a non-personalized case where all communications
are public and the agent type is privately known. As we will show,
while due to the presence of private information, engagement has to
necessarily be shorter, the principal can trade this off with quality
of information and thus still capture the full surplus from the agent.

\subsubsection*{General Structure of Optimal Communication }

Our key departure from the personalized news case is that communication
between the principal and agent is public and is thus not fully targeted.
This assumption implies that we can formulate optimal communication
simply as a pure recommendation mechanism. 

At first glance, this appears to be a complicated game with persistent
private information. Specifically, if an equilibrium calls for an
agent of type $j$ to quit at $t$, upon deviation and staying, she
persistently holds superior information over the principal, and her
future incentives need to be respected. However, due to the particular
principal's objective, such deviations need not be considered. That
is because if staying instead of quitting is profitable, the principal's
initial strategy is suboptimal since the principal prefers staying
to quitting. Thus, we can describe the best equilibrium for the principal
as a recommendation strategy that only needs to be obedient. This
leads to the following lemma:
\begin{lem}
\label{lem:The-best-equilibrium}The best equilibrium of the game
from the principal's perspective can be described by a signal structure
$\left(S_{\infty},\mathcal{F},\mathbb{P}_{P},\left\{ \mathcal{F}_{t,P}\right\} _{t\geq0}\right)$
together with a recommendation strategy for each type such that:
\begin{enumerate}
\item If type $j$ is recommended to quit following signal history $s^{t}$,
the value of staying engaged for $j$ is not higher than $v\left(\mu_{A}^{j}\left(s^{t}\right)\right)$,
\item If type $j$ is recommended to stay following signal history $s^{t}$,
the value of staying engaged for $j$ is not lower than $v\left(\mu_{A}^{j}\left(s^{t}\right)\right)$,
\end{enumerate}
where $\mu_{A}^{j}\left(s^{t}\right)$ is the agent of type $j$'s
belief induced by the signal structure $\left(S_{\infty},\mathcal{F},\mathbb{P}_{P},\left\{ \mathcal{F}_{t,P}\right\} _{t\geq0}\right)$.
\end{lem}
Lemma \ref{lem:The-best-equilibrium} allows us to significantly simplify
the principal's constrained optimization by only imposing obedience
constraint for each type, similar to those in Equation (\ref{eq: Obedience}). 

To characterize non-personalized optimal communication, we leverage
the solution of the personalized case. Lemma \ref{lem:The-best-equilibrium}
implies that once one type quits, the solution for the remaining type
reverts to the personalized solution. Recall that the relative likelihood
ratios, $\ell^{j}=\mu_{A}^{j}\left(1-\mu_{P}\right)/\mu_{P}\left(1-\mu_{A}^{j}\right)$,
remain constant. Hence, the principal's value in the personalized
game for type $j$ agent is $\alpha^{j}V(\mu_{P}(t),\ell^{j})$, as
in Equation (\ref{eq:P}). 

This formulation allows us to focus on the principal's strategy when
both agent types are engaged, as the principal's value upon type $-j$
quitting is $\alpha_{j}V(\mu_{P}(t),\ell^{j})$. Moreover, analogous
to the personalized case, the value for the remaining agent of type
$j$ will be $v(\mu_{A}^{j}(t))$, while the value for the quitting
agent of type $-j$ will be $v(\mu_{A}^{-j}(t))$. Thus, the payoff
of agent of type $j$ is always $v\left(\mu_{A}^{j}\left(t\right)\right)$
upon realization of a transition signal.

With two types, a recommendation mechanism needs at most 4 recommendations
per period, each associated with a pair of actions from the two types
of agents. Moreover, if both types of agents are recommended to quit
in a period, then an equivalent of Lemma \ref{lem:Suppose-that-Assumption}
holds: it is still optimal to fully reveal the state to keep them
engaged longer. However, if only one type is quitting, full revelation
might be sub-optimal as it could cause the other type to quit as well.
Let us define $\left\{ p_{\sigma}\left(t\right),\mu_{\sigma,P}\left(t\right)\right\} _{\sigma\in\Sigma}$
to be the distribution of posteriors induced by the realization of
$\sigma\in\Sigma$ for the principal at the beginning of phase 2,
i.e., conditional on transition.

The above discussion implies that the principal's strategy can be
thought of as having two phases:

1. \emph{Full Engagement Phase} (Phase 1): Both types are engaged
until a transition signal arrives. 

2. \emph{Partial Engagement Phase} (Phase 2): Transition to phase
2 happens when it is recommended that only one type stays. Each transitional
signal realization $\sigma$ is associated with $j(\sigma)\in{\left\{ L,H\right\} }$,
the type recommended to stay engaged. When recommending both agents
to quit, the game ends at the end of phase 1, while with recommending
only one type to stay we transition to phase 2. With one type engaged,
we revert to the personalized case. 

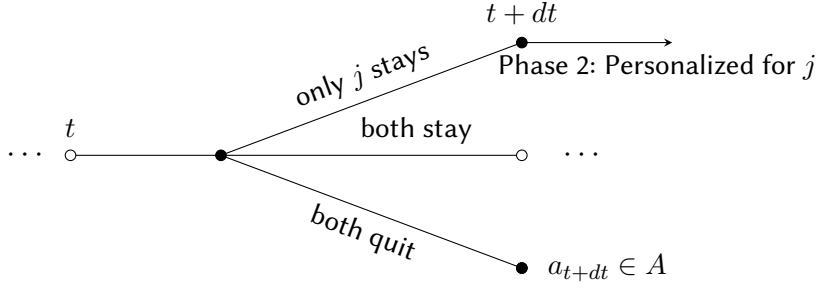
\begin{figure}
\centering{}\begin{tikzpicture} 
\sffamily{
\draw (-1,1) node[right]{$\cdots$};
\draw (0.07,1) -- (2,1) ;
\draw (2.,1) -- (6.,2.5)  node[midway, above,sloped] {\small only $j$ stays}; 
\draw (2.,1) -- (6.,1.)   node[pos=.65, above,sloped] {\small  both stay};
\draw (2.,1) -- (6.,-0.5)   node[midway, below,sloped] {\small both quit};
\draw[radius=2pt,black] (0,1) circle node[above=.1cm]{{\small $t$}};
\filldraw[radius=2pt,black] (6,2.5) circle node[above=.1cm]{\small $t+dt$};;
\filldraw[radius=2pt,black] (6,-0.5) circle;
\filldraw[radius=2pt,black] (2,1) circle;
\filldraw[radius=2pt,black,fill=white] (6.,1) circle;
\filldraw[radius=2pt,black] (6.,-.5) circle node[right=.2cm]{\small $a_{t+dt}\in A$};
\draw (6.4,1) node[right]{$\cdots$};
\draw[-stealth] (6,2.5) -- (8,2.5) node[pos=.9,below]{\small Phase 2: Personalized for $j$};
}
\end{tikzpicture} \caption{The two stage structure of optimal communication\label{fig:The-two-stage}}
\end{figure}

Figure \ref{fig:The-two-stage} depicts the two phases of communication. 

\subsubsection*{Stationary Engagement}

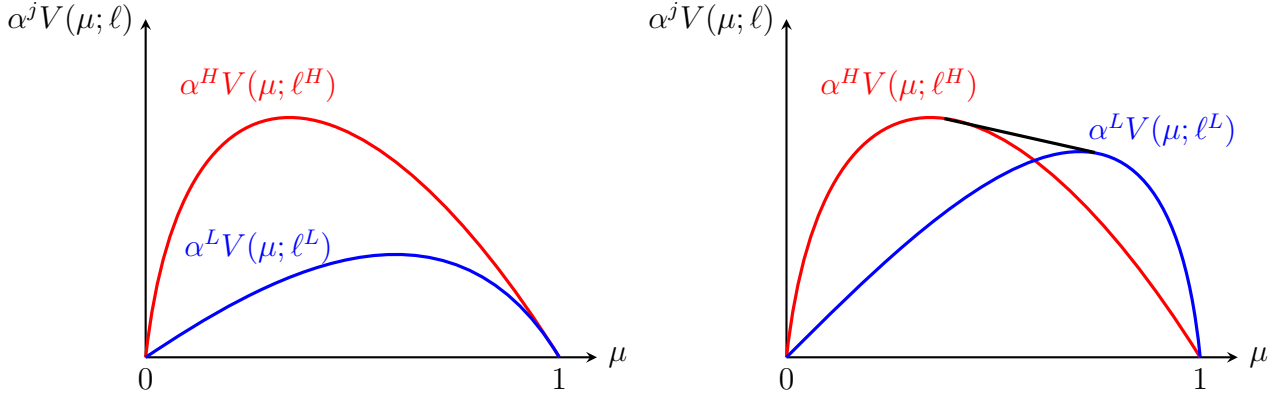
\begin{figure}
\begin{tikzpicture}[scale=.5]
\draw[-stealth,thick] (0,0) -- (12,0) node[right] {$\mu$};
\draw[-stealth,thick] (0,0) -- (0,9) node[left] {$\alpha^j V(\mu;\ell)$};
\draw[red,very thick] (0,0) .. controls (1,9) and (6,8) .. (11,0);
\draw[blue,very thick] (0,0) .. controls (3,2) and (8,5) .. (11,0);
\draw[red] (3,6.5) node[above]{$\alpha^H V(\mu;\ell^H)$};
\draw[blue] (3,2.2) node[above]{$\alpha^L V(\mu;\ell^L)$};
\draw (11,0) node[below]{$1$};
\draw (0,0) node[below]{$0$};
\end{tikzpicture}\begin{tikzpicture}[scale=.5]
\draw[-stealth,thick] (0,0) -- (12,0) node[right] {$\mu$};
\draw[-stealth,thick] (0,0) -- (0,9) node[left] {$\alpha^j V(\mu;\ell)$};
\draw[red,very thick] (0,0) .. controls (1,9) and (6,8) .. (11,0);
\draw[blue,very thick] (0,0) .. controls (4,4) and (10,10) .. (11,0);
\draw[red] (3,6.5) node[above]{$\alpha^H V(\mu;\ell^H)$};
\draw[blue] (10,5.3) node[above]{$\alpha^L V(\mu;\ell^L)$};
\draw (11,0) node[below]{$1$};
\draw (0,0) node[below]{$0$};
\draw[very thick] (4.2,6.35) -- (8.2,5.45);
\end{tikzpicture}

\caption{Value of the principal in phase 2\label{fig:Value-of-the}}
\end{figure}

Having established the above general structure and those of the relevant
incentive constraints, we can analyze optimal communication similar
to that in section \ref{sec:Catering-to-the}. Before providing this
characterization, we discuss one more property that will prove useful:
\begin{lem}
\label{lem: maxV} In any period at the beginning of phase 2 and for
any signal realization $\sigma$, $j\left(\sigma\right)=H$ if and
only if $\alpha^{H}V\left(\mu^{\sigma};\ell^{H}\right)\geq\alpha^{L}V\left(\mu^{\sigma};\ell^{L}\right)$
and vice versa. Moreover, it is sufficient to restrict the number
of transition signals,$\left|\Sigma\right|$, to at most 4.
\end{lem}
The first part of Lemma \ref{lem: maxV} is simply a result of the
fact that when the agent of type $j\in\left\{ L,H\right\} $ exits,
her payoff is always her outside option of $v\left(\mu_{A}^{j}\left(t\right)\right)$.
This implies that if a type is chosen to stay engaged, it should be
the type that delivers the higher value to the principal. Figure \ref{fig:Value-of-the}
depicts two possible cases that can occur. In the first case (left
panel), type $L$ is never chosen since their fraction $\alpha^{L}$
is small enough while in the second case, $L$ is chosen for high
beliefs and $H$ is chosen for low values. We can thus use $\hat{V}\left(\mu\right)$
instead of $\alpha^{j\left(\mu\right)}V\left(\mu;\ell^{j\left(\mu\right)}\right)$
in the principal's payoff, where
\[
\hat{V}\left(\mu\right)=\max\left\{ \alpha^{H}V\left(\mu;\ell^{H}\right),\alpha^{L}V\left(\mu;\ell^{L}\right)\right\} .
\]
Moreover, an application of Fenchel and Bunt's theorem (see \citet{kamenica2010bayesian}
or Theorem 1.3.7. in \citet{hiriart2004fundamentals}) implies that
we only need 4 signals.\footnote{This can be seen by considering the convex hull of the set $S=\left\{ \left(\mu_{P},\hat{V}\left(\mu_{P}\right),v\left(\mu_{A}^{H}\right),v\left(\mu_{A}^{L}\right)\right):\mu_{P}\in\left[0,1\right]\right\} \subset\mathbb{R}^{4}$
where $\mu_{A}^{j}/\left(1-\mu_{A}^{j}\right)=\ell^{j}\mu_{P}/\left(1-\mu_{P}\right)$.
This set is connected since any two convex combinations of these points
can be connected with a path which connects the weights in the appropriate
simplex that contains them. Fenchel and Bunt's theorem then implies
that any points in the convex hull can be written as a convex combination
of 4 points in $S$.}

To draw a parallel with the personalized model of section \ref{sec:Catering-to-the},
we can start by characterizing the point of long-run engagement or
the level of beliefs for the principal $\mu_{P}^{*}$, for which the
strategy of the principal is stationary. Note that a stationary strategy
for the principal involves an arrival rate of the transition signal
$\lambda^{*}$ from phase 1, and a distribution of posteriors, $\left\{ p_{\sigma}^{*},\mu_{\sigma}^{*}\right\} _{\sigma\in\Sigma}$.
Let $\mu_{P,E}$ be the beliefs at the beginning of phase 2 -- following
the realization of transition signal -- but before the realization
of $\sigma$ which determines which type stays. Bayes plausibility
implies that
\[
\mu_{P,E}=\sum_{\sigma\in\Sigma}p_{\sigma}^{*}\mu_{\sigma}^{*}
\]
Moreover, since $\mu_{P}$ is to stay constant over time, we must
have that $\mu_{P,E}^{*}=\mu_{P}^{*}$. Finally, an argument akin
to duality shown in Theorem \ref{thm: Luenberger} in the Appendix
implies that Lagrange multipliers $\Lambda_{j}\geq0$ should exist
so that $\left\{ p_{\sigma}^{*},\mu_{\sigma}^{*}\right\} _{\sigma\in\Sigma}$
should maximize
\begin{equation}
\sum_{\sigma\in\Sigma}p_{\sigma}\left[\hat{V}\left(\mu_{\sigma}\right)+\sum_{j=L,H}\Lambda_{j}v_{j}\left(\mu_{\sigma}\right)\right]\label{eq: concav}
\end{equation}
where $v_{j}\left(\mu\right)=\left(\ell^{j}\mu_{\sigma}+1-\mu_{\sigma}\right)v\left(\frac{\ell^{j}\mu_{\sigma}}{\ell^{j}\mu_{\sigma}+1-\mu_{\sigma}}\right)$
is a change in payoff function that allows us to write everything
from the perspective of the principal. Since $\mu_{P,E}$ satisfies
the Bayes plausibility above, the distribution of posteriors upon
transition to phase 2 is given by a concavification a la \citet{10.1257/aer.101.6.2590}
of the function $\hat{V}+\sum_{j=L,H}\Lambda_{j}v_{j}$.\footnote{Our problem is essentially a constrained Bayesian Persuasion problem
-- constrained by the agent's incentive constraints. For a related
paper, see \citet{doval2024constrained} for a general class of constrained
Bayesian Persuasion problems.} Let us refer to the concavification of the above as $\hat{V}^{\text{cav}}\left(\mu_{P,E};\left\{ \Lambda_{j}\right\} \right)$.
Finally, optimality of choice of transition probabilities from phase
1 and optimality $\mu_{P,E}$ implies that
\begin{align}
\delta_{P}\frac{\partial\hat{V}^{\text{cav}}}{\partial\mu_{P,E}}\left(\mu_{P}^{*};\Lambda_{L},\Lambda_{H}\right) & =\left(\delta_{A}-\delta_{P}\right)\sum_{j=L,H}\Lambda_{j}v_{j}'\left(\mu_{P}^{*}\right),\label{eq: Belief Smoothing}\\
1-\delta_{P}\hat{V}^{\text{cav}}\left(\mu_{P}^{*};\Lambda_{L}^{*},\Lambda_{H}^{*}\right) & =\left(\delta_{A}-\delta_{P}\right)\sum_{j=L,H}\Lambda_{j}v_{j}\left(\mu_{P}^{*}\right)\label{eq: OptimalTrans}
\end{align}
The above are cost and benefits of changes in belief at the start
of phase 2, $\mu_{P,E}$, and probability of exit from phase 1. In
the equation \ref{eq: Belief Smoothing}, the right-hand-side (RHS)
is the marginal cost of raising $\mu_{P}\left(t\right)$ and the left-hand-side
(LHS) is marginal benefit of increase $\mu_{P,E}$ in the future.
Note that an increase in future $\mu_{P,E}$ should be accompanied
by an increase in current $\mu_{P}$ since beliefs have to be a martingale.
One can think about this as a form of belief smoothing. Similarly,
in \ref{eq: OptimalTrans}, the left hand side is the benefit of lengthening
phase 1, while the right hand side is the incentive cost of doing
so. We summarize this in the following Proposition:
\begin{prop}
\label{prop:Let--be}The steady state level of belief for the principal
$\mu_{P}^{*}$ is either 0 or 1 and is achieved in finite time, or
$\mu_{P}^{*}\in\left(0,1\right)$ and $\lambda^{*},\Lambda_{L}^{*},\Lambda_{H}^{*}\geq0$
exists that satisfy: 

(1) The Belief Smoothing equation \ref{eq: Belief Smoothing} holds,

(2) The phase 1 optimality \ref{eq: OptimalTrans} hold, 

(3) The following incentive compatibility and complementary slackness
conditions are satisfied:
\begin{align*}
\frac{\lambda^{*}}{\lambda^{*}+\delta_{A}}\sum_{\sigma}p_{\sigma}^{*}v_{j}\left(\mu_{\sigma}^{*}\right) & \geq v_{j}\left(\mu_{P}^{*}\right),\text{with equality if }\Lambda_{j}^{*}>0.
\end{align*}
\end{prop}
The above Proposition allows us to characterize the steady state of
the non-personalized model. In the following example, we compare the
model with personalized and non-personalized news in order to shed
light on the key trade-offs.
\begin{example}
\label{exa:Ex} Suppose that $v\left(\mu\right)=1-2\mu\left(1-\mu\right)$,
$\delta_{P}=0$ and $\delta_{A}=0.2$. In order to compare with the
personalized model of section \ref{sec:Catering-to-the}, we fix $\ell^{L}$
at the value of 0.3 and vary $\ell^{H}\geq0.3$ and find the steady
state. We assume that the fraction of high types is 0.6.

Figure \ref{fig:Speed-vs.-Quality} illustrates the idea of speed
versus quality. In Figure \ref{fig:Quality-of-Information}, for given
likelihood ratios $\ell^{L}=0.3$ and $\ell^{H}\geq0.3$, we depict
the stationary level of principal's belief $\mu_{P}^{*}$ (the green-crossed
line) together with the beliefs that are induced in phase 2 (red and
blue lines), $\mu_{P}^{\sigma}$. For comparison, we have also included
the stationary belief of the principal in the personalized case for
$\ell^{H}\geq0.3$. Not surprisingly, the non-personalized stationary
belief is higher -- since the existence of low type agent has to
be taken into account. More importantly, the beliefs induced upon
transition to phase 2 are different from certainty and sometimes fairly
close to highest level of uncertainty. This implies that the type
that receives the exit recommendation (High type for the red line
and low type for the blue line) is exiting with a low quality of information,
especially when $\ell^{H}$ is low. In other words, in order to keep
the high type engaged, low type may exit at the end of phase 1 with
fairly poor information.

In Figure \ref{fig:Speed-of-Arrival}, we depict $\lambda^{*}$, the
rate at which phase 1 ends in the non-personalized model together
with the rate of arrival of exit in the personalized model. As this
figure illustrates, the speed with which transition to phase 2 happens
is significantly higher in the presence of private information about
beliefs.

\begin{figure}
\begin{centering}
\subfloat[Quality of Information\label{fig:Quality-of-Information}]{\includegraphics[scale=0.15]{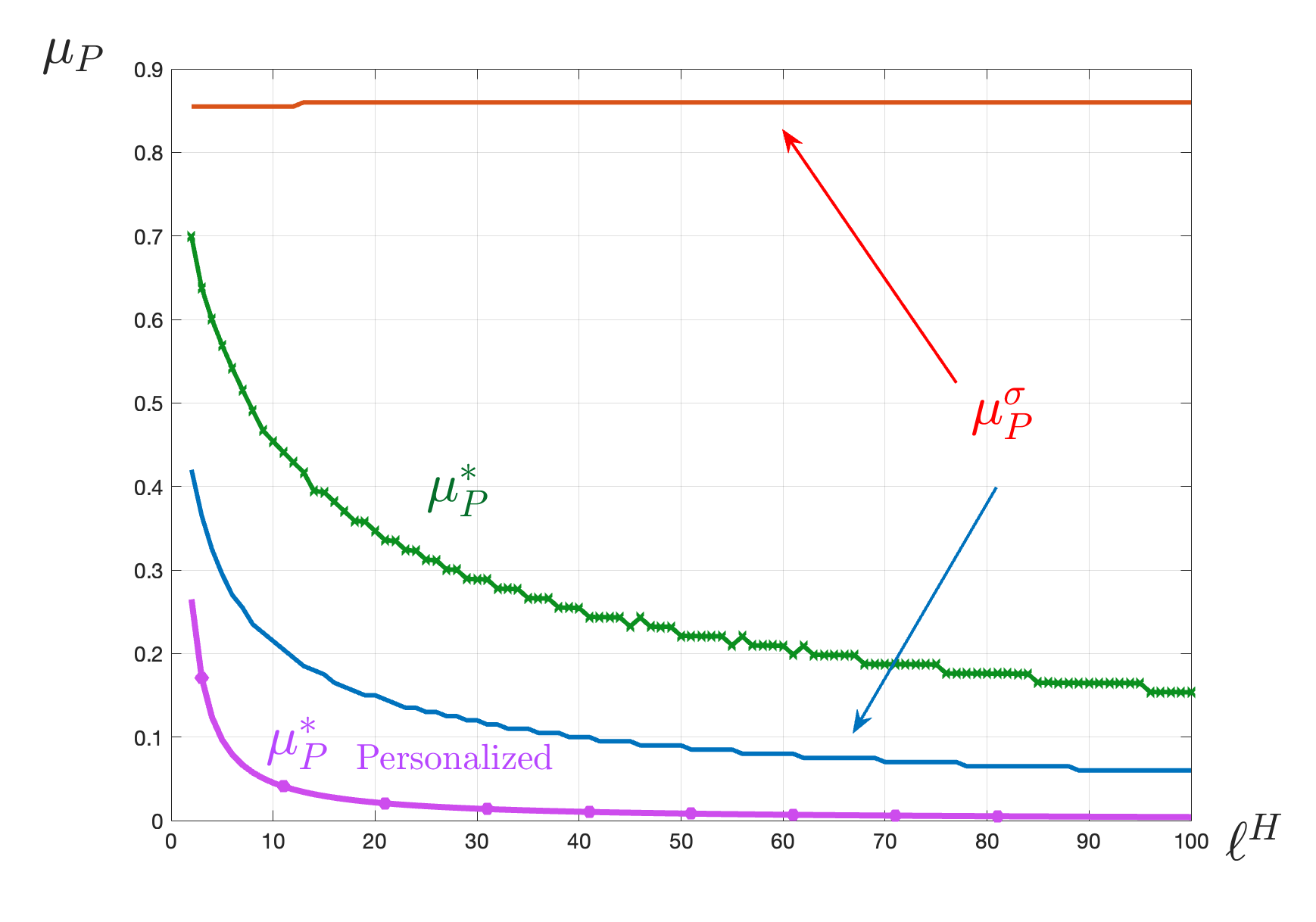}

}\subfloat[Speed of Arrival of Information\label{fig:Speed-of-Arrival}]{\includegraphics[scale=0.15]{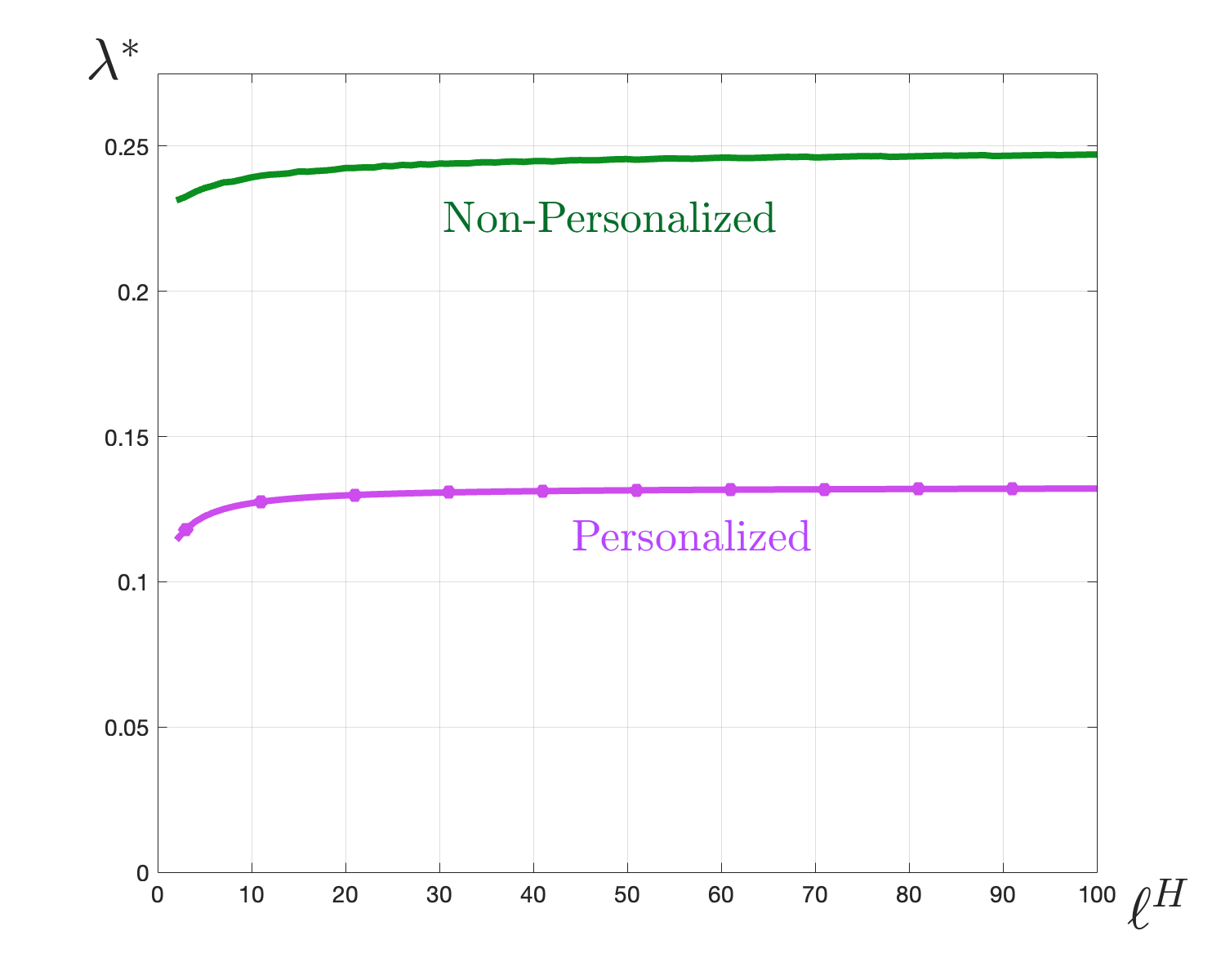}

}
\par\end{centering}
\caption{Speed vs. Quality\label{fig:Speed-vs.-Quality}}
\end{figure}
\end{example}
The above discussions highlight the key differences and similarities
between personalized and non-personalized communication strategies.
In the personalized scenario, the principal can exert perfect control
over the agent's beliefs upon exit. In contrast, under non-personalized
communication, while phase 1 of the interaction bears similarities
to the personalized model, it differs in that the value of quitting
from this phase is no longer zero for the principal. Furthermore,
to prevent one type from quitting, it may sometimes be optimal to
allow some types to leave without perfect information. Despite these
distinctions, the fundamental principle governing the non-personalized
optimal mechanism remains similar to that of the personalized one.

\section{\label{sec:Extensions-and-Applications}Extensions and Applications}

In this section, we explore several extensions to our base model.
We begin by incorporating random exit into our framework. We then
investigate the implications of personalization on belief polarization,
using our model to shed light on the ongoing debate about algorithmic
news feeds and their impact on political discourse. Finally, we consider
more general forms of time preferences, demonstrating how our model
can accommodate various discounting structures beyond simple exponential
discounting. 

\subsection{Random Exit}

Throughout our analysis, our assumption has been that the agent can
perfectly control her acquisition of information. In reality, however,
attention is not perfect, and it is possible that audiences of news
randomly exit. This is also in line with the extensive literature
on limited attention or \textit{rational inattention} that models
attention as a noisy process where reducing the noise is costly; see
\citet{sims2003implications} and the extensive literature following
it.\footnote{\citet{morris2019wald} provide a foundation for the static costly
noise reduction model of \citet{sims2003implications} with a dynamic
decision maker. Our agent essentially behaves like theirs and solves
an optimal stopping problem. }

To allow for random exit, suppose that if the agent is engaged at
$t$, with probability $e^{-\rho\left(\tau-t\right)}$ she is still
engaged at $\tau>t$. That is, she will exogenously exit in an interval
$\left[\tau,\tau+d\tau\right]$ with probability $e^{-\rho\left(\tau-t\right)}-e^{-\rho\left(\tau+d\tau-t\right)}$.
Under this modification and absent any discounting by the principal,
his payoff is given by
\[
-\int_{0}^{\infty}td\left(e^{-\rho t}G_{P}\left(t\right)\right)=\int_{0}^{\infty}e^{-\rho t}G_{P}\left(t\right)dt
\]
where $G_{P}$ is the decumulative probability that the principal
recommends staying. In this case, unlike the baseline model, exit
before $t$ can also occur exogenously and with probability $1-e^{-\rho t}$
. Hence, the probability has been adjusted accordingly.

The payoff of the agent is somewhat more complicated, as we need to
take into account the occasions in which the agent exits exogenously
and her payoff upon such exit. If the payoff of the agent upon recommended
exit is $v_{A}\left(t\right)$, then her payoff at time $t$ conditional
on being engaged is given by
\[
-\int_{t}^{\infty}e^{-\left(\delta_{A}+\rho\right)\left(\tau-t\right)}v_{A}\left(\tau\right)d\frac{G_{A}\left(\tau\right)}{G_{A}\left(t\right)}+\rho\int_{t}^{\infty}e^{-\left(\delta_{A}+\rho\right)\left(\tau-t\right)}v\left(\mu_{A}\left(\tau\right)\right)\frac{G_{A}\left(\tau\right)}{G_{A}\left(t\right)}d\tau.
\]
When recommendation is personalized, we have $v_{A}\left(\tau\right)=v\left(1\right)$
which then implies that the above can be written as:
\[
v\left(1\right)-\int_{t}^{\infty}e^{-\left(\delta_{A}+\rho\right)\left(\tau-t\right)}\left[\left(\rho+\delta_{A}\right)v\left(1\right)-\rho v\left(\mu_{A}\left(\tau\right)\right)\right]\frac{G_{A}\left(\tau\right)}{G_{A}\left(t\right)}d\tau.
\]
Therefore, the incentive constraint of the agent ensures that the
agent's payoff above is higher than $v\left(\mu_{A}\left(t\right)\right)$.

The following Proposition establishes that the case with exit is similar
(but not identical) to that of the unequal discounting case with $\delta_{P}>\delta_{A}$:
\begin{prop}
\label{prop: Prop1Exit}Optimal communication with exit satisfies
the following properties:
\begin{enumerate}
\item There exists $\mu_{P}^{*}\in\left[0,1\right]$ such that $G_{P,\omega}\left(t\right)=G_{P,\omega}\left(0\right)e^{-\lambda t}$
and $\mu_{A}\left(t\right)=\mu_{A}$ for all $t$ where $\lambda$
and $\mu_{P}^{*}$ satisfy
\begin{align*}
\mu_{P}^{*} & =\max\left\{ \min\left\{ \mu_{A}+\frac{\left(\delta_{A}-\rho\right)\mu_{A}\left(1-\mu_{A}\right)v'\left(\mu_{A}\right)}{\left(\delta_{A}-\rho\right)v\left(\mu_{A}\right)+\rho v\left(1\right)},1\right\} ,0\right\} \\
\lambda & =\frac{\delta_{A}v\left(\mu_{A}\right)}{v\left(1\right)-v\left(\mu_{A}\right)}
\end{align*}
\item If $\delta_{A}>\rho$, then if $\mu_{P}>\mu_{P}^{*}$ ($\mu_{P}<\mu_{P}^{*}$),
$\omega=1$ ($\omega=0$) is initially revealed to the agent until
$t=t^{*}$ where $\mu_{P}^{*}\left(\mu_{A}\left(t^{*}\right)\right)=\mu_{A}\left(t^{*}\right)$.
\item If $\delta_{A}<\rho$, then if $\mu_{P}>\mu_{P}^{*}$ ($\mu_{P}<\mu_{P}^{*}$),
$\omega=0$ ($\omega=1$) is initially revealed to the agent until
$t=t^{*}$ where $\mu_{P}^{*}\left(\mu_{A}\left(t^{*}\right)\right)=\mu_{A}\left(t^{*}\right)$.
\item When $t\geq t^{*}$, $\mu_{A}\left(t\right)=\mu_{A}\left(t^{*}\right)$
and both states are revealed at a rate $\delta_{A}v\left(\mu_{A}\left(t^{*}\right)\right)/\left(v\left(1\right)-v\left(\mu_{A}\left(t^{*}\right)\right)\right)$.
\end{enumerate}
\end{prop}

\subsection{Personalization and Belief Polarization\label{subsec:Personalization-and-Belief}}

One of the much debated issues related to political polarization is
the conflict of interest between algorithmic news feeds and news consumers.
The concern is that social media platforms, whose main revenue source
is targeted advertising, aim to maximize engagement rather than provide
the most accurate information, potentially leading to political polarization.\footnote{The term \textit{filter bubble }was coined by \citet{pariser2011filter}
to describe content controlled by algorithms that can create bias.}

Using an example, we demonstrate how our model can be used as a framework
to study the effect of targeting news. We consider the model with
random exit discussed in the previous section and compare the optimal
personalized communication outcomes with those of non-personalized
communication outcomes. This comparison allows us to examine the potential
impacts of news targeting strategies on information dissemination.

More specifically, we assume the following parametric assumptions.
The agent's payoff as a function of her belief is $v\left(\mu\right)=1-2\mu\left(1-\mu\right)$,
her discount rate is $\delta_{A}=0.25$, and her exit rate is $\rho=0.05$.
Moreover, we consider two types of agents with different prior beliefs:
high type and low type. We assume that their prior beliefs are $\mu_{A}^{H}=0.647,\mu_{A}^{L}=0.196$,
respectively, and the fraction of high types is $\alpha^{H}=0.6$.
The prior belief of the principal is assumed to be $\mu_{P}=0.55$
and its discount rate to be zero.

For this specification, we compare the evolution of beliefs under
personalized and non-personalized communication. In this example,
the principal's prior belief falls between those of the high and low
agent types, specifically being less than the high type's prior and
greater than the low type's prior. In the personalized case, the principal's
strategy differs for each type. For the high type, the principal initially
reveals information about $\omega=1$, as this is the state toward
which the high type is more biased. Conversely, for the low type,
the principal initially reveals information about $\omega=0$, reflecting
the low type's bias. If no signals arrive, this strategy gradually
shifts the agents' beliefs toward the opposite state. The steady state
points for these two types are 0.71 for the low type and 0.43 for
the high type. Once the agents reach their respective steady state
points, the principal reveals both states at the same rate. This process
continues until either a signal arrives or the agent exits due to
the exogenous exit probability.

The optimal non-personalized communication can be calculated using
the arguments in Section \ref{sec:personalized}. As mentioned before,
there are two phases in this process. In phase 1, the principal attempts
to keep both agents engaged and during this phase the beliefs merge
to a steady state level. This phase continues until a transition signal
arrives which results in at least one of the agents leaving. We first
numerically solve the constrained optimization problem that characterizes
the steady state in phase 1. Next, we guess with an initial value
for first order derivative of transition signals at time zero. Using
the optimality conditions and this initial guess, we can solve the
rest of parameters in the model. Finally, we checked that the resulting
solution satisfies the time 0 incentive constraints, if not we modify
the guess and reiterate. 

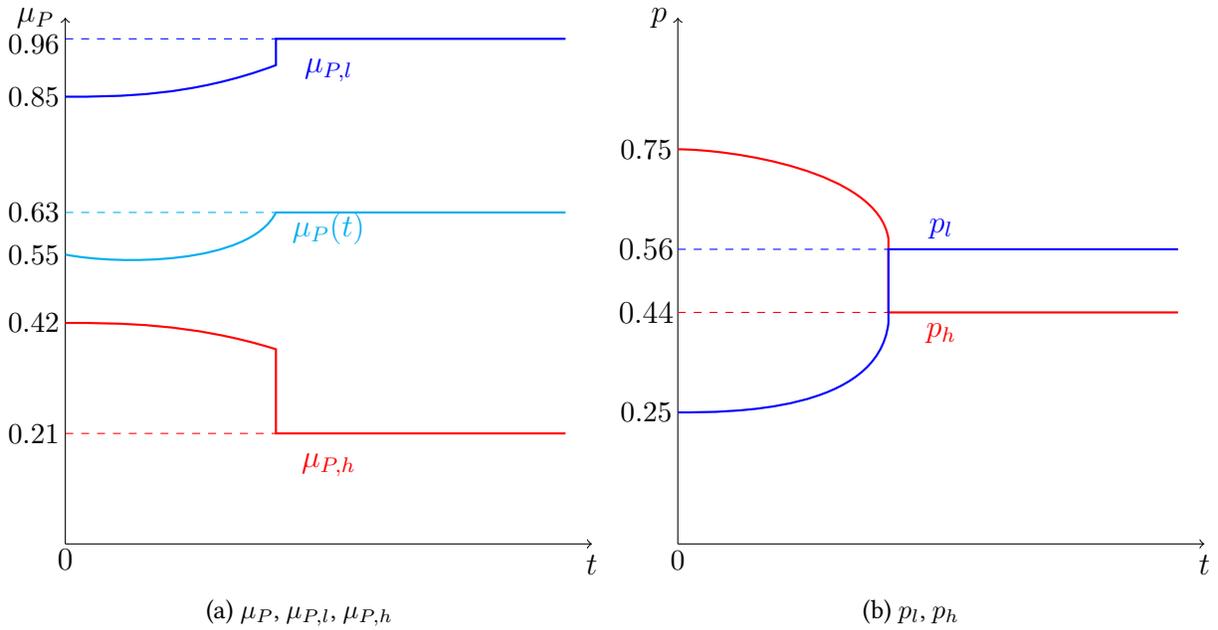
\begin{figure}
\begin{centering}
\subfloat[$\mu_{P}$, $\mu_{P,l}$, $\mu_{P,h}$ \label{fig:G-mu}]{\begin{tikzpicture}[scale=.7]
\draw[->] (0,0) -- (10,0) node[anchor=north] {$t$};     \draw[->] (0,0) -- (0,10) node[anchor=east] {$\mu_P$};     \draw[thick,color=blue] (0,8.5) .. controls (1,8.5) and (2.4,8.5) .. (4,9.1) -- (4,9.6) -- (9.5, 9.6);     \draw[dashed,color=blue] (0,9.6) -- (4,9.6) ;     \draw[thick,color=red] (0,4.2) .. controls (1,4.2) and (2.4,4.2) .. (4,3.7) -- (4,2.1) -- (9.5, 2.1);     \draw[dashed,color=red] (0,2.1) -- (4,2.1) ;
    \draw[thick,color=cyan] (0,5.5) .. controls (1,5.3) and (3.5,5.3) .. (4,6.3) -- (9.5, 6.3);     \draw[dashed,color=cyan] (0,6.3) -- (4,6.3) ;          \draw[color=red] (5,1.5) node{{$\mu_{P,h}$}};     \draw[color=blue] (5,9) node{{$\mu_{P,l}$}};     \draw[color=cyan] (5,6) node{{$\mu_P(t)$}};     
    \draw (-0.6,5.5) node{{\small $0.55$}};     \draw (-0.6,4.2) node{{\small $0.42$}};     \draw (-0.6,8.5) node{{\small$0.85$}};     \draw (-0.6,6.3) node{{\small$0.63$}};     \draw (-0.6,9.5) node{{\small$0.96$}};     \draw (-0.6,2.1) node{{\small$0.21$}};   
    \draw (0,-0.3) node{{$0$}};
\end{tikzpicture}}\subfloat[$p_{l}$, $p_{h}$ \label{fig:plh10}]{\begin{tikzpicture}[scale=.7]
 \draw[->] (0,0) -- (10,0) node[anchor=north] {$t$};     \draw[->] (0,0) -- (0,10) node[anchor=east] {$p$};     \draw[thick,color=red] (0,7.5) .. controls (1,7.5) and (3.8,7.1) .. (4,5.8) -- (4,4.4) -- (9.5, 4.4);     \draw[dashed,color=red] (0,4.4) -- (4,4.4) ;     \draw[thick,color=blue] (0,2.5) .. controls (1,2.5) and (3.8,2.5) .. (4,4.2) -- (4,5.6) -- (9.5, 5.6);     \draw[dashed,color=blue] (0,5.6) -- (4,5.6) ;          \draw[color=red] (5,4) node{{$p_h$}};     \draw[color=blue] (5,6) node{{$p_l$}};     
    \draw (-0.6,7.5) node{{\small$0.75$}};     \draw (-0.6,2.5) node{{\small$0.25$}};     \draw (-0.6,4.4) node{{$\small 0.44$}};     \draw (-0.6,5.6) node{{$\small 0.56$}};     
    \draw (0,-0.3) node{{\small$0$}};
\end{tikzpicture}}\caption{Evolution of beliefs and engagement probabilities}
\par\end{centering}
\noindent \textit{\small{}Notes}\textit{\emph{\small{}: (a)}}{\small{}
The variable $\mu_{P}\left(t\right)$ is principal's belief if phase
1 continues. Variable $\mu_{P,l}$ ($\mu_{P,h}$) is principal's belief
if the low-type (high-type) agent stays for phrase 2. (b) The variable
$p_{l}$($p_{h})$ is the probability that the low-type (high-type)
agent stays for phase 2 (personalized phase).}{\small\par}
\end{figure}

During phase 1, the principal's belief changes until it converges
to a steady state level, as depicted in Figure \ref{fig:G-mu}. This
change occurs because the transitional signal may arrive with different
probabilities given the state. The principal's belief increases from
0.55 to a long-run rate of 0.63 in phase 1. The beliefs of the two
types of agents also shift, as the likelihood ratio of the beliefs
remains constant.

The transition signal can take different forms, either instructing
only one type to stay or delivering full information and allowing
both types of agents to leave. Figure \ref{fig:plh10} illustrates
the probability of these different outcomes. Given the parameters
in this example, it is never optimal to reveal the state and let both
types leave. The probabilities for which type of agent stays and which
leaves vary by the arrival time of the signal, reflecting changes
in beliefs. These probabilities are calculated using the concavification
method mentioned in Section \ref{sec:personalized}. They are associated
with the arrival of one of two different types of signals that discontinuously
change the principal's and agents' beliefs. Figure \ref{fig:plh10}
also illustrates the principal's beliefs after the arrival of each
of these signals. The signal associated with the exit of the high
type and the stay of the low type is indicated by the belief $\mu_{P,l}$.
After this signal arrives, the belief of the high type becomes very
close to one, while the belief of the low type falls below the depicted
line.\footnote{To keep the figures tidy, we only depict the beliefs of the principal,
the likelihood ratio of the beliefs stay the same, so beliefs of high
type is always above the belief of principal and belief of low type
is always below it.} After receiving this signal, the high-type agent will quit while
the low-type agent chooses to stay and gather more information and
we arrive to phase 2. The opposite occurs when the signal for keeping
the high type arrives. 

\begin{figure}
\begin{centering}
\includegraphics[scale=0.55]{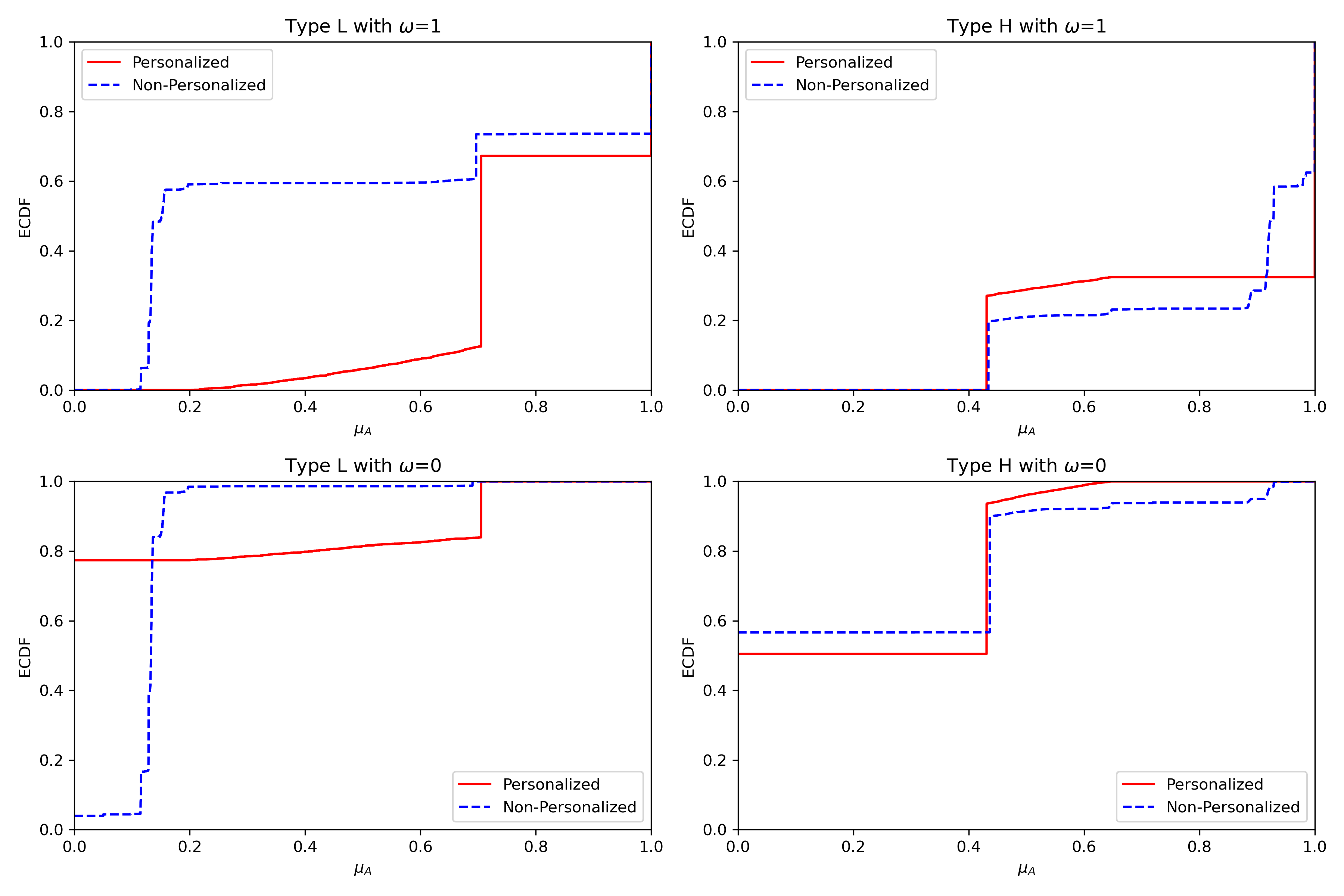}\caption{Distribution of exit beliefs under optimal personalized and non-personalized
communication\label{fig:Evolution-of-beliefs-1}}
\par\end{centering}
\noindent \textit{\small{}Notes: }{\small{}The red solid line is the
empirical cumulative distribution function (CDF) of agents' beliefs
under personalized communication strategies. The blue dashed line
is the empirical CDF of agents' beliefs under non-personalized communication
strategies. The left two panels are for the low-type agent and the
right two panels are for the high-type agent. The upper two figures
are under true state $\omega=1$ and the lower two figures are under
true state $\omega=0$. The empirical CDF comes from simulations with
10,000 belief paths for each case.}{\small\par}
\end{figure}

\begin{figure}
\begin{centering}
\includegraphics[scale=0.4]{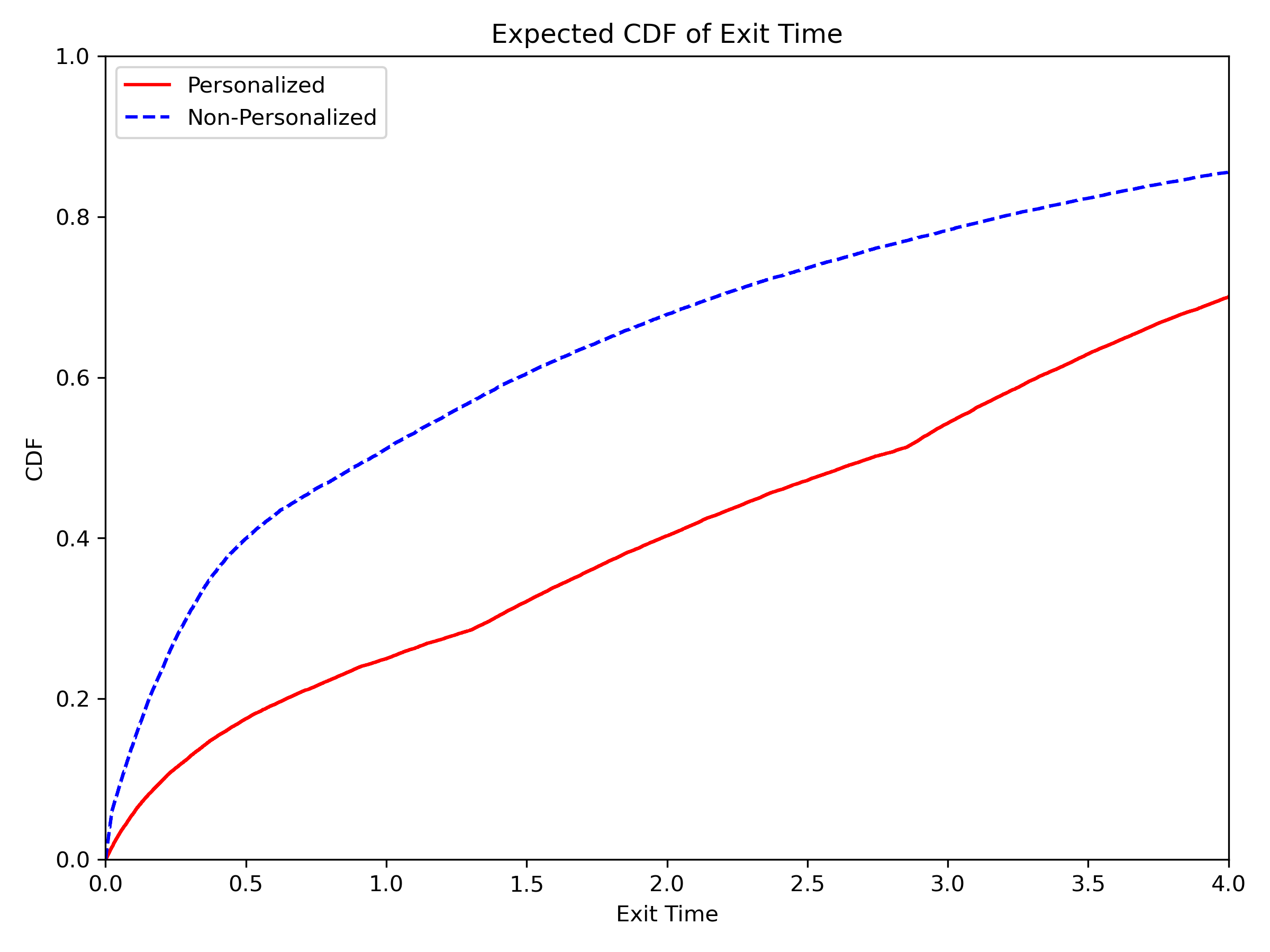}\caption{Distribution of exit time under optimal personalized and non-personalized
communication\label{fig:Evolution-of-time-Total}}
\par\end{centering}
\noindent \textit{\small{}Notes: }{\small{}Assuming that the distribution
of the true state matches the principal's prior $\mu_{p}=0.55$, the
red solid line is the CDF of exit time of agents under personalized
recommendation strategies, and the blue dashed line is the CDF of
exit time of agents under non-personalized recommendation strategies.}{\small\par}
\end{figure}

Figure \ref{fig:Evolution-of-beliefs-1} depicts the CDF of beliefs
for each state and agent type as time approaches infinity. Due to
exogenous exit rates, some agents may leave without full information
even in the personalized case, resulting in non-degenerate belief
distributions. In the personalized case, prolonged interactions may
result in agents leaving without perfect information. We observe that
if the state is 1, 70\% of high-type agents exit with this information
compared to 30\% of low-type agents, while if the state is 0, 80\%
of low-type agents leave with this information compared to 50\% of
high-type agents. This asymmetry arises from the principal catering
to agents' biases. If the true state aligns with an agent's bias,
they are likely to receive this information quickly. Otherwise, they
may exit before receiving a signal revealing the opposite state and
stay uninformed and biased toward the other state. 

In the non-personalized case, alongside exogenous exits, the principal
may choose to cater to only one agent type, prompting the other type
to leave without full information. Comparing the distribution of agents
across the four scenarios, we observe that in the non-personalized
case, agents' beliefs tend to be more dispersed and generally less
aligned with the true state. For instance, when the state is 1, approximately
60\% of low-type agents still strongly believe that the low state
is likely ($\mu<0.2$), and only about 25\% of these agents learn
that the state is actually 1. Similarly, fewer high-type agents discover
that the state is 1 compared to the personalized case. These observations
indicate that personalization leads to higher quality information
dissemination, as agents are more likely to learn the true state under
personalized communication strategies.

Another significant difference between personalized and non-personalized
cases lies in the speed of information provision to agents. Figure
\ref{fig:Evolution-of-time-Total} illustrates the CDF of exit times
for agents in both personalized and non-personalized scenarios. We
consistently observe earlier and faster exits in the non-personalized
case, driven by two key forces. First, in the non-personalized case,
during phase 1, the principal must keep both agent types engaged simultaneously,
unable to cater to them specifically. In contrast, the personalized
case allows the principal to guide each agent to a point where the
cost of maintaining engagement is lowest, thus prolonging interaction.
This strategy is generally not feasible when dealing with two distinct
agent types simultaneously. Second, in the personalized case, the
arrival of a signal leads to state revelation, increasing the reward
for staying engaged. In the non-personalized case the state is not
going to be perfectly revealed as it prompts both types to quit. Consequently,
given the lower reward in the non-personalized scenario, the principal
must increase the frequency of rewards and, therefore, the speed of
information arrival. This strategy ultimately results in quicker agent
exits in the non-personalized case.

These findings highlight the opposing forces present in our model,
illustrating a trade-off between information quality and speed of
delivery. Personalization results in slower information transmission
but higher quality, while non-personalized communication leads to
faster transmission but lower quality. In both scenarios, we observe
that agents starting with lower priors tend to maintain their belief
that their preferred state is more likely more often than those agents
who initially believe the opposite state is more likely. This persistence
of beliefs suggests that polarization can occur in both personalized
and non-personalized communication settings. Importantly, our results
indicate that personalized communication does not necessarily lead
to more polarization than non-personalized communication. This insight
may help explain the mixed results found in studies of the ``filter
bubble'' phenomenon.\footnote{Recent research offers conflicting evidence on this topic. For instance,
\citet{jones2024can} use survey data to show that social media, in
contrast to traditional media, leads to less polarization. Conversely,
\citet{bakshy2015exposure} conduct a randomized experiment and find
that algorithmically ranked news feeds result in 15\% lower exposure
to cross-cutting content. These contrasting findings underscore the
complexity of the issue and the need for further research in this
area.} However, it is crucial to note that a more comprehensive quantitative
investigation is needed to fully understand these findings.

\subsection{Other forms of Discounting}

It is possible to extend our analysis to more general cases of time
preferences. Specifically, suppose that principal's preferences are
given by $T$ and that of the agent is $D\left(T\right)\hat{u}\left(\omega,a\right)$.
Below, we show how this encompasses several popular models of time
preferences.

\subsubsection{The Nature of Time Cost for the Agent\label{subsec:The-Nature-of}}

Here, we show how different commonly used setups for inter-temporal
discounting map into the setting described above:
\begin{enumerate}
\item \textbf{Exponential Discounting. }The case of exponential discounting
that we have so far studied is a special case of this extension. In
this case the principal payoff is $\delta_{P}\int_{0}^{\hat{T}}e^{-\delta_{p}t}dt$
and the agent payoff is $e^{-\delta_{a}\hat{T}}\hat{u}\left(\omega,a\right)$,
where $\hat{T}$ is the engagement time by the agent.We can let $T$
be the payoff of the principal and in this case the payoff of the
agent is given by 
\[
u\left(T,\omega,a\right)=\left(1-T\right)^{\frac{\delta_{a}}{\delta_{p}}}\hat{u}\left(\omega,a\right)
\]
whose concavity and convexity is determined by how $\delta_{A}/\delta_{P}$
compares to 1.
\item \textbf{Habit Formation a la \citet{becker1988theory}: }Another example
of inter-temporal preferences is the habit formation model of \citet{becker1988theory}
(also used in \Citet{allcott2022digital}). In this model, preferences
are a function of habit -- a stock variable that increases exponentially
with consumption. In the context of digital addiction setting of \citet{allcott2022digital},
consumption can be interpreted as social media engagement. Let $H_{t}$
be the stock of habit which satisfies 
\[
\frac{d}{dt}H_{t}=\lambda H_{t}-h\rightarrow H_{t}=e^{\lambda t}H_{0}+h/\lambda
\]
and suppose that the agent's utility is given by
\[
H_{\hat{T}}e^{-\delta_{A}\hat{T}}=e^{\left(\lambda-\delta_{A}\right)\hat{T}}H_{0}+e^{-\delta_{A}\hat{T}}h/\lambda
\]
When $\lambda<\delta_{A}$, this is equivalent to a mixed-discounting
model and can similarly be expressed in terms of the payoff of the
principal $T=1-e^{-\delta_{P}\hat{T}}$. As depicted in Figure \ref{fig:Habit-Formation-1},
when habits are weak and the agent is less patient than the principal
($\delta_{P}<\delta_{A}-\lambda$), MCE is decreasing over time while
when when the agent is less patient than the principal, MCE is increasing
over time. In the intermediate case when habits are strong enough
to make $\delta_{A}-\lambda<\delta_{P}$, MCE initially decreases
over time and as time passes habit formation becomes dominant and
MCE starts increasing over time.
\item \textbf{A Reduced Form Model of Boredom: }Another way to think about
time preferences for the agent is through the lens of a reduced form
version of the dual-self control model of \citet{fudenberg2006dual}.
We can think of the agent as consisted of two selves: a long-run planner
and a short-run decider. The decider's costly choice depends on the
total time spent being engaged which can be a stock variable associated
with boredom. The long-run planner in turn has to make contingent
planning taking into account the cost of boredom for the short-run
player. For example, the agent could have a utility of the form
\[
D\left(T\right)=e^{-\delta_{A}T-\gamma T^{2}}
\]
where $\gamma T^{2}$ captures the cost of boredom. In this case,
MCE initially increases while after a certain period it decreases.
This is depicted in Figure \ref{fig:A-Model-of-1}.\footnote{See also \citet{wojtowicz2019boredom} for an extended version of
this model with information acquisition by the dual selves.}
\end{enumerate}
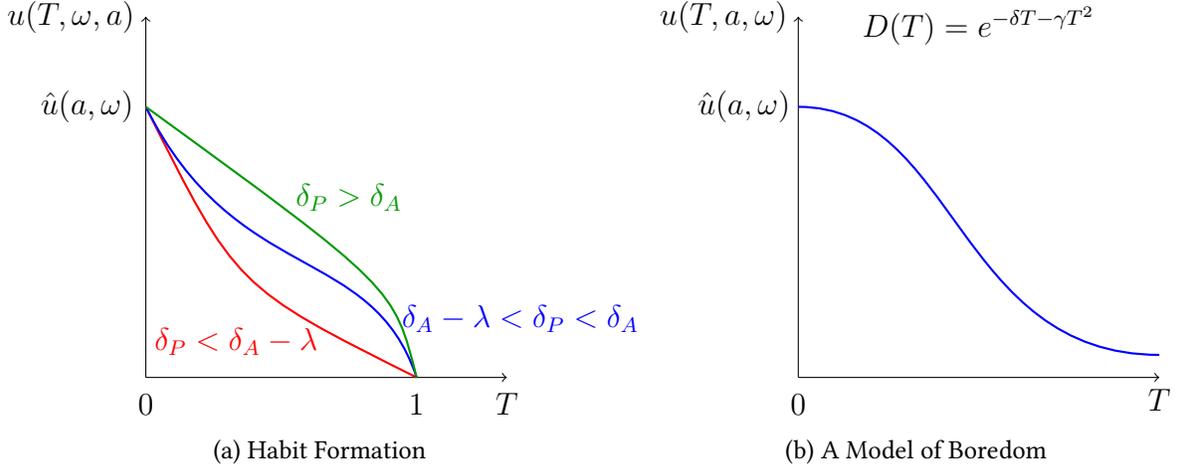
\begin{figure}[h]
\centering{}\subfloat[Habit Formation \label{fig:Habit-Formation-1}]{\begin{tikzpicture}[scale=.6]
\draw[->] (0,0) -- (8,0) node[below = .06cm] {$T$};
\draw[->] (0,0) -- (0,8) node[anchor=east] {$u(T,\omega,a)$};     \draw[thick,color=red] (0,6) .. controls (2,2) .. (6,0);
\draw[thick,color=blue] (0,6) .. controls (2,2) and (5,3) .. (6,0);     \draw[thick,color=darkgreen] (0,6) .. controls  (5.5,2) .. (6,0);
\draw[color=red] (2.,.8) node{{$\delta_P < \delta_A - \lambda$}};     \draw[color=blue] (8.3,1.3) node{{$\delta_A - \lambda <\delta_P <\delta_A$}};     \draw[color=darkgreen] (4.5,4) node{{$\delta_P>\delta_A$}};     
\draw (-1.3,6) node{{$\hat{u}(a,\omega)$}};
\draw (6,-0.6) node{{$1$}};
\draw (0,-0.6) node{{$0$}};
\end{tikzpicture}}\subfloat[A Model of Boredom \label{fig:A-Model-of-1}]{\begin{tikzpicture}[scale=.6]     
\draw[->] (0,0) -- (8,0) node[anchor=north] {$T$};     
\draw[->] (0,0) -- (0,8) node[anchor=east] {$u(T,a,\omega)$};     \draw[thick,color=blue] (0,6) .. controls (3.5,6) and (3.5,.5) .. (8,.5);          
\draw (4,7.8) node{$D(T)=e^{-\delta T -\gamma T^2}$};          
\draw (-1.2,6) node{{$\hat{u}(a,\omega)$}};
\draw (0,-0.6) node{{$0$}};
\end{tikzpicture} 

}\caption{Examples of Relative Marginal Cost of Engagement\label{fig:Examples-of-Relative-1}}
\end{figure}

\subsubsection{\label{subsec:general_discount}Optimal Communication with General
Discounting}

The case of concave and convex $D\left(T\right)$ are qualitatively
the same as their associated cases with exponential discounting. Here,
we describe what happens when $D\left(T\right)$ is either a concave-convex
or a convex-concave function. In these cases, perhaps not surprisingly,
the solution is often determined by superimposing the Poisson gradual
revelation and abrupt revelations. In what follows, we describe the
nature of this superimposition and key properties of optimal revelation.
\begin{prop}
\textbf{\label{prop:Convex-Concave-Discounting.}Convex-Concave Discounting.
}Suppose that there exists a threshold $t^{i}$ such that above $D'\left(t\right)$
is increasing for values of $t\leq t^{i}$ and decreasing for values
of $t>t^{i}$. Then optimal information provision has three phases:
\begin{enumerate}
\item A phase for $0\leq t<t_{1}^{*}<t^{i}$ where only one state is revealed
(towards which the agent is more optimistic) gradually, i.e. according
to a time varying Poisson process. When $\mu_{0}>1/2$, the evolution
of beliefs is given by the ODE:
\[
\frac{\mu'\left(t\right)}{\mu\left(t\right)\left(1-\mu\left(t\right)\right)}\frac{v\left(1\right)-v\left(\mu\left(t\right)\right)-v'\left(\mu\left(t\right)\right)\left(1-\mu\left(t\right)\right)}{v\left(\mu\left(t\right)\right)}=\frac{D'\left(t\right)}{D\left(t\right)}.
\]
\item A second phase for $t\in\left[t_{1}^{*},t_{2}^{*}\right]$ with $t_{2}^{*}<t^{i}$
where both states are gradually revealed according to a Poisson process
with arrival rate $-D'\left(t\right)/D\left(t\right)$.
\item A final phase \textup{for} $t\in\left[t_{2}^{*},t_{3}^{*}\right]$
with $t^{i}<t_{3}^{*}$ in which no information is provided and all
information is revealed at $t_{3}^{*}$.
\end{enumerate}
\end{prop}
As Proposition \ref{prop:Convex-Concave-Discounting.} illustrates,
the convex-concave case is a superimposition of the two cases discussed
before. Considering the habit formation example of section \ref{subsec:The-Nature-of},
when $0<\delta_{A}-\lambda<\delta_{P}<\delta_{A}$, as the agent is
initially impatient relative to the principal, information must be
revealed gradually and in the second phase at rate
\[
-\frac{D'\left(t\right)}{D\left(t\right)}=\delta_{A}+\left(\lambda-\delta_{A}\right)\frac{\alpha}{\alpha+\beta e^{-\lambda t}}
\]
for some $\alpha,\beta>0$. This gradual revelation is slowed down
as time goes by and the agent accumulates habit. Once the force of
habit formation becomes strong enough, the value of gradual revelation
disappears and instead abrupt revelation is used.

\section{Conclusion}

This paper examines the strategic interaction between an expert (principal)
seeking to maximize engagement and an agent aiming to collect information
swiftly. Our analysis reveals several key insights into optimal information
provision strategies in such environments.

First, we find that the relative patience of the agent and principal
is a crucial determinant of the optimal information revelation strategy.
In the case of common priors, a more patient agent leads to gradual
information disclosure, while an impatient agent results in more abrupt
revelation. This underscores the importance of considering time preferences
in designing engagement strategies.

Second, our model demonstrates that when the principal and agent have
different prior beliefs, optimal disclosure often involves ``catering
to the bias.'' The principal tends to initially reveal information
aligned with the agent's bias, exploiting this asymmetry to prolong
engagement. This finding has significant implications for understanding
how information providers may strategically utilize belief disagreements.

Third, in scenarios with private agent beliefs, we identify two phases
in communication strategies. During phase 1, the principal attempts
to keep both agent types engaged. A transition signal during this
phase determines which type of agent should stay and which should
leave. Notably, the principal may allow one agent to depart without
full information, contrasting with the personalized case, to keep
the other agent engaged. Once only one agent type remains, the strategy
reverts to the personalized case.

Our comparison of personalized and non-personalized information revelation
strategies highlights important trade-offs. Personalization enables
longer engagement and slower information delivery, while non-personalized
approaches lead to faster information transmission but may result
in lower quality information overall.

\bibliographystyle{ecta}
\bibliography{recommender}

\newpage{}

\appendix

\section{Proofs}

\subsection{Proof of Theorem \ref{thm: thm1}}
\begin{proof}
We prove a more general version of the theorem. As we discussed in
Section \ref{subsec:general_discount}, our model is equivalent to
one in which time is given by $T\in\left[0,\overline{T}\right]$ with
$\overline{T}<\infty$ and the payoff of the principal is given by
$T$ while the payoff of the agent is given by
\[
D\left(T\right)v\left(\mu_{A}\left(T\right)\right)
\]
where $D\left(T\right)$ is a strictly decreasing and positive function
with possibly satisfying $D\left(\overline{T}\right)=0$. Note that
when exponential discounting, $T=\frac{1-e^{-\delta_{P}t}}{\delta_{P}}$
and $D\left(T\right)=\left(1-\delta_{P}T\right)^{\frac{\delta_{A}}{\delta_{P}}}$
with $\overline{T}=1/\delta_{P}$.

For completeness and easier reference, let us restate Theorem 1 on
page 217 of \citet{luenberger1997optimization}:
\begin{thm}
\label{thm: Luenberger}Let $X$ be a linear vector space, $Z$ a
normed space, $\Omega$ a convex subset of $X$, and $P$ a positive
cone in $Z$. Suppose that $P$ contains an interior point. Let $f$
be a real-valued convex functional on $\Omega$ and $H$ a convex
mapping from $\Omega$ into $Z$. Assume the existence of $x_{1}\in\Omega$
for which $H\left(x_{1}\right)<0$, i.e., $H\left(x_{1}\right)$ is
an interior point of $-P$. Let
\begin{equation}
\mu_{0}=\inf_{x\in\Omega}f\left(x\right)\text{ subject to }H\left(x\right)\leq0\label{eq: 2L}
\end{equation}
and assume $\mu_{0}$ is finite. Then there is an element $z_{0}^{*}\geq0$
in $Z^{*}$ such that
\begin{equation}
\mu_{0}=\inf_{x\in\Omega}f\left(x\right)+\left\langle z_{0}^{*},H\left(x\right)\right\rangle .\label{eq:3L}
\end{equation}
Furthermore, if the infimum is achieved in (\ref{eq: 2L}) by an $x_{0}\in\Omega,H\left(x_{0}\right)\leq0$,
it is achieved by $x_{0}$ in (\ref{eq:3L}) and
\[
\left\langle z_{0}^{*},H\left(x_{0}\right)\right\rangle =0
\]
\end{thm}
We will verify that our optimization problem satisfies the conditions
in Theorem \ref{thm: Luenberger}. 

A key assumption in Theorem \ref{thm: Luenberger} is the existence
of $x_{1}$; existence of $x_{1}$ $-H\left(x_{1}\right)\in\text{int}\left(P\right)$.
This proves difficult in general. Therefore, we consider two cases:
1. when $D\left(\overline{T}\right)>0$ and 2. when $D\left(\overline{T}\right)=0$.
In the first case, we can use Theorem \ref{thm: Luenberger}, and
in the second case, we use a limiting argument.

\textit{Case 1. }Suppose that $D\left(\overline{T}\right)<\infty$.
We define the objects in Theorem \ref{thm: Luenberger} as follows:

\begin{align*}
X= & \left\{ x=\left(G_{P,1}\left(T\right),G_{P,0}\left(T\right)\right)|T\in\left[0,\overline{T}\right],G_{P,\omega}:\text{ bounded}\right\} =L^{\infty}\left(\left[0,\overline{T}\right]\right)^{2}\\
Z= & L^{\infty}\left(\left[0,\overline{T}\right]\right)\times\mathbb{R}^{2}\\
\Omega= & \left\{ G_{P,\omega}\text{: non-increasing},G_{P,\omega}\geq0\right\} \\
P= & \left\{ h\left(T\right)\geq0,\forall T\in\left[0,\overline{T}\right]\right\} \times\mathbb{R}_{+}^{2}\\
H\left(x\right)= & \left(H_{1}\left(x\right),G_{P,1}\left(0\right)-\mu_{P},G_{P,0}\left(0\right)-1+\mu_{P}\right)\\
H_{1}\left(x\right)\left(T\right)= & D\left(T\right)\left(\ell G_{P,1}\left(T\right)+G_{P,0}\left(T\right)\right)\left[v\left(\frac{\ell G_{P,1}\left(T\right)}{\ell G_{P,1}\left(T\right)+G_{P,0}\left(T\right)}\right)-v\left(1\right)\right]\\
 & -v\left(1\right)\int_{T}^{\overline{T}}\left(\ell G_{P,1}\left(s\right)+G_{P,0}\left(s\right)\right)D'\left(s\right)ds\\
f\left(x\right) & =-\int_{0}^{\overline{T}}\left[G_{P,1}\left(T\right)+G_{P,0}\left(T\right)\right]dT
\end{align*}
where in the above $\mathbb{R}_{+}$ is the set of real numbers that
are non-negative. Trivially, $X$ and $Z$ satisfy the hypothesis
in Theorem \ref{thm: Luenberger}. The set $\Omega$ is a convex set,
and $P$ is a convex cone. Finally, since $v$ is a convex function,
\begin{align*}
\frac{\frac{1}{2}\left(x+\ell y\right)v\left(\frac{x}{x+\ell y}\right)+\frac{1}{2}\left(x'+\ell y'\right)v\left(\frac{x'}{x'+\ell'y'}\right)}{\frac{1}{2}\left(x+\ell y\right)+\frac{1}{2}\left(x'+\ell y'\right)} & \geq v\left(\frac{\frac{1}{2}\left(x+\ell y\right)\frac{x}{x+\ell y}+\frac{1}{2}\left(x'+\ell y'\right)\frac{x'}{x'+\ell'y'}}{\frac{1}{2}\left(x+\ell y\right)+\frac{1}{2}\left(x'+\ell y'\right)}\right)\\
 & =v\left(\frac{\frac{1}{2}x+\frac{1}{2}x'}{\frac{1}{2}\left(x+\ell y\right)+\frac{1}{2}\left(x'+\ell y'\right)}\right)
\end{align*}
which then implies that
\begin{align*}
 & \frac{1}{2}\left(x+\ell y\right)v\left(\frac{x}{x+\ell y}\right)+\frac{1}{2}\left(x'+\ell y'\right)v\left(\frac{x'}{x'+\ell'y'}\right)\geq\\
 & \qquad\qquad\qquad\left(\frac{1}{2}\left(x+\ell y\right)+\frac{1}{2}\left(x'+\ell y'\right)\right)v\left(\frac{\frac{1}{2}x+\frac{1}{2}x'}{\frac{1}{2}\left(x+\ell y\right)+\frac{1}{2}\left(x'+\ell y'\right)}\right).
\end{align*}
In other words, the function $\left(x+\ell y\right)v\left(\frac{x}{x+\ell y}\right)$
is convex. As a result, the function $H\left(x\right)$ is convex.

Let $G_{P,1}\left(T\right)=\left(1-\varepsilon\right)\mu_{P}D\left(T\right)^{\lambda},G_{P,0}\left(T\right)=\left(1-\varepsilon\right)\left(1-\mu_{P}\right)D\left(T\right)^{\lambda}$
where
\[
\lambda=\frac{v\left(\mu_{A}\right)}{v\left(1\right)-v\left(\mu_{A}\right)}>0
\]
and the strict concavity of $v$ implies that if $\mu_{A},\mu_{P}\in\left(0,1\right)$,
the above is positive. Let $x_{1}$ be the associated member of $L^{\infty}\left(\left[0,\overline{T}\right]\right)$.
We have that: 
\begin{align*}
G_{P,1}\left(0\right)-\mu_{P}= & -\varepsilon\mu_{P},\\
G_{P,0}\left(0\right)+1-\mu_{P}= & -\varepsilon\left(1-\mu_{P}\right)\\
H_{1}\left(x_{1}\right)\left(T\right)= & D\left(T\right)^{1+\lambda}\left(1-\varepsilon\right)\left(\ell\mu_{P}+1-\mu_{P}\right)\left[v\left(\mu_{A}\right)-v\left(1\right)\right]\\
 & -v\left(1\right)\left(1-\varepsilon\right)\left(\ell\mu_{P}+1-\mu_{P}\right)\int_{T}^{\overline{T}}D\left(s\right)^{\lambda}D'\left(s\right)ds\\
= & \left(1-\varepsilon\right)\frac{1-\mu_{P}}{1-\mu_{A}}\left(\left[v\left(\mu_{A}\right)-v\left(1\right)\right]D\left(T\right)^{1+\lambda}-v\left(1\right)\frac{D\left(\overline{T}\right)^{1+\lambda}-D\left(T\right)^{1+\lambda}}{1+\lambda}\right)\\
= & \left(1-\varepsilon\right)\frac{1-\mu_{P}}{1-\mu_{A}}\left(-\frac{v\left(1\right)}{1+\lambda}D\left(T\right)^{\lambda}-v\left(1\right)\frac{D\left(\overline{T}\right)^{1+\lambda}-D\left(T\right)^{1+\lambda}}{1+\lambda}\right)\\
= & -\left(1-\varepsilon\right)\frac{1-\mu_{P}}{1-\mu_{A}}v\left(1\right)\frac{D\left(\overline{T}\right)^{1+\lambda}}{1+\lambda}
\end{align*}
If we let $\overline{\varepsilon}=\min\left\{ \varepsilon\mu_{P},\varepsilon\left(1-\mu_{P}\right),\left(1-\varepsilon\right)v\left(1\right)\frac{1-\mu_{P}}{1-\mu_{A}}\frac{D\left(\overline{T}\right)^{1+\lambda}}{1+\lambda}\right\} >0$,
then we have that
\[
\max\left\{ \sup_{T\leq\overline{T}}H_{1}\left(x_{1}\right)\left(T\right),G_{P,1}\left(0\right)-\mu_{P},G_{P,0}\left(0\right)-1+\mu_{P}\right\} \leq-\overline{\varepsilon}
\]
and therefore, a ball of radius $\overline{\varepsilon}/2$ around
$x_{1}$ is a subset of $-P$. Moreover, $f$ is linear in $x$ and
its value is bounded below by 0 and above by $\overline{T}<\infty$.
Finally, since the functions $G_{P,\omega}$ are bounded, the set
of $x$'s such that $x\in\Omega,H\left(x\right)\leq0$ is compact.
This means that the optimization $\inf_{x\in\Omega,H\left(x\right)\leq0}f\left(x\right)$
achieves its value, and this confirms all the hypotheses of Theorem
\ref{thm: Luenberger}. 

By standard Representation Theorems (Riesz-Kakutani-Markov), see Theorem
IV.16 in \citet{dunford1988linear} as an example, the dual of the
space of bounded functions is isomorphic to the space of finitely
additive measures of bounded variation or in turn the space of functions
of bounded variations that vanish at $t=0$. This implies that there
must exists $z^{*}\in Z^{*}=\left(\Lambda\left(T\right),\lambda_{1},\lambda_{0}\right)$
where $\lambda_{1},\lambda_{0}\geq0$, $\Lambda\left(t\right)$ is
an increasing function of bounded variation\footnote{The fact that $\Lambda^{T}\left(t\right)$ must be (weakly) increasing
is implied by $z^{*}\geq0$ which means that $z^{*}\in P^{*}$, the
dual of the cone $P$ which is defined as members of $Z^{*}$, such
that $\left\langle z^{*},z\right\rangle \geq0$ for all $z\in P$. } with $\Lambda\left(0\right)=0$ such that $\hat{x}=\left(\hat{G}_{P,1},\hat{G}_{P,0}\right)$
is the solution to $\max_{x\in\Omega,H\left(x\right)\leq0}-f\left(x\right)$
if
\begin{align*}
\hat{x} & \in\arg\max_{x\in\Omega}-f\left(x\right)-\int_{0}^{\overline{T}}H_{1}\left(x\right)\left(T\right)d\Lambda\left(T\right)-\lambda_{1}\left(G_{P,1}\left(0\right)-\mu_{P}\right)-\lambda_{0}\left(G_{P,0}\left(0\right)-1+\mu_{P}\right)\\
 & =\arg\max_{x\in\Omega}L\left(x\right)
\end{align*}

Note that $\Omega$ is the set of positive, bounded, and non-increasing
functions. Thus, it is a convex cone in $X$. Hence, we can apply
the following Lemma from \citet{luenberger1997optimization} -- Lemma
1, page 227, section 8.7: 
\begin{lem}
\label{lem: Luen} Let $f$ be a Fréchet differentiable convex functional
on a real normed space $X$. Let $P$ be a convex cone in $X$. A
necessary and sufficient condition for $x_{0}\in P$ to minimize $f$
over $P$ is that
\begin{align*}
\partial f\left(x_{0};x\right) & \geq0,\forall x\in P\\
\partial f\left(x_{0};x_{0}\right) & =0.
\end{align*}
\end{lem}
Applying the above to $L$, we have
\begin{align}
\partial L\left(\hat{x};x\right)= & \int_{0}^{\overline{T}}\left(G_{P,1}\left(T\right)+G_{P,0}\left(T\right)\right)dT-\label{eq: FOCL}\\
 & \int_{0}^{\overline{T}}\left[v\left(\mu_{A}\left(T\right)\right)-v\left(1\right)\right]\left(\ell G_{P,1}\left(T\right)+G_{P,0}\left(T\right)\right)D\left(T\right)d\Lambda\left(T\right)-\nonumber \\
 & \int_{0}^{\overline{T}}\left[v'\left(\mu_{A}\left(T\right)\right)\left(1-\mu_{A}\left(T\right)\right)\ell G_{P,1}\left(T\right)-v'\left(\mu_{A}\left(T\right)\right)\mu_{A}\left(T\right)G_{P,0}\left(T\right)\right]D\left(T\right)d\Lambda\left(T\right)+\nonumber \\
 & v\left(1\right)\int_{0}^{\overline{T}}\int_{T}^{\overline{T}}D'\left(s\right)\left(\ell G_{P,1}\left(s\right)+G_{P,0}\left(s\right)\right)dsd\Lambda\left(T\right)-\nonumber \\
 & \lambda_{1}G_{P,1}\left(0\right)-\lambda_{0}G_{P,0}\left(0\right)\leq0\nonumber \\
\partial L\left(\hat{x};\hat{x}\right)= & 0\label{eq:FOCeq}
\end{align}
where in the above $x=\left(G_{P,1}\left(t\right),G_{P,0}\left(t\right)\right)$
is an arbitrary member of $\Omega$ and $\mu_{A}\left(T\right)$ is
the belief of the agent given by $\ell G_{P,1}\left(T\right)/\left(\ell G_{P,1}\left(T\right)+G_{P,0}\left(T\right)\right)$
calculated at the optimum. This proves the claim for case 1.

To show the result for case 2, i.e., $D\left(\overline{T}\right)=0$,
we use continuity arguments. Since $D\left(\hat{T}\right)>0$, we
can apply the above logic to $T\in\left[0,\hat{T}\right]$ with $\hat{T}<\overline{T}$.
We then consider the solution $\hat{x}_{\hat{T}}$ and its associated
multiplier $\Lambda_{\hat{T}}$ that satisfy the above condition.
We extend $\Lambda_{\hat{T}}$ to the entire interval $\left[0,\overline{T}\right]$
by setting $\Lambda_{\hat{T}}\left(T\right)=\Lambda_{\hat{T}}\left(\hat{T}\right),\forall T\geq\hat{T}$.
We use the following lemma to characterize the properties of a limit
point of the solutions $\hat{x}_{\hat{T}}$ and multipliers $\Lambda_{\hat{T}}$.
\begin{lem}
\label{lem:There-exists-a}There exists a constant $C>0$ such that
for all values of $\hat{T}<\overline{T}$, $D\left(\hat{T}\right)^{\frac{v\left(1\right)}{v\left(1\right)-v\left(1/2\right)}}\Lambda_{\hat{T}}\left(\hat{T}\right)\leq C$.
\end{lem}
We relegate the proof to Section \ref{subsec:Proof-of-Lemma}.

Since each function $\Lambda_{\hat{T}}\left(\cdot\right)$ is increasing,
Lemma \ref{lem:There-exists-a} together with the compactness of bounded
functions in the space of functions with bounded variations\footnote{By the Banach-Alaoglu theorem.}
imply that there exists a bounded function with bounded variation
$\overline{\Gamma}\left(T\right)$ and a subsequence of $\Lambda_{\hat{T}}$'s,
given by$\Lambda_{k}$'s, so that
\begin{equation}
\lim_{k\rightarrow\infty}\sup_{T\in\left(0,\overline{T}\right)}\left|D\left(T\right)^{\frac{v\left(1\right)}{v\left(1\right)-v\left(1/2\right)}}\Lambda_{k}\left(T\right)-\overline{\Gamma}\left(T\right)\right|=0\label{eq: normlim}
\end{equation}
Moreover, since each $\Lambda_{k}$ is increasing so is $\overline{\Gamma}\left(T\right)D\left(T\right)^{-\frac{v\left(1\right)}{v\left(1\right)-v\left(1/2\right)}}$. 

Additionally, note that $\frac{v\left(1\right)}{v\left(1\right)-v\left(1/2\right)}>1$,
$\int_{0}^{\overline{T}}D\left(T\right)^{-\frac{v\left(1\right)}{v\left(1\right)-v\left(1/2\right)}}dT<\infty$.
This implies that the measure $\overline{\Lambda}$ associated with
the function $\overline{\Gamma}\left(T\right)D\left(T\right)^{-\frac{v\left(1\right)}{v\left(1\right)-v\left(1/2\right)}}$,
i.e., $\overline{\Gamma}\left(T\right)D\left(T\right)^{-\frac{v\left(1\right)}{v\left(1\right)-v\left(1/2\right)}}$
is the CDF of $\overline{\Lambda}$, is a finite measure. This implies
that $\int fd\Lambda_{k}\rightarrow\int fd\overline{\Lambda}$ for
all functions with bounded derivatives.

Since $x_{\hat{T}}$'s are bounded, they must converge as $\hat{T}\rightarrow\overline{T}$.
By Berge's maximum theorem, the limit of $x_{\hat{T}}$ should be
an optimizer for $\overline{T}$. The result then follows from the
continuity of the expressions in (\ref{eq: FOCL}).
\end{proof}
\newpage{}

\subsection{Proof of Proposition \ref{prop:When--is}}
\begin{proof}
We use Theorem \ref{thm: thm1} and construct multiplier so that abrupt
revelation is optimal. 
\begin{align*}
\Lambda\left(t\right)= & \begin{cases}
0 & t=0\\
\frac{e^{\left(\delta_{A}-\delta_{P}\right)t^{*}}}{v\left(1\right)\delta_{A}} & t>0
\end{cases}.\\
\lambda_{1}= & \frac{1-e^{-\delta_{P}t^{*}}}{\delta_{P}}+\frac{e^{\left(\delta_{A}-\delta_{P}\right)t^{*}}}{\delta_{A}v\left(1\right)}\left(v\left(1\right)e^{-\delta_{A}t^{*}}-v\left(\mu_{A}\right)-v'\left(\mu_{A}\right)\left(1-\mu_{A}\right)\right)\\
\lambda_{0}= & \frac{1-e^{-\delta_{P}t^{*}}}{\delta_{P}}+\frac{e^{\left(\delta_{A}-\delta_{P}\right)t^{*}}}{\delta_{A}v\left(1\right)}\left(v\left(1\right)e^{-\delta_{A}t^{*}}-v\left(\mu_{A}\right)+v'\left(\mu_{A}\right)\mu_{A}\right)
\end{align*}
where 
\[
t^{*}=\frac{1}{\delta_{A}}\log\frac{v\left(1\right)}{v\left(\mu_{A}\right)}
\]
Under abrupt revelation, the distribution of exit rates are 
\begin{align*}
\hat{G}_{P,1}\left(t\right) & =\mu_{P}\mathbf{1}\left[t<t^{*}\right]\\
\hat{G}_{P,0}\left(t\right) & =\left(1-\mu_{P}\right)\mathbf{1}\left[t<t^{*}\right]
\end{align*}
As we show in the proof of Theorem \ref{thm: thm1}, $\partial L\left(\hat{x};\hat{x}\right)=0$
is equivalent to
\[
\int_{0}^{t^{*}}e^{-\delta_{P}t}dt-\lambda_{1}\mu_{P}-\lambda_{0}\left(1-\mu_{P}\right)=0
\]
which is satisfied by construction.

Next, we need to show that for all non-increasing positive functions
$x=\left(G_{P,0}\left(t\right),G_{P,1}\left(t\right)\right)$, $\partial L\left(\hat{x};x\right)\leq0$
is satisfied. We have
\begin{align*}
\partial L\left(\hat{x};x\right)= & \int_{0}^{\infty}\left(G_{P,1}\left(t\right)+G_{P,0}\left(t\right)\right)e^{-\delta_{P}t}dt-\\
 & \left[v\left(\mu_{A}\right)-v\left(1\right)\right]\left(G_{P,1}\left(0\right)+G_{P,0}\left(0\right)\right)\Lambda\left(0+\right)-\\
 & \left[v'\left(\mu_{A}\right)\left(1-\mu_{A}\right)G_{P,1}\left(0\right)-v'\left(\mu_{A}\right)\mu_{A}\left(0\right)G_{P,0}\left(0\right)\right]\Lambda\left(0+\right)+\\
 & -v\left(1\right)\int_{0}^{\infty}\delta_{A}e^{-\delta_{A}s}\left(G_{P,1}\left(s\right)+G_{P,0}\left(s\right)\right)ds\Lambda\left(0+\right)-\\
 & \lambda_{1}G_{P,1}\left(0\right)-\lambda_{0}G_{P,0}\left(0\right)
\end{align*}
The above is linear in $x$. Thus, if we show that for any positive
and decreasing function $f\left(t\right)$, $\partial L\left(\hat{x};\left(f,0\right)\right)\leq0$
and $\partial L\left(\hat{x};\left(0,f\right)\right)\leq0$, this
proves the desired result. We will show that $\partial L\left(\hat{x};\left(f,0\right)\right)\leq0$;
the second inequality can be shown similarly. We have
\begin{align*}
\partial L\left(\hat{x};x\right)= & \int_{0}^{\infty}f\left(t\right)e^{-\delta_{P}t}dt-\\
 & f\left(0\right)\left[v\left(\mu_{A}\right)-v\left(1\right)+v'\left(\mu_{A}\right)\left(1-\mu_{A}\right)\right]\Lambda\left(0+\right)-\\
 & -v\left(1\right)\int_{0}^{\infty}\delta_{A}e^{-\delta_{A}s}f\left(s\right)ds\Lambda\left(0+\right)-\lambda_{1}f\left(0\right)\\
= & \int_{0}^{\infty}f\left(t\right)e^{-\delta_{P}t}dt-\\
 & f\left(0\right)\left[v\left(\mu_{A}\right)-v\left(1\right)+v'\left(\mu_{A}\right)\left(1-\mu_{A}\right)\right]\Lambda\left(0+\right)-\\
 & -v\left(1\right)\int_{0}^{\infty}\delta_{A}e^{-\delta_{A}s}f\left(s\right)ds\Lambda\left(0+\right)\\
 & -f\left(0\right)\frac{1-e^{-\delta_{P}t^{*}}}{\delta_{P}}-f\left(0\right)\Lambda\left(0+\right)\left(v\left(1\right)e^{-\delta_{A}t^{*}}-v\left(\mu_{A}\right)-v'\left(\mu_{A}\right)\left(1-\mu_{A}\right)\right)\\
= & \int_{0}^{\infty}f\left(t\right)e^{-\delta_{P}t}dt-f\left(0\right)\frac{1-e^{-\delta_{P}t^{*}}}{\delta_{P}}\\
 & +\Lambda\left(0+\right)\left[v\left(1\right)f\left(0\right)-v\left(1\right)e^{-\delta_{A}t^{*}}f\left(0\right)-v\left(1\right)\int_{0}^{\infty}\delta_{A}e^{-\delta_{A}s}f\left(s\right)ds\right]\\
\leq & \int_{0}^{\infty}\left(f\left(t\right)-f\left(0\right)\right)e^{-\delta_{P}t}dt+f\left(0\right)\frac{e^{\delta_{P}t^{*}}}{\delta_{P}}\\
 & +\Lambda\left(0+\right)\left[v\left(1\right)f\left(0\right)-v\left(1\right)e^{-\delta_{A}t^{*}}f\left(0\right)-v\left(1\right)f\left(0\right)\right]\\
= & \int_{0}^{\infty}\left(f\left(t\right)-f\left(0\right)\right)e^{-\delta_{P}t}dt+\left[\frac{e^{\delta_{P}t^{*}}}{\delta_{P}}-\Lambda\left(0+\right)v\left(1\right)e^{-\delta_{A}t^{*}}\right]f\left(0\right)\\
= & \int_{0}^{\infty}\left(f\left(t\right)-f\left(0\right)\right)e^{-\delta_{P}t}dt+\left[\frac{e^{\delta_{P}t^{*}}}{\delta_{P}}-v\left(1\right)e^{-\delta_{A}t^{*}}\frac{e^{\left(\delta_{A}-\delta_{P}\right)t^{*}}}{v\left(1\right)\delta_{A}}\right]f\left(0\right)\\
= & \int_{0}^{\infty}\left(f\left(t\right)-f\left(0\right)\right)e^{-\delta_{P}t}dt+\left[\frac{e^{-\delta_{P}t^{*}}}{\delta_{P}}-v\left(1\right)e^{-\delta_{A}t^{*}}\frac{e^{\left(\delta_{A}-\delta_{P}\right)t^{*}}}{v\left(1\right)\delta_{A}}\right]f\left(0\right)\\
= & \int_{0}^{\infty}\left(f\left(t\right)-f\left(0\right)\right)e^{-\delta_{P}t}dt+e^{-\delta_{P}t^{*}}\left[\frac{1}{\delta_{P}}-\frac{1}{\delta_{A}}\right]f\left(0\right)\leq0
\end{align*}
where the above is negative since $\delta_{P}\geq\delta_{A}$ and
$f\left(t\right)\leq f\left(0\right)$. Finally, we should note that
$\Lambda\left(\cdot\right)$ satisfies the complementary slackness
condition as $d\Lambda\left(t\right)=0$ for all $t>0$ and the incentive
constraint is slack, while $d\Lambda\left(0\right)=\Lambda\left(0+\right)>0$
and the incentive constraint is binding. This concludes the proof.
\end{proof}
\newpage{}

\subsection{Proof of Proposition \ref{prop:When--is-1}}
\begin{proof}
For the proof, we will focus on the $\mu_{P}\geq1/2$ case. The proof
for the case with $\mu_{P}<1/2$ is almost identical. Let $\mu_{P}^{*}\left(t\right)$
be the solution to the following differential equation:
\[
-\delta_{A}=\frac{\mu_{P}'\left(t\right)}{v\left(\mu_{P}\left(t\right)\right)}\left[\frac{v\left(1\right)-v\left(\mu_{P}\left(t\right)\right)}{1-\mu_{P}\left(t\right)}-v'\left(\mu_{P}\left(t\right)\right)\right]
\]
with initial value given by $\mu_{P}^{*}\left(0\right)=\mu_{P}\geq1/2$.
This differential equation is derived from the binding incentive constraint
\ref{eq: Obedience} using the fact that $G_{P,0}\left(t\right)=1-\mu_{P}$
in the first phase. Simple calculations allow us to show that the
solution of the above satisfies
\[
-\delta_{A}t=\int_{\mu_{P}}^{\mu_{P}^{*}\left(t\right)}\frac{v\left(1\right)-v\left(\mu\right)-\left(1-\mu\right)v'\left(\mu\right)}{\left(1-\mu\right)v\left(\mu\right)}d\mu
\]
and thus $\mu_{P}^{*}\left(t\right)$ exists, and it is unique and
decreasing. Moreover, we can define $t^{*}$ as the first time that
$\mu_{P}^{*}\left(t^{*}\right)=1/2$ given by
\[
\delta t^{*}=\int_{1/2}^{\mu_{P}}\frac{v\left(1\right)-v\left(\mu\right)-\left(1-\mu\right)v'\left(\mu\right)}{\left(1-\mu\right)v\left(\mu\right)}d\mu.
\]
We extend $\mu_{P}^{*}\left(t\right)$ beyond $t^{*}$ by setting
it equal to $1/2$. In addition, the function $\mu_{P}^{*}\left(t\right)$
defines the engagement probability functions given by
\begin{align*}
G_{P,1}^{*}\left(t\right)= & \begin{cases}
\frac{\mu_{P}^{*}\left(t\right)}{\mu_{P}^{*}\left(t\right)+1-\mu_{P}} & t\leq t^{*}\\
\left(1-\mu_{P}\right)e^{-\lambda^{*}\left(t-t^{*}\right)} & t\geq t^{*}
\end{cases}\\
G_{P,0}^{*}\left(t\right)= & \begin{cases}
1-\mu_{P} & t\leq t^{*}\\
\left(1-\mu_{P}\right)e^{-\lambda^{*}\left(t-t^{*}\right)} & t\geq t^{*}
\end{cases}
\end{align*}
where $\lambda^{*}=\delta_{A}v\left(1\right)/v\left(1/2\right)-\delta_{A}$.
The variable $\lambda^{*}$ is the arrival rate of revelation in the
second phase and is calculated using the \ref{eq: Obedience}. Additionally,
for all values of $t$, let $\Lambda\left(t\right)$ be defined by
\[
\Delta_{1}\left(t\right)\Lambda'\left(t\right)-\delta_{A}v\left(1\right)\Lambda\left(t\right)=-e^{-\left(\delta_{P}-\delta_{A}\right)t}
\]
where $\Delta_{1}\left(t\right)=v\left(1\right)-v\left(\mu_{P}^{*}\left(t\right)\right)-\left(1-\mu_{P}^{*}\left(t\right)\right)v'\left(\mu_{P}^{*}\left(t\right)\right)$.
This is a simple ordinary differential equation whose solution is
given by

\begin{align}
\Lambda\left(t\right) & =\Gamma\left(t\right)\Lambda\left(0+\right)-\Gamma\left(t\right)\int_{0}^{t}\frac{e^{\left(\delta_{A}-\delta_{P}\right)s}}{\Gamma\left(s\right)\Delta_{1}\left(s\right)}ds,\label{eq: Lambda}\\
\log\Gamma\left(t\right) & =\delta_{A}v\left(1\right)\int_{0}^{t}\frac{ds}{\Delta_{1}\left(s\right)}.\nonumber 
\end{align}
Finally, we set
\begin{align*}
\Lambda\left(0+\right)= & \int_{0}^{\infty}\frac{e^{\left(\delta_{A}-\delta_{P}\right)t}}{\Gamma\left(t\right)\Delta_{1}\left(t\right)}dt\\
\lambda_{1}= & \Lambda\left(0+\right)\Delta_{1}\left(0\right)>0\\
\lambda_{0}= & \Lambda\left(0+\right)\Delta_{0}\left(0\right)\\
 & +\int_{0}^{t^{*}}\left(e^{-\delta_{P}t}+e^{-\delta_{A}t}\Delta_{0}\left(t\right)\Lambda'\left(t\right)-\delta_{A}e^{-\delta_{A}t}v\left(1\right)\Lambda\left(t\right)\right)dt
\end{align*}
where $\Lambda\left(0+\right)$ exists because for large values of
$t$, $\frac{d}{dt}\log\Gamma\left(t\right)=\frac{\delta_{A}v\left(1\right)}{v\left(1\right)-v\left(1/2\right)}>\delta_{A}$
and $\Delta_{1}\left(t\right)=v\left(1\right)-v\left(1/2\right)$. 

We can rewrite the derivative of the Lagrangian using integration
by part as:
\begin{align*}
\partial L\left(\hat{x};x\right)= & \int_{0}^{\infty}\left(G_{P,1}\left(t\right)+G_{P,0}\left(t\right)\right)e^{-\delta_{P}t}dt+\\
 & \int_{0}^{\infty}e^{-\delta_{A}t}\left(\Delta_{1}\left(t\right)G_{P,1}\left(t\right)+\Delta_{0}\left(t\right)G_{P,0}\left(t\right)\right)d\Lambda\left(t\right)\\
 & -v\left(1\right)\int_{0}^{\infty}\delta_{A}e^{-\delta_{A}t}\left(G_{P,1}\left(t\right)+G_{P,0}\left(t\right)\right)\Lambda\left(t\right)dt\\
 & -\lambda_{1}G_{P,1}\left(0\right)-\lambda_{0}G_{P,0}\left(0\right)
\end{align*}
where in the above $\Delta_{0}\left(t\right)=v\left(1\right)-v\left(\mu_{P}^{*}\left(t\right)\right)+\mu_{P}^{*}\left(t\right)v'\left(\mu_{P}^{*}\left(t\right)\right)$.
Since $\mu_{P}^{*}\left(t\right)=1/2$ for all $t\geq t^{*}$ which
implies that $\Delta_{0}\left(t\right)=\Delta_{1}\left(t\right)=v\left(1\right)-v\left(1/2\right)$
for all values of $t\geq t^{*}$. We can thus write
\begin{align*}
\partial L\left(\hat{x};x\right)= & \int_{0}^{\infty}G_{P,1}\left(t\right)\left[e^{-\delta_{P}t}+e^{-\delta_{A}t}\Delta_{1}\left(t\right)\Lambda'\left(t\right)-v\left(1\right)\delta_{A}e^{-\delta_{A}t}\Lambda\left(t\right)\right]dt\\
 & \int_{0}^{\infty}G_{P,0}\left(t\right)\left[e^{-\delta_{P}t}+e^{-\delta_{A}t}\Delta_{0}\left(t\right)\Lambda'\left(t\right)-v\left(1\right)\delta_{A}e^{-\delta_{A}t}\Lambda\left(t\right)\right]dt\\
 & +\Delta_{1}\left(0\right)G_{P,1}\left(0\right)\Lambda\left(0+\right)+\Delta_{0}\left(0\right)G_{P,0}\left(0\right)\Lambda\left(0+\right)\\
 & -\lambda_{1}G_{P,1}\left(0\right)-\lambda_{0}G_{P,0}\left(0\right)\\
= & \int_{0}^{t^{*}}G_{P,0}\left(t\right)\left[e^{-\delta_{P}t}+e^{-\delta_{A}t}\Delta_{0}\left(t\right)\Lambda'\left(t\right)-v\left(1\right)\delta_{A}e^{-\delta_{A}t}\Lambda\left(t\right)\right]dt\\
 & +\Delta_{1}\left(0\right)G_{P,1}\left(0\right)\Lambda\left(0+\right)+\Delta_{0}\left(0\right)G_{P,0}\left(0\right)\Lambda\left(0+\right)-\lambda_{1}G_{P,1}\left(0\right)-\lambda_{0}G_{P,0}\left(0\right)\\
= & \int_{0}^{t^{*}}G_{P,0}\left(t\right)\left[e^{-\delta_{P}t}+e^{-\delta_{A}t}\Delta_{0}\left(t\right)\Lambda'\left(t\right)-v\left(1\right)\delta_{A}e^{-\delta_{A}t}\Lambda\left(t\right)\right]dt\\
 & -G_{P,0}\left(0\right)\int_{0}^{t^{*}}\left(e^{-\delta_{P}t}+e^{-\delta_{A}t}\Delta_{0}\left(t\right)\Lambda'\left(t\right)-\delta_{A}e^{-\delta_{A}t}v\left(1\right)\Lambda\left(t\right)\right)dt\\
= & \int_{0}^{t^{*}}\left(G_{P,0}\left(t\right)-G_{P,0}\left(0\right)\right)\left[e^{-\delta_{P}t}+e^{-\delta_{A}t}\Delta_{0}\left(t\right)\Lambda'\left(t\right)-v\left(1\right)\delta_{A}e^{-\delta_{A}t}\Lambda\left(t\right)\right]dt\\
\leq & \int_{0}^{t^{*}}\left(G_{P,0}\left(t\right)-G_{P,0}\left(0\right)\right)\underbrace{\left[e^{-\delta_{P}t}+e^{-\delta_{A}t}\Delta_{1}\left(t\right)\Lambda'\left(t\right)-v\left(1\right)\delta_{A}e^{-\delta_{A}t}\Lambda\left(t\right)\right]}_{=0}dt\\
= & 0
\end{align*}
In the above, the inequality follows from two observations: first,
since $G_{P,0}\left(t\right)$ is non-increasing, it must be that
$G_{P,0}\left(t\right)-G_{P,0}\left(0\right)\leq0$; second, $\mu_{P}^{*}\left(t\right)\geq1/2$
which then implies that $\Delta_{0}\left(t\right)\geq\Delta_{1}\left(t\right)$
and as we will show later $\Lambda'\left(t\right)\geq0$.

To finish the proof, we need to show that $\Lambda'\left(t\right),\lambda_{0}\geq0$.
To see these, recall from (\ref{eq: Lambda}) that
\begin{align*}
\Lambda'\left(t\right) & =\Gamma'\left(t\right)\Lambda\left(0+\right)-\Gamma'\left(t\right)\int_{0}^{t}\frac{e^{\left(\delta_{A}-\delta_{P}\right)s}}{\Gamma\left(s\right)\Delta_{1}\left(s\right)}ds-\frac{e^{\left(\delta_{A}-\delta_{P}\right)t}}{\Delta_{1}\left(t\right)}\\
\frac{\Lambda'\left(t\right)}{\Gamma\left(t\right)}\Delta_{1}\left(t\right) & =\delta_{A}v\left(1\right)\Lambda\left(0+\right)-\delta_{A}v\left(1\right)\int_{0}^{t}\frac{e^{\left(\delta_{A}-\delta_{P}\right)s}}{\Gamma\left(s\right)\Delta_{1}\left(s\right)}ds-\frac{e^{\left(\delta_{A}-\delta_{P}\right)t}}{\Gamma\left(t\right)}\\
 & =\delta_{A}v\left(1\right)\int_{0}^{\infty}\frac{e^{\left(\delta_{A}-\delta_{P}\right)s}}{\Gamma\left(s\right)\Delta_{1}\left(s\right)}ds-\delta_{A}v\left(1\right)\int_{0}^{t}\frac{e^{\left(\delta_{A}-\delta_{P}\right)s}}{\Gamma\left(s\right)\Delta_{1}\left(s\right)}ds-\frac{e^{\left(\delta_{A}-\delta_{P}\right)t}}{\Gamma\left(t\right)}\\
 & =\delta_{A}v\left(1\right)\int_{t}^{\infty}\frac{e^{\left(\delta_{A}-\delta_{P}\right)s}}{\Gamma\left(s\right)\Delta_{1}\left(s\right)}ds-\frac{e^{\left(\delta_{A}-\delta_{P}\right)t}}{\Gamma\left(t\right)}\\
 & =\int_{t}^{\infty}\frac{\Gamma'\left(s\right)e^{\left(\delta_{A}-\delta_{P}\right)s}}{\Gamma\left(s\right)^{2}}ds-\frac{e^{\left(\delta_{A}-\delta_{P}\right)t}}{\Gamma\left(t\right)}=-\int_{t}^{\infty}e^{\left(\delta_{A}-\delta_{P}\right)s}d\left(\frac{1}{\Gamma\left(s\right)}\right)-\frac{e^{\left(\delta_{A}-\delta_{P}\right)t}}{\Gamma\left(t\right)}\\
 & =-\lim_{s\rightarrow\infty}\frac{e^{\left(\delta_{A}-\delta_{P}\right)s}}{\Gamma\left(s\right)}+\frac{e^{\left(\delta_{A}-\delta_{P}\right)t}}{\Gamma\left(t\right)}+\left(\delta_{A}-\delta_{P}\right)\int_{t}^{\infty}\frac{e^{\left(\delta_{A}-\delta_{P}\right)s}}{\Gamma\left(s\right)}ds-\frac{e^{\left(\delta_{A}-\delta_{P}\right)t}}{\Gamma\left(t\right)}\\
 & =\left(\delta_{A}-\delta_{P}\right)\int_{t}^{\infty}\frac{e^{\left(\delta_{A}-\delta_{P}\right)s}}{\Gamma\left(s\right)}ds>0
\end{align*}
where in the above, we have used the definition of $\Gamma\left(t\right)$
and the fact that as $t\rightarrow\infty$, $\Gamma\left(t\right)\propto e^{\delta_{A}\frac{v\left(1\right)}{v\left(1\right)-v\left(1/2\right)}t}$.
Finally,
\begin{align*}
\lambda_{0} & =\Lambda\left(0+\right)\Delta_{0}\left(0\right)+\int_{0}^{t^{*}}\left(e^{-\delta_{P}t}+e^{-\delta_{A}t}\Delta_{0}\left(t\right)\Lambda'\left(t\right)-\delta_{A}e^{-\delta_{A}t}v\left(1\right)\Lambda\left(t\right)\right)dt\\
 & \geq\int_{0}^{t^{*}}\left(e^{-\delta_{P}t}+e^{-\delta_{A}t}\Delta_{1}\left(t\right)+\left(\Delta_{0}\left(t\right)-\Delta_{1}\left(t\right)\right)\Lambda'\left(t\right)-\delta_{A}e^{-\delta_{A}t}v\left(1\right)\Lambda\left(t\right)\right)dt\\
 & =\int_{0}^{t^{*}}\left(\Delta_{0}\left(t\right)-\Delta_{1}\left(t\right)\right)\Lambda'\left(t\right)dt>0
\end{align*}
where the inequality follows from the fact that $\Delta_{0}\left(t\right)\geq\Delta_{1}\left(t\right)\geq0$
for all values of $\mu_{P}^{*}\left(t\right)\geq1/2$. This concludes
the proof.
\end{proof}
\newpage{}

\subsection{Proof of Proposition \ref{thm2: Bias}}
\begin{proof}
As in the statement of the Proposition, we focus on $\ell<1$; the
case with $\ell>1$ is the mirror of this case. To prove the claim
of the proposition, we first define a class of exit probability functions
parametrized by $\mu_{1}\in\left[\mu_{A},1\right]$ -- the agent's
belief at the end of phase 1. Given $\mu_{1}$, the length of phase
1, $t_{1}$, and end of engagement, $t_{2}$ as well as the agent's
belief is given by 
\begin{align}
\hat{\mu}_{A}\left(t\right) & =\mu_{A}:0\leq t<t_{1}\nonumber \\
e^{-\delta_{A}t_{1}}\left(\frac{\mu_{A}}{\mu_{1}}v\left(\mu_{1}\right)+\left(1-\frac{\mu_{A}}{\mu_{1}}\right)v\left(0\right)\right) & =v\left(\mu_{A}\right)\nonumber \\
\frac{d\hat{\mu}_{A}\left(t\right)}{dt}\frac{v\left(1\right)-v\left(\hat{\mu}_{A}\left(t\right)\right)+\hat{\mu}_{A}\left(t\right)v'\left(\hat{\mu}_{A}\left(t\right)\right)}{\hat{\mu}_{A}\left(t\right)v\left(\hat{\mu}_{A}\left(t\right)\right)} & =\delta_{A},t_{2}>t\geq t_{1}\label{eq: ODEmu}\\
\hat{\mu}_{1}\left(t_{1}\right)=\mu_{1},\hat{\mu}_{A}\left(t_{2}\right) & =1.\nonumber 
\end{align}
When $\mu_{1}=\mu_{A}$ the length of phase 1 is 0 while its maximum
length is $t^{*}=\frac{1}{\delta_{A}}\log\frac{v\left(1\right)}{v\left(\mu_{A}\right)}$
when $\mu_{1}=1$. These beliefs are associated with engagement probability
functions given by
\begin{align*}
\hat{G}_{P,1}\left(t\right) & =\begin{cases}
\mu_{P} & 0\leq t<t_{2}\\
0 & t_{2}\leq t
\end{cases}\\
\hat{G}_{P,0}\left(t\right) & =\begin{cases}
1-\mu_{P} & 0\leq t<t_{1}\\
\ell\mu_{P}\frac{1-\hat{\mu}_{A}\left(t\right)}{\hat{\mu}_{A}\left(t\right)} & t_{1}\leq t\leq t_{2}
\end{cases}
\end{align*}
For each parameter value $\mu_{1}$, we also define the following
Lagrange multipliers given by
\[
\Lambda\left(t\right)=\begin{cases}
0 & t=0\\
\frac{e^{\left(\delta_{A}-\delta_{P}\right)t_{1}}}{\delta_{A}v\left(1\right)} & 0<t\leq t_{1}
\end{cases}
\]
and for values of $t\in\left[t_{1},t_{2}\right]$ the ODE
\begin{equation}
e^{\left(\delta_{A}-\delta_{P}\right)t}+\Delta_{0}\left(t\right)\Lambda'\left(t\right)-\delta_{A}v\left(1\right)\Lambda\left(t\right)=0\label{eq:ODE}
\end{equation}
holds where $\Delta_{0}\left(t\right)=v\left(1\right)-v\left(\hat{\mu}_{A}\left(t\right)\right)+\hat{\mu}_{A}\left(t\right)v'\left(\hat{\mu}_{A}\left(t\right)\right)$
and finally, we set $\Lambda\left(t\right)=\Lambda\left(t_{2}\right)$
for values of $t\geq t_{2}$. Let $\mu_{1}^{*}$ be the value of the
parameter $\mu_{1}$ such that either the above multiplier satisfies
$\Lambda\left(t_{2}\right)=e^{\left(\delta_{A}-\delta_{P}\right)t_{2}}/\left(\ell\delta_{A}v\left(1\right)\right)$
and if such value does not exist, we set $\mu_{1}^{*}=\mu_{A}$. Later,
we will show that there exists at most one such value of $\mu_{1}$
and that $\mu_{1}^{*}<1$ has to hold. In other words, if $\mu_{1}=1$,
we will show that $\Lambda\left(t_{2}\right)<e^{\left(\delta_{A}-\delta_{P}\right)t_{2}}/\left(\ell\delta_{A}v\left(1\right)\right)$.
Let $t_{1}^{*}$ and $t_{2}^{*}$ be the associated end of phase 1
and phase 2. Finally, we define $\lambda_{1},\lambda_{0}$ as
\begin{align*}
\lambda_{0}= & \int_{0}^{t_{1}^{*}}e^{-\delta_{P}s}ds+\Lambda\left(0+\right)\left(v\left(1\right)e^{-\delta_{A}t_{1}^{*}}-v\left(\mu_{A}\right)+\mu_{A}v'\left(\mu_{A}\right)\right)\\
\lambda_{1}= & \int_{0}^{t_{2}^{*}}e^{-\delta_{P}s}ds+\ell\Lambda\left(0+\right)\left(v\left(1\right)-v\left(\mu_{A}\right)-\left(1-\mu_{A}\right)v'\left(\mu_{A}\right)\right)\\
 & +\ell\int_{0}^{t_{2}^{*}}e^{-\delta_{A}s}\left(\Delta_{1}\left(s\right)\Lambda'\left(s\right)-\delta_{A}v\left(1\right)\Lambda\left(s\right)\right)ds
\end{align*}

For this allocation, we show that $\partial L\left(\hat{x};\hat{x}\right)=0$
and $\partial L\left(\hat{x};x\right)\leq0$ for all non-increasing
and positive pair of functions $x=\left(G_{P,0}\left(t\right),G_{P,1}\left(t\right)\right)$.
Moreover, the multipliers $\Lambda,\lambda_{1},\lambda_{0}$ need
to satisfy the complementary slackness and $\Lambda\left(t\right)$
has to be non-decreasing.
\begin{enumerate}
\item \textbf{Complementary Slackness (CS): }The incentive constraint is
binding when $t=0$ and when $t\in\left[t_{1}^{*},t_{2}^{*}\right]$.
Since $\Lambda\left(t\right)$ is flat for $t\in\left(0,t_{1}^{*}\right)$,
for CS to be satisfied, we need $\Lambda\left(0+\right)=\frac{e^{\left(\delta_{A}-\delta_{P}\right)t_{1}^{*}}}{\delta_{A}v\left(1\right)}$
to be positive. which holds, and $\Lambda\left(t\right)$ to be increasing
when $t\in\left[t_{1}^{*},t_{2}^{*}\right]$. The ODE in \ref{eq:ODE}
has an exact solution given by
\begin{align}
\Lambda\left(t\right) & =\Gamma\left(t\right)\left[\frac{e^{\left(\delta_{A}-\delta_{P}\right)t_{1}^{*}}}{\delta_{A}v\left(1\right)}-\int_{t_{1}}^{t}\frac{e^{\left(\delta_{A}-\delta_{P}\right)s}}{\Delta_{0}\left(s\right)\Gamma\left(s\right)}ds\right]\label{eq: LamSol}\\
\log\Gamma\left(t\right) & =\int_{t_{1}}^{t}\frac{\delta_{A}v\left(1\right)}{\Delta_{0}\left(s\right)}ds\nonumber 
\end{align}
Differentiating $\Lambda\left(t\right)$, we have
\begin{align*}
\Lambda'\left(t\right)= & \frac{\delta_{A}v\left(1\right)}{\Delta_{0}\left(t\right)}\Gamma\left(t\right)\left[\frac{e^{\left(\delta_{A}-\delta_{P}\right)t_{1}^{*}}}{\delta_{A}v\left(1\right)}-\int_{t_{1}}^{t}\frac{e^{\left(\delta_{A}-\delta_{P}\right)s}}{\Delta_{0}\left(s\right)\Gamma\left(s\right)}ds\right]\\
 & -\frac{e^{\left(\delta_{A}-\delta_{P}\right)t}}{\Delta_{0}\left(t\right)}\\
\rightarrow\Lambda'\left(t\right)\frac{\Delta_{0}\left(t\right)}{\Gamma\left(t\right)\delta_{A}v\left(1\right)}= & \frac{e^{\left(\delta_{A}-\delta_{P}\right)t_{1}^{*}}}{\delta_{A}v\left(1\right)}-\int_{t_{1}}^{t}\frac{e^{\left(\delta_{A}-\delta_{P}\right)s}}{\Delta_{0}\left(s\right)\Gamma\left(s\right)}ds-\frac{e^{\left(\delta_{A}-\delta_{P}\right)t}}{\Gamma\left(t\right)\delta_{A}v\left(1\right)}\\
\rightarrow\frac{d}{dt}\left(\Lambda'\left(t\right)\frac{\Delta_{0}\left(t\right)}{\Gamma\left(t\right)\delta_{A}v\left(1\right)}\right)= & -\frac{e^{\left(\delta_{A}-\delta_{P}\right)t}}{\Delta_{0}\left(t\right)\Gamma\left(t\right)}-\left(\delta_{A}-\delta_{P}\right)\frac{e^{\left(\delta_{A}-\delta_{P}\right)t}}{\Gamma\left(t\right)\delta_{A}v\left(1\right)}+\frac{\Gamma'\left(t\right)}{\Gamma\left(t\right)}\frac{e^{\left(\delta_{A}-\delta_{P}\right)t}}{\Gamma\left(t\right)\delta_{A}v\left(1\right)}\\
= & -\frac{e^{\left(\delta_{A}-\delta_{P}\right)t}}{\Delta_{0}\left(t\right)\Gamma\left(t\right)}-\left(\delta_{A}-\delta_{P}\right)\frac{e^{\left(\delta_{A}-\delta_{P}\right)t}}{\Gamma\left(t\right)\delta_{A}v\left(1\right)}+\frac{\delta_{A}v\left(1\right)}{\Delta_{0}\left(t\right)}\frac{e^{\left(\delta_{A}-\delta_{P}\right)t}}{\Gamma\left(t\right)\delta_{A}v\left(1\right)}\\
= & \left(\delta_{P}-\delta_{A}\right)\frac{e^{\left(\delta_{A}-\delta_{P}\right)t}}{\Gamma\left(t\right)\delta_{A}v\left(1\right)}\geq0
\end{align*}
Since $\Delta_{0}\left(t\right),\Gamma\left(t\right)\geq0$ for all
values of $t$, the above implies that if $\Lambda'\left(t_{1}^{*}\right)\geq0$,
then for all values of $t\geq t_{1}$, $\Lambda'\left(t\right)\geq0$.
At $t=t_{1}^{*}$ we have
\begin{align*}
e^{\left(\delta_{A}-\delta_{P}\right)t_{1}^{*}}+\Delta_{0}\left(t_{1}^{*}\right)\Lambda'\left(t_{1}^{*}\right)-\delta_{A}v\left(1\right)\Lambda\left(t_{1}^{*}\right) & =0\\
\rightarrow e^{\left(\delta_{A}-\delta_{P}\right)t_{1}^{*}}+\Delta_{0}\left(t_{1}^{*}\right)\Lambda'\left(t_{1}^{*}\right)-\delta_{A}v\left(1\right)\frac{e^{\left(\delta_{A}-\delta_{P}\right)t_{1}^{*}}}{\delta_{A}v\left(1\right)} & =0\\
\Delta_{0}\left(t_{1}^{*}\right)\Lambda'\left(t_{1}^{*}\right) & =0
\end{align*}
Since $\mu_{1}^{*}\geq\mu_{A}>0$, we must have that $\Delta_{0}\left(t_{1}^{*}\right)>0$
and hence $\Lambda'\left(t_{1}^{*}\right)=0$.\\
Additionally, we have to show that $\lambda_{1},\lambda_{0}\geq0$.
We have
\begin{align*}
\lambda_{0} & =\int_{0}^{t_{1}^{*}}e^{-\delta_{P}s}ds+\Lambda\left(0+\right)\left(v\left(1\right)e^{-\delta_{A}t_{1}^{*}}-v\left(\mu_{A}\right)+\mu_{A}v'\left(\mu_{A}\right)\right)\\
 & =\frac{1-e^{-\delta_{P}t_{1}^{*}}}{\delta_{P}}+\frac{e^{\left(\delta_{A}-\delta_{P}\right)t_{1}^{*}}}{\delta_{A}v\left(1\right)}\left(v\left(1\right)e^{-\delta_{A}t_{1}^{*}}-v\left(1\right)+v\left(1\right)-v\left(\mu_{A}\right)+\mu_{A}v'\left(\mu_{A}\right)\right)\\
 & =\frac{1-e^{-\delta_{P}t_{1}^{*}}}{\delta_{P}}+\frac{e^{-\delta_{P}t_{1}^{*}}-e^{\left(\delta_{A}-\delta_{P}\right)t_{1}^{*}}}{\delta_{A}}+e^{\left(\delta_{A}-\delta_{P}\right)t_{1}^{*}}\frac{v\left(1\right)-v\left(\mu_{A}\right)+\mu_{A}v'\left(\mu_{A}\right)}{\delta_{A}v\left(1\right)}\\
 & =\frac{1-e^{-\delta_{P}t_{1}^{*}}}{\delta_{P}}+e^{-\delta_{P}t_{1}^{*}}\frac{1-e^{\delta_{A}t_{1}^{*}}}{\delta_{A}}+e^{\left(\delta_{A}-\delta_{P}\right)t_{1}^{*}}\frac{v\left(1\right)-v\left(\mu_{A}\right)+\mu_{A}v'\left(\mu_{A}\right)}{\delta_{A}v\left(1\right)}\\
 & =e^{-\delta_{P}t_{1}^{*}}\left(\frac{e^{\delta_{P}t_{1}^{*}}-1}{\delta_{P}}-\frac{e^{\delta_{A}t_{1}^{*}}-1}{\delta_{A}}\right)+e^{\left(\delta_{A}-\delta_{P}\right)t_{1}^{*}}\frac{v\left(1\right)-v\left(\mu_{A}\right)+\mu_{A}v'\left(\mu_{A}\right)}{\delta_{A}v\left(1\right)}
\end{align*}
The first expression above is positive since $\delta_{P}>\delta_{A}$
and the second expression is positive since $v$ is convex and positive.
Therefore, $\lambda_{0}\geq0$. Additionally,
\begin{align*}
\lambda_{1}= & \int_{0}^{t_{2}^{*}}e^{-\delta_{P}s}ds+\ell\Lambda\left(0+\right)\left(v\left(1\right)-v\left(\mu_{A}\right)-\left(1-\mu_{A}\right)v'\left(\mu_{A}\right)\right)\\
 & +\ell\int_{0}^{t_{2}^{*}}e^{-\delta_{A}s}\left(\Delta_{1}\left(s\right)\Lambda'\left(s\right)-\delta_{A}v\left(1\right)\Lambda\left(s\right)\right)ds\\
= & \int_{0}^{t_{1}^{*}}\left(e^{-\delta_{P}s}-\ell\delta_{A}v\left(1\right)\Lambda\left(0+\right)e^{-\delta_{A}s}\right)ds+\ell\Lambda\left(0+\right)\Delta_{1}\left(0\right)\\
 & +\int_{t_{1}^{*}}^{t_{2}^{*}}\left[e^{-\delta_{P}s}+\ell e^{-\delta_{A}s}\left(\Delta_{1}\left(s\right)\Lambda'\left(s\right)-\delta_{A}v\left(1\right)\Lambda\left(s\right)\right)\right]ds\\
= & e^{-\delta_{P}t_{1}^{*}}\int_{0}^{t_{1}^{*}}\left(\underbrace{e^{\delta_{P}\left(t_{1}^{*}-s\right)}-\ell e^{\delta_{A}\left(t_{1}^{*}-s\right)}}_{\geq0}\right)ds+\ell\Lambda\left(0+\right)\Delta_{1}\left(0\right)\\
 & +\int_{t_{1}^{*}}^{t_{2}^{*}}e^{-\delta_{P}s}\left[1+\ell e^{\left(\delta_{P}-\delta_{A}\right)s}\left(\Delta_{1}\left(s\right)\Lambda'\left(s\right)-\delta_{A}v\left(1\right)\Lambda\left(s\right)\right)\right]ds
\end{align*}
The first two terms in the above expression are positive since $\delta_{P}>\delta_{A},\ell<1$.
Additionally, since $\Lambda$ satisfies the ODE (\ref{eq:ODE}),
we can write the integrand in the last integral as 
\begin{align}
1+\ell e^{\left(\delta_{P}-\delta_{A}\right)s}\left(\Delta_{1}\left(s\right)\Lambda'\left(s\right)-\delta_{A}v\left(1\right)\Lambda\left(s\right)\right) & =\nonumber \\
1+\ell e^{\left(\delta_{P}-\delta_{A}\right)s}\left(\frac{\Delta_{1}\left(s\right)}{\Delta_{0}\left(s\right)}\left(\delta_{A}v\left(1\right)\Lambda\left(s\right)-e^{\left(\delta_{A}-\delta_{P}\right)s}\right)-\delta_{A}v\left(1\right)\Lambda\left(s\right)\right) & =\nonumber \\
1-\ell+\ell e^{\left(\delta_{P}-\delta_{A}\right)s}\left(\frac{\Delta_{1}\left(s\right)}{\Delta_{0}\left(s\right)}\left(\delta_{A}v\left(1\right)\Lambda\left(s\right)-e^{\left(\delta_{A}-\delta_{P}\right)s}\right)+e^{\left(\delta_{P}-\delta_{A}\right)s}-\delta_{A}v\left(1\right)\Lambda\left(s\right)\right) & =\nonumber \\
1-\ell+\ell\frac{\Delta_{1}\left(s\right)-\Delta_{0}\left(s\right)}{\Delta_{0}\left(s\right)}\left(e^{\left(\delta_{P}-\delta_{A}\right)s}\delta_{A}v\left(1\right)\Lambda\left(s\right)-1\right) & =\nonumber \\
1-\ell-\ell\frac{v'\left(\hat{\mu}_{A}\left(s\right)\right)}{\Delta_{0}\left(s\right)}\left(e^{\left(\delta_{P}-\delta_{A}\right)s}\delta_{A}v\left(1\right)\Lambda\left(s\right)-1\right)\label{eq: integrand}
\end{align}
In the above, since $\delta_{P}>\delta_{A}$, $\Lambda'\left(s\right)\geq0$,
and $\Lambda\left(t_{1}\right)\delta_{A}v\left(1\right)e^{\left(\delta_{P}-\delta_{A}\right)s}=1$,
$e^{\left(\delta_{P}-\delta_{A}\right)s}\delta_{A}v\left(1\right)\Lambda\left(s\right)-1$
is positive and increasing. If $\hat{\mu}_{A}\leq1/2$, then $v'\leq0$
which implies that the above is positive. Finally, when $\hat{\mu}_{A}\geq1/2$,
$v'\geq0$
\[
\frac{d}{ds}\frac{v'\left(\hat{\mu}_{A}\left(s\right)\right)}{\Delta_{0}\left(s\right)}=\frac{v''\left(\hat{\mu}_{A}\left(s\right)\right)\left(v\left(1\right)-v\left(\hat{\mu}_{A}\left(s\right)\right)\right)}{\Delta_{0}\left(s\right)^{2}}\hat{\mu}_{A}'\left(s\right)\geq0.
\]
As a result, $\frac{v'\left(\hat{\mu}_{A}\left(s\right)\right)}{\Delta_{0}\left(s\right)}\left(e^{\left(\delta_{P}-\delta_{A}\right)s}\delta_{A}v\left(1\right)\Lambda\left(s\right)-1\right)$
is increasing, and therefore the integrand in (\ref{eq: integrand})
is decreasing in $s$. Its value at $t_{2}^{*}$ is given by
\[
1-\ell-\ell\frac{v'\left(1\right)}{v'\left(1\right)}\left(e^{\left(\delta_{P}-\delta_{A}\right)t_{2}^{*}}\delta_{A}v\left(1\right)\Lambda\left(t_{2}^{*}\right)-1\right)=1-\ell e^{\left(\delta_{P}-\delta_{A}\right)t_{2}^{*}}\delta_{A}v\left(1\right)\Lambda\left(t_{2}^{*}\right).
\]
By construction of $\mu_{1}^{*}$, we must have that $1-\ell e^{\left(\delta_{P}-\delta_{A}\right)t_{2}^{*}}\delta_{A}v\left(1\right)\Lambda\left(t_{2}^{*}\right)\geq0$.
Since the integrand in (\ref{eq: integrand}) is decreasing in $s$,
it must always be positive. As a result, the last integral in the
expression for $\lambda_{1}$ is positive and $\lambda_{1}\geq0$. 
\item \textbf{Optimality Condition (\ref{eq: FOCL}): }Since $\partial L\left(\hat{x};x\right)$
is linear in $x$, it is sufficient to show $\partial L\left(\hat{x};x\right)\leq0$
for $x=\left(f\left(t\right),0\right)$ and $x=\left(0,f\left(t\right)\right)$
with $f$ an arbitrary non-increasing and non-negative function. For
such values, we have
\begin{align*}
\partial L\left(\hat{x};\left(f\left(t\right),0\right)\right)= & \int_{0}^{\infty}f\left(t\right)e^{-\delta_{P}t}dt+\int_{0}^{\infty}e^{-\delta_{A}t}\Delta_{0}\left(t\right)f\left(t\right)d\Lambda\left(t\right)\\
 & -v\left(1\right)\int_{0}^{\infty}\delta_{A}e^{-\delta_{A}t}f\left(t\right)\Lambda\left(t\right)dt-\lambda_{0}f\left(0\right)
\end{align*}
Given our construction of the allocation and the multipliers, we can
rewrite the above as
\begin{align*}
\partial L\left(\hat{x};\left(f\left(t\right),0\right)\right)= & \int_{0}^{\infty}f\left(t\right)e^{-\delta_{P}t}dt+\Delta_{0}\left(0\right)f\left(0\right)\Lambda\left(0+\right)+\\
 & \int_{t_{1}^{*}}^{t_{2}^{*}}e^{-\delta_{A}t}\Lambda'\left(t\right)\Delta_{0}\left(t\right)f\left(t\right)dt-\delta_{A}v\left(1\right)\int_{0}^{\infty}e^{-\delta_{A}t}f\left(t\right)\Lambda\left(t\right)dt\\
 & -\Lambda\left(0+\right)\left(v\left(1\right)e^{-\delta_{A}t_{1}^{*}}-v\left(\mu_{A}\right)+\mu_{A}v'\left(\mu_{A}\right)\right)f\left(0\right)-\int_{0}^{t_{1}^{*}}e^{-\delta_{P}s}dsf\left(0\right)\\
= & \int_{0}^{t_{2}^{*}}f\left(t\right)e^{-\delta_{P}t}dt+\left(v\left(1\right)-v\left(\mu_{A}\right)+\mu_{A}v'\left(\mu_{A}\right)\right)f\left(0\right)\Lambda\left(0+\right)\\
 & +\int_{t_{1}^{*}}^{t_{2}^{*}}e^{-\delta_{A}t}\Lambda'\left(t\right)\Delta_{0}\left(t\right)f\left(t\right)dt-\delta_{A}v\left(1\right)\int_{t_{1}^{*}}^{t_{2}^{*}}e^{-\delta_{A}t}f\left(t\right)\Lambda\left(t\right)dt\\
 & -\delta_{A}v\left(1\right)\int_{0}^{t_{1}^{*}}e^{-\delta_{A}t}f\left(t\right)dt\Lambda\left(0+\right)\\
 & -\Lambda\left(t_{1}\right)\left(v\left(1\right)e^{-\delta_{A}t_{1}^{*}}-v\left(\mu_{A}\right)+\mu_{A}v'\left(\mu_{A}\right)\right)f\left(0\right)\\
 & +\int_{t_{2}^{*}}^{\infty}f\left(t\right)\left[e^{-\delta_{P}t}-e^{-\delta_{A}t}\delta_{A}v\left(1\right)\Lambda\left(t\right)\right]dt\\
= & \int_{0}^{t_{1}^{*}}\left[e^{-\delta_{P}t}\left(f\left(t\right)-f\left(0\right)\right)+\Lambda\left(0+\right)e^{-\delta_{A}t}\left(\delta_{A}v\left(1\right)f\left(0\right)-\delta_{A}v\left(1\right)f\left(t\right)\right)\right]dt\\
 & +\int_{t_{1}^{*}}^{t_{2}^{*}}f\left(t\right)\left[\underbrace{e^{-\delta_{P}t}+e^{-\delta_{A}t}\left(\Lambda'\left(t\right)\Delta_{0}\left(t\right)-\delta_{A}v\left(1\right)\Lambda\left(t\right)\right)}_{=0\text{ by ODE \eqref{eq:ODE}}}\right]dt\\
 & +\int_{t_{2}^{*}}^{\infty}e^{-\delta_{P}t}f\left(t\right)\left[1-e^{\left(\delta_{P}-\delta_{A}\right)t}\delta_{A}v\left(1\right)\underbrace{\Lambda\left(t\right)}_{=\Lambda\left(t_{2}^{*}\right)}\right]dt\\
= & \int_{0}^{t_{1}^{*}}\left(f\left(t\right)-f\left(0\right)\right)e^{-\delta_{P}t}\left[1-\frac{e^{\left(\delta_{A}-\delta_{P}\right)t_{1}^{*}}}{\delta_{A}v\left(1\right)}e^{\left(\delta_{P}-\delta_{A}\right)t}\delta_{A}v\left(1\right)\right]dt+\\
 & \int_{t_{2}^{*}}^{\infty}e^{-\delta_{P}t}f\left(t\right)\left[1-e^{\left(\delta_{P}-\delta_{A}\right)t}\delta_{A}v\left(1\right)\frac{e^{\left(\delta_{A}-\delta_{P}\right)t_{2}^{*}}}{\ell\delta_{A}v\left(1\right)}\right]dt\\
= & \int_{0}^{t_{1}^{*}}\left(\underbrace{f\left(t\right)-f\left(0\right)}_{\leq0}\right)e^{-\delta_{P}t}\left[1-\underbrace{e^{\left(\delta_{P}-\delta_{A}\right)\left(t-t_{1}^{*}\right)}}_{\leq1}\right]dt\\
 & +\int_{t_{2}^{*}}^{\infty}e^{-\delta_{P}t}f\left(t\right)\left[1-\underbrace{\frac{e^{\left(\delta_{P}-\delta_{A}\right)\left(t-t_{2}^{*}\right)}}{\ell}}_{\geq1}\right]dt\leq0
\end{align*}
where the last inequality follows from $\delta_{P}>\delta_{A}$ and
$\ell<1$.\\
Similarly, for $x=\left(0,f\left(t\right)\right)$, we can write 
\begin{align*}
\partial L\left(\hat{x};\left(0,f\left(t\right)\right)\right)= & \int_{0}^{\infty}f\left(t\right)e^{-\delta_{P}t}dt+\\
 & \ell\Delta_{1}\left(0\right)f\left(0\right)\Lambda\left(0+\right)+\ell\int_{t_{1}^{*}}^{t_{2}^{*}}e^{-\delta_{A}t}\Lambda'\left(t\right)\Delta_{1}\left(t\right)f\left(t\right)dt\\
 & -\ell\delta_{A}v\left(1\right)\int_{0}^{\infty}e^{-\delta_{A}t}f\left(t\right)\Lambda\left(t\right)dt\\
 & -f\left(0\right)\int_{0}^{t_{2}^{*}}e^{-\delta_{P}s}ds-f\left(0\right)\ell\Lambda\left(0+\right)\left(v\left(1\right)-v\left(\mu_{A}\right)-\left(1-\mu_{A}\right)v'\left(\mu_{A}\right)\right)\\
 & -f\left(0\right)\int_{0}^{t_{2}^{*}}\left[e^{-\delta_{P}s}+\ell e^{-\delta_{A}s}\left(\Delta_{1}\left(s\right)\Lambda'\left(s\right)-\delta_{A}v\left(1\right)\Lambda\left(s\right)\right)\right]ds\\
= & \int_{0}^{t_{2}^{*}}e^{-\delta_{P}t}\left(f\left(t\right)-f\left(0\right)\right)dt+\\
 & \ell\int_{0}^{t_{2}^{*}}e^{-\delta_{A}t}\left[\Lambda'\left(t\right)\Delta_{1}\left(t\right)-\delta_{A}v\left(1\right)\Lambda\left(t\right)\right]\left(f\left(t\right)-f\left(0\right)\right)dt\\
 & +\int_{t_{2}^{*}}^{\infty}e^{-\delta_{P}t}f\left(t\right)\left[1-\ell e^{\left(\delta_{P}-\delta_{A}\right)t}\delta_{A}v\left(1\right)\Lambda\left(t\right)\right]dt
\end{align*}
Since $\Lambda\left(t\right)=\Lambda\left(t_{2}\right)=e^{\left(\delta_{A}-\delta_{P}\right)t_{2}^{*}}/\left(\delta_{A}\ell v\left(1\right)\right)$,
the last integral is negative. It is thus sufficient to show that
the first two terms add up to a negative value. We can use integration
by parts to write
\begin{align*}
 & \int_{0}^{t_{2}^{*}}e^{-\delta_{P}t}\left(f\left(t\right)-f\left(0\right)\right)dt+\\
 & \ell\int_{0}^{t_{2}^{*}}e^{-\delta_{A}t}\left[\Lambda'\left(t\right)\Delta_{1}\left(t\right)-\delta_{A}v\left(1\right)\Lambda\left(t\right)\right]\left(f\left(t\right)-f\left(0\right)\right)dt=\\
 & \int_{0}^{t_{2}^{*}}\left(f\left(0\right)-f\left(t\right)\right)d\left(\int_{t}^{t_{2}^{*}}\left[e^{-\delta_{P}s}+\ell e^{-\delta_{A}s}\left(\Lambda'\left(s\right)\Delta_{1}\left(s\right)-\delta_{A}v\left(1\right)\Lambda\left(s\right)\right)\right]ds\right)=\\
 & \int_{0}^{t_{2}^{*}}\int_{t}^{t_{2}^{*}}\left[e^{-\delta_{P}s}+\ell e^{-\delta_{A}s}\left(\Lambda'\left(s\right)\Delta_{1}\left(s\right)-\delta_{A}v\left(1\right)\Lambda\left(s\right)\right)\right]dsdf\left(t\right)
\end{align*}
Earlier we have shown that $e^{-\delta_{P}s}+\ell e^{-\delta_{A}s}\left(\Lambda'\left(s\right)\Delta_{1}\left(s\right)-\delta_{A}v\left(1\right)\Lambda\left(s\right)\right)\geq0$,
see the discussion following (\ref{eq: integrand}). Hence, since
$df\left(t\right)\leq0$, that is, $f$ is decreasing, the above integral
is negative, which is the desired result.
\item \textbf{Optimality Condition (\ref{eq:FOCeq}): }In the proof of Theorem
\ref{thm: thm1}, we showed that Equation (\ref{eq:FOCeq}) boils
down to checking $\lambda_{1}\mu_{P}+\lambda_{0}\left(1-\mu_{P}\right)=\int_{0}^{t_{2}^{*}}e^{-\delta_{P}t}G_{P}\left(t\right)dt$.
Given the values of $\lambda_{1}$ and $\lambda_{0}$, we have\textbf{
\begin{align*}
\lambda_{0}= & \int_{0}^{t_{1}^{*}}e^{-\delta_{P}s}ds+\Lambda\left(0+\right)\left(v\left(1\right)-v\left(\mu_{A}\right)+\mu_{A}v'\left(\mu_{A}\right)\right)-\Lambda\left(0+\right)\delta_{A}v\left(1\right)\int_{0}^{t_{1}^{*}}e^{-\delta_{A}s}ds\\
\lambda_{1}= & \int_{0}^{t_{2}^{*}}e^{-\delta_{P}s}ds+\ell\Lambda\left(0+\right)\left(v\left(1\right)-v\left(\mu_{A}\right)-\left(1-\mu_{A}\right)v'\left(\mu_{A}\right)\right)\\
 & +\ell\int_{0}^{t_{2}^{*}}e^{-\delta_{A}s}\left(\Delta_{1}\left(s\right)\Lambda'\left(s\right)-\delta_{A}v\left(1\right)\Lambda\left(s\right)\right)ds
\end{align*}
}Since $G_{P,0}\left(t\right)=1-\mu_{P},\forall t\leq t_{1}^{*},G_{P,1}\left(t\right)=\mu_{P},\forall t\leq t_{2}^{*}$,
we can use the above to write
\begin{align*}
\left(1-\mu_{P}\right)\lambda_{0}+\mu_{P}\lambda_{1}= & \int_{0}^{t_{1}^{*}}G_{P,0}\left(s\right)e^{-\delta_{P}s}ds+\Lambda\left(0+\right)\left(v\left(1\right)-v\left(\mu_{A}\right)+\mu_{A}v'\left(\mu_{A}\right)\right)\left(1-\mu_{P}\right)\\
 & -\Lambda\left(0+\right)\delta_{A}v\left(1\right)\int_{0}^{t_{1}^{*}}G_{P,0}\left(s\right)e^{-\delta_{A}s}ds\\
 & +\int_{0}^{t_{2}^{*}}G_{P,1}\left(s\right)e^{-\delta_{P}s}ds+\ell\Lambda\left(0+\right)\left(v\left(1\right)-v\left(\mu_{A}\right)-\left(1-\mu_{A}\right)v'\left(\mu_{A}\right)\right)\mu_{P}\\
 & +\ell\int_{0}^{t_{2}^{*}}G_{P,1}\left(s\right)e^{-\delta_{A}s}\left(\Delta_{1}\left(s\right)\Lambda'\left(s\right)-\delta_{A}v\left(1\right)\Lambda\left(s\right)\right)ds
\end{align*}
Recall the definition of $\ell=\frac{\mu_{A}}{1-\mu_{A}}\frac{1-\mu_{P}}{\mu_{P}}$
which implies that $\ell\left(1-\mu_{A}\right)\mu_{P}=\left(1-\mu_{P}\right)\mu_{A}$
and hence, we can write the above as
\begin{align*}
\left(1-\mu_{P}\right)\lambda_{0}+\mu_{P}\lambda_{1}= & \int_{0}^{t_{1}^{*}}G_{P,0}\left(s\right)e^{-\delta_{P}s}ds+\Lambda\left(0+\right)\left(v\left(1\right)-v\left(\mu_{A}\right)\right)\left(\ell G_{P,1}\left(0\right)+G_{P,0}\left(0\right)\right)\\
 & -\Lambda\left(0+\right)\delta_{A}v\left(1\right)\int_{0}^{t_{1}^{*}}G_{P,0}\left(s\right)e^{-\delta_{A}s}ds+\int_{0}^{t_{2}^{*}}G_{P,1}\left(s\right)e^{-\delta_{P}s}ds\\
 & +\ell\int_{0}^{t_{2}^{*}}G_{P,1}\left(s\right)e^{-\delta_{A}s}\left(\Delta_{1}\left(s\right)\Lambda'\left(s\right)-\delta_{A}v\left(1\right)\Lambda\left(s\right)\right)ds
\end{align*}
Now, given that $\Lambda\left(t\right)$ satisfies the ODE (\ref{eq:ODE}),
we can multiply the ODE by $G_{P,0}\left(t\right)$ and integrate
over time. The resulting expression is 0 and thus can be added to
the above. Hence,
\begin{align*}
\left(1-\mu_{P}\right)\lambda_{0}+\mu_{P}\lambda_{1}= & \int_{0}^{t_{1}^{*}}G_{P,0}\left(s\right)e^{-\delta_{P}s}ds+\Lambda\left(0+\right)\left(v\left(1\right)-v\left(\mu_{A}\right)\right)\left(\ell G_{P,1}\left(0\right)+G_{P,0}\left(0\right)\right)\\
 & -\Lambda\left(0+\right)\delta_{A}v\left(1\right)\int_{0}^{t_{1}^{*}}G_{P,0}\left(s\right)e^{-\delta_{A}s}ds+\int_{0}^{t_{2}^{*}}G_{P,1}\left(s\right)e^{-\delta_{P}s}ds\\
 & +\ell\int_{0}^{t_{2}^{*}}G_{P,1}\left(s\right)e^{-\delta_{A}s}\left(\Delta_{1}\left(s\right)\Lambda'\left(s\right)-\delta_{A}v\left(1\right)\Lambda\left(s\right)\right)ds+\\
 & \int_{t_{1}^{*}}^{t_{2}^{*}}G_{P,0}\left(s\right)e^{-\delta_{P}s}ds+\int_{t_{1}^{*}}^{t_{2}^{*}}G_{P,0}\left(s\right)e^{-\delta_{A}s}\left(\Delta_{0}\left(s\right)\Lambda'\left(s\right)-\delta_{A}v\left(1\right)\Lambda\left(s\right)\right)ds\\
= & \int_{0}^{t_{2}^{*}}G_{P}\left(s\right)e^{-\delta_{A}s}ds\\
 & +\int_{0}^{t_{2}^{*}}\left(\ell G_{P,1}\left(s\right)+G_{P,0}\left(s\right)\right)e^{-\delta_{A}s}\left(v\left(1\right)-v\left(\hat{\mu}_{A}\left(s\right)\right)\right)d\Lambda\left(s\right)\\
 & -\delta_{A}v\left(1\right)\int_{0}^{t_{2}^{*}}e^{-\delta_{A}s}\Lambda\left(s\right)\left(\ell G_{P,1}\left(s\right)+G_{P,0}\left(s\right)\right)ds\\
= & \int_{0}^{t_{2}^{*}}G_{P}\left(s\right)e^{-\delta_{A}s}ds\\
 & +\int_{0}^{t_{2}^{*}}\left(\ell G_{P,1}\left(s\right)+G_{P,0}\left(s\right)\right)e^{-\delta_{A}s}\left(v\left(1\right)-v\left(\hat{\mu}_{A}\left(s\right)\right)\right)d\Lambda\left(s\right)\\
 & -\delta_{A}v\left(1\right)\int_{0}^{t_{2}^{*}}\int_{t}^{t_{2}^{*}}e^{-\delta_{A}s}\left(\ell G_{P,1}\left(s\right)+G_{P,0}\left(s\right)\right)dsd\Lambda\left(t\right)
\end{align*}
\end{enumerate}
Using integration by part the above simplifies to:

\begin{align*}
 & \int_{0}^{t_{2}^{*}}G_{P}\left(s\right)e^{-\delta_{A}s}ds+\\
 & \underbrace{\int_{0}^{t_{2}^{*}}\left[G_{A}\left(t\right)e^{-\delta_{A}t}\left(v\left(1\right)-v\left(\hat{\mu}_{A}\left(s\right)\right)\right)-\delta_{A}v\left(1\right)\int_{t}^{t_{2}^{*}}G_{A}\left(s\right)e^{-\delta_{A}s}ds\right]d\Lambda\left(s\right)}_{=0\text{ by complementary slackness}}\\
= & \int_{0}^{t_{2}^{*}}G_{P}\left(s\right)e^{-\delta_{A}s}ds
\end{align*}

Finally, to finish the proof, we will show that there is at most one
$\mu_{1}$ for which $\Lambda\left(t_{2}\right)=\frac{e^{\left(\delta_{A}-\delta_{P}\right)t_{2}}}{\delta_{A}\ell v\left(1\right)}$
holds and that if such $\mu_{1}$ exist it must be less than 1. To
see this, suppose that $\mu_{1}=1$. In this case, $t_{1}=t_{2}=\frac{1}{\delta_{A}}\log\frac{v\left(1\right)}{v\left(\mu_{A}\right)}$.
Note under this assumption, engagement strategy in phase 2 does not
exist and hence we must have that $\Lambda\left(t_{1}\right)=\Lambda\left(t_{2}\right)=\frac{e^{\left(\delta_{A}-\delta_{P}\right)t_{1}}}{\delta_{A}v\left(1\right)}$.
Since $\ell<1$, this is less than $\frac{e^{\left(\delta_{A}-\delta_{P}\right)t_{2}}}{\delta_{A}\ell v\left(1\right)}$
which implies that if $\mu_{1}^{*}$ exists, it cannot be equal to
1. In order to show the existence of $\mu_{1}^{*}$, we will show
that as we increase $\mu_{1}$, $\frac{e^{\left(\delta_{A}-\delta_{P}\right)t_{2}}}{\delta_{A}\ell v\left(1\right)}-\Lambda\left(t_{2}\right)$
decreases. As a result, either there is a unique value $\mu_{1}^{*}$
under which $\frac{e^{\left(\delta_{A}-\delta_{P}\right)t_{2}}}{\delta_{A}\ell v\left(1\right)}-\Lambda\left(t_{2}\right)=0$
or that for all values of $\mu_{1}$, $\frac{e^{\left(\delta_{A}-\delta_{P}\right)t_{2}}}{\delta_{A}\ell v\left(1\right)}-\Lambda\left(t_{2}\right)>0$
in which case $\mu_{1}^{*}=\mu_{A}$. To avoid further lengthening
the proof of this proposition, we show this in the proof of Corollary
\ref{cor: 1} below.
\end{proof}
\pagebreak{}

\subsection{Proof of Corollary \ref{cor: 1}}
\begin{proof}
We first focus on the case with $\ell<1$ or $\mu_{A}<\mu_{P}$. In
this case, we first show that $\frac{1}{\delta_{A}\ell v\left(1\right)}-\frac{\Lambda\left(t_{2}\right)}{e^{\left(\delta_{A}-\delta_{P}\right)t_{2}}}$
is strictly decreasing in $\mu_{1}$. As a result, to have no phase
1, we would need $\frac{e^{\left(\delta_{A}-\delta_{P}\right)t_{2}}}{\delta_{A}\ell v\left(1\right)}-\Lambda\left(t_{2}\right)>0$
for $\mu_{1}=\mu_{A}$. For a given $\mu_{1}$, since $\hat{\mu}_{A}$
is strictly increasing over phase 2, we can do a change of variable
and write the Lagrange multiplier $\Lambda\left(t\right)$ as a function
of beliefs so that $L\left(\hat{\mu}_{A}\left(t\right)\right)=\Lambda\left(t\right)$.
The law of motion for $\hat{\mu}_{A}\left(t\right)$ in phase 2 can
be written as:
\[
\frac{d\mu}{dt}\frac{v\left(1\right)-v\left(\mu\right)+\mu v'\left(\mu\right)}{\mu v\left(\mu\right)}=\delta_{A}.
\]
This implies that 
\[
\Psi\left(\mu\right)=\int_{\mu}^{1}\frac{v\left(1\right)-v\left(x\right)+xv'\left(x\right)}{xv\left(x\right)}dx=\delta_{A}\left(t_{2}-t\right)\Rightarrow t=t_{2}-\frac{\Psi\left(\mu\right)}{\delta_{A}}
\]
Given the above, we have 
\[
e^{\delta_{A}t_{1}}=e^{\delta_{A}t_{2}-\Psi\left(\mu_{1}\right)}=\frac{\left(\mu_{1}-\mu_{A}\right)v\left(1\right)+\mu_{A}v\left(\mu_{1}\right)}{\mu_{1}v\left(\mu_{A}\right)}
\]
We can use this change of variable and write the solution for $\Lambda\left(t\right)$
in (\ref{eq: LamSol}) as
\begin{align*}
L\left(\mu\right) & =\hat{\Gamma}\left(\mu\right)\left[\frac{e^{\left(\delta_{A}-\delta_{P}\right)\left(t_{2}-\Psi\left(\mu_{1}\right)/\delta_{A}\right)}}{\delta_{A}v\left(1\right)}-\int_{\mu_{1}}^{\mu}\frac{e^{\left(\delta_{A}-\delta_{P}\right)\left(t_{2}-\Psi\left(x\right)/\delta_{A}\right)}}{v\left(1\right)-v\left(x\right)+xv'\left(x\right)}\frac{1}{\hat{\Gamma}\left(x\right)}d\left(-\frac{\Psi\left(x\right)}{\delta_{A}}\right)\right]\\
\log\hat{\Gamma}\left(\mu\right) & =\int_{\mu_{1}}^{\mu}\frac{\delta_{A}v\left(1\right)}{v\left(1\right)-v\left(x\right)+xv'\left(x\right)}d\left(-\frac{\Psi\left(x\right)}{\delta_{A}}\right)
\end{align*}
Using the definition of $\Psi$, we can rewrite the above as
\begin{align*}
L\left(\mu\right) & =\hat{\Gamma}\left(\mu\right)\left[\frac{e^{\left(\delta_{A}-\delta_{P}\right)\left(t_{2}-\Psi\left(\mu_{1}\right)/\delta_{A}\right)}}{\delta_{A}v\left(1\right)}-\int_{\mu_{1}}^{\mu}\frac{e^{\left(\delta_{A}-\delta_{P}\right)\left(t_{2}-\Psi\left(x\right)/\delta_{A}\right)}}{\delta_{A}xv\left(x\right)}\frac{1}{\hat{\Gamma}\left(x\right)}dx\right]\\
\log\hat{\Gamma}\left(\mu\right) & =\int_{\mu_{1}}^{\mu}\frac{v\left(1\right)}{xv\left(x\right)}dx
\end{align*}
Hence,
\[
\frac{\Lambda\left(t_{2}\right)}{e^{\left(\delta_{A}-\delta_{P}\right)t_{2}}}=\frac{L\left(1\right)}{e^{\left(\delta_{A}-\delta_{P}\right)t_{2}}}=\hat{\Gamma}\left(1\right)\left[\frac{e^{-\left(\delta_{A}-\delta_{P}\right)\Psi\left(\mu_{1}\right)/\delta_{A}}}{\delta_{A}v\left(1\right)}-\int_{\mu_{1}}^{1}\frac{e^{-\left(\delta_{A}-\delta_{P}\right)\Psi\left(x\right)/\delta_{A}}}{\delta_{A}xv\left(x\right)}\frac{1}{\hat{\Gamma}\left(x\right)}dx\right]
\]
Taking a derivative of the above with respect to $\mu_{1}$, we have
\begin{align*}
\frac{d}{d\mu_{1}}\frac{\Lambda\left(t_{2}\right)}{e^{\left(\delta_{A}-\delta_{P}\right)t_{2}}}= & -\frac{v\left(1\right)}{\mu_{1}v\left(\mu_{1}\right)}\hat{\Gamma}\left(1\right)\left[\frac{e^{-\left(\delta_{A}-\delta_{P}\right)\Psi\left(\mu_{1}\right)/\delta_{A}}}{\delta_{A}v\left(1\right)}-\int_{\mu_{1}}^{1}\frac{e^{-\left(\delta_{A}-\delta_{P}\right)\Psi\left(x\right)/\delta_{A}}}{\delta_{A}xv\left(x\right)}\frac{1}{\hat{\Gamma}\left(x\right)}dx\right]\\
 & +\hat{\Gamma}\left(1\right)\frac{v\left(1\right)-v\left(\mu_{1}\right)+\mu_{1}v'\left(\mu_{1}\right)}{\mu_{1}v\left(\mu_{1}\right)}\frac{\delta_{A}-\delta_{P}}{\delta_{A}}\frac{e^{-\left(\delta_{A}-\delta_{P}\right)\Psi\left(\mu_{1}\right)/\delta_{A}}}{\delta_{A}v\left(1\right)}\\
 & +\hat{\Gamma}\left(1\right)\frac{e^{-\left(\delta_{A}-\delta_{P}\right)\Psi\left(\mu_{1}\right)/\delta_{A}}}{\delta_{A}\mu_{1}v\left(\mu_{1}\right)}\\
 & -\hat{\Gamma}\left(1\right)\frac{v\left(1\right)}{\mu_{1}v\left(\mu_{1}\right)}\int_{\mu_{1}}^{1}\frac{e^{-\left(\delta_{A}-\delta_{P}\right)\Psi\left(x\right)/\delta_{A}}}{\delta_{A}xv\left(x\right)}\frac{1}{\hat{\Gamma}\left(x\right)}dx\\
= & \hat{\Gamma}\left(1\right)\frac{v\left(1\right)-v\left(\mu_{1}\right)+\mu_{1}v'\left(\mu_{1}\right)}{\mu_{1}v\left(\mu_{1}\right)}\underbrace{\frac{\delta_{A}-\delta_{P}}{\delta_{A}}}_{<0}\frac{e^{-\left(\delta_{A}-\delta_{P}\right)\Psi\left(\mu_{1}\right)/\delta_{A}}}{\delta_{A}v\left(1\right)}<0
\end{align*}
Hence, either there exists a unique $\mu_{1}^{*}\in\left(\mu_{A},1\right)$
such that $\frac{\Lambda\left(t_{2}\right)}{e^{\left(\delta_{A}-\delta_{P}\right)t_{2}}}=\frac{1}{\delta_{A}\ell v\left(1\right)}$
or that 
\[
\frac{\Lambda\left(t_{2}\right)}{e^{\left(\delta_{A}-\delta_{P}\right)t_{2}}}\leq\frac{1}{\delta_{A}\ell v\left(1\right)}\text{ at }\mu_{1}=\mu_{A}
\]
in which case $\mu_{1}^{*}=\mu_{A}$ and phase 1 is of length 0. That
is
\[
e^{\int_{\mu_{A}}^{1}\frac{v\left(1\right)}{xv\left(x\right)}dx}\left[\frac{e^{-\left(\delta_{A}-\delta_{P}\right)\Psi\left(\mu_{A}\right)/\delta_{A}}}{\delta_{A}v\left(1\right)}-\int_{\mu_{A}}^{1}\frac{e^{-\left(\delta_{A}-\delta_{P}\right)\Psi\left(x\right)/\delta_{A}}}{\delta_{A}xv\left(x\right)}e^{-\int_{\mu_{A}}^{x}\frac{v\left(1\right)}{zv\left(z\right)}dz}dx\right]\leq\frac{1}{\delta_{A}\ell v\left(1\right)}
\]
or
\begin{equation}
e^{\int_{\mu_{A}}^{1}\frac{v\left(1\right)}{xv\left(x\right)}dx}e^{-\left(\delta_{A}-\delta_{P}\right)\Psi\left(\mu_{A}\right)/\delta_{A}}-v\left(1\right)\int_{\mu_{A}}^{1}\frac{e^{-\left(\delta_{A}-\delta_{P}\right)\Psi\left(x\right)/\delta_{A}}}{xv\left(x\right)}e^{\int_{x}^{1}\frac{v\left(1\right)}{zv\left(z\right)}dz}dx\leq1/\ell\label{eq: ineq}
\end{equation}
The derivative of the above with respect to $\mu_{A}$ is
\begin{align*}
\left(\frac{\delta_{A}-\delta_{P}}{\delta_{A}}\frac{v\left(1\right)-v\left(\mu_{A}\right)+\mu_{A}v'\left(\mu_{A}\right)}{\mu_{A}v\left(\mu_{A}\right)}-\frac{v\left(1\right)}{\mu_{A}v\left(\mu_{A}\right)}\right)e^{\int_{\mu_{A}}^{1}\frac{v\left(1\right)}{xv\left(x\right)}dx}e^{-\left(\delta_{A}-\delta_{P}\right)\Psi\left(\mu_{A}\right)/\delta_{A}}+\\
v\left(1\right)\frac{e^{-\left(\delta_{A}-\delta_{P}\right)\Psi\left(\mu_{A}\right)/\delta_{A}}}{\mu_{A}v\left(\mu_{A}\right)}e^{\int_{\mu_{A}}^{1}\frac{v\left(1\right)}{zv\left(z\right)}dz} & =\\
\frac{\delta_{A}-\delta_{P}}{\delta_{A}}\frac{v\left(1\right)-v\left(\mu_{A}\right)+\mu_{A}v'\left(\mu_{A}\right)}{\mu_{A}v\left(\mu_{A}\right)}e^{\int_{\mu_{A}}^{1}\frac{v\left(1\right)}{xv\left(x\right)}dx}e^{-\left(\delta_{A}-\delta_{P}\right)\Psi\left(\mu_{A}\right)/\delta_{A}} & <0
\end{align*}
Additionally an increase in $\mu_{A}$ decreases $\ell$. This implies
that the inequality (\ref{eq: ineq}) holds for low values of $\mu_{A}$.

When $\mu_{A}>\mu_{P}$ or $\ell>1$, the law of motion for $\hat{\mu}_{A}\left(t\right)$
in phase 2 can be written as
\[
\frac{d\mu}{dt}\frac{v\left(1\right)-v\left(\mu\right)-\left(1-\mu\right)v'\left(\mu\right)}{\left(1-\mu\right)v\left(\mu\right)}=-\delta_{A}
\]
This implies that 
\[
\Phi\left(\mu\right)=\int_{0}^{\mu}\frac{v\left(1\right)-v\left(x\right)-\left(1-x\right)v'\left(x\right)}{\left(1-x\right)v\left(x\right)}dx=\delta_{A}\left(t_{2}-t\right)\Rightarrow t=t_{2}-\frac{\Phi\left(\mu\right)}{\delta_{A}}
\]
Given the above, we have 
\[
e^{\delta_{A}t_{1}}=e^{\delta_{A}t_{2}-\Phi\left(\mu_{1}\right)}=\frac{\left(\mu_{A}-\mu_{1}\right)v\left(1\right)+\left(1-\mu_{A}\right)v\left(\mu_{1}\right)}{\left(1-\mu_{1}\right)v\left(\mu_{A}\right)}
\]
At the same time, the ODE characterizing the Lagrange multiplier in
phase 2 is given by
\[
e^{\left(\delta_{A}-\delta_{P}\right)t}+\ell\Lambda'\left(t\right)\Delta_{1}\left(t\right)-\ell\delta_{A}v\left(1\right)\Lambda\left(t\right)=0
\]
with $\Lambda\left(t_{1}\right)=e^{\left(\delta_{A}-\delta_{P}\right)t_{1}}/\left(\ell\delta_{A}v\left(1\right)\right)$.
In this case, the solution of this ODE is given by
\begin{align*}
\Lambda\left(t\right) & =\Omega\left(t\right)\left[\frac{e^{\left(\delta_{A}-\delta_{P}\right)t_{1}}}{\ell\delta_{A}v\left(1\right)}-\int_{t_{1}}^{t}\frac{1}{\Omega\left(s\right)}\frac{e^{\left(\delta_{A}-\delta_{P}\right)s}}{\ell\Delta_{1}\left(s\right)}ds\right]\\
\log\Omega\left(t\right) & =\int_{t_{1}}^{t}\frac{\delta_{A}v\left(1\right)}{\Delta_{1}\left(s\right)}ds.
\end{align*}
We can use a change of variable as above and write
\begin{align*}
\frac{L\left(\mu\right)}{e^{\left(\delta_{A}-\delta_{P}\right)t_{2}}} & =\hat{\Omega}\left(\mu\right)\left[\frac{e^{-\left(\delta_{A}-\delta_{P}\right)\Phi\left(\mu_{1}\right)/\delta_{A}}}{\ell\delta_{A}v\left(1\right)}-\int_{\mu}^{\mu_{1}}\frac{1}{\hat{\Omega}\left(x\right)}\frac{e^{-\left(\delta_{A}-\delta_{P}\right)\Phi\left(x\right)/\delta_{A}}}{\ell\delta_{A}\left(1-x\right)v\left(x\right)}dx\right]\\
\log\hat{\Omega}\left(\mu\right) & =\int_{\mu}^{\mu_{1}}\frac{v\left(1\right)}{\left(1-x\right)v\left(x\right)}dx.
\end{align*}
A similar argument as before establishes that $\frac{L\left(0\right)}{e^{\left(\delta_{A}-\delta_{P}\right)t_{2}}}$
is increasing in $\mu_{1}$ and thus for phase 1 to be of length 0,
we need
\begin{align*}
\mu_{1}=\mu_{A}:\frac{L\left(0\right)}{e^{\left(\delta_{A}-\delta_{P}\right)t_{2}}} & \leq\frac{1}{\delta_{A}v\left(1\right)}\Rightarrow\\
e^{\int_{0}^{\mu_{A}}\frac{v\left(1\right)}{\left(1-x\right)v\left(x\right)}dx}\left[\frac{e^{-\left(\delta_{A}-\delta_{P}\right)\Phi\left(\mu_{A}\right)/\delta_{A}}}{\ell\delta_{A}v\left(1\right)}-\int_{0}^{\mu_{A}}\frac{1}{\hat{\Omega}\left(x\right)}\frac{e^{-\left(\delta_{A}-\delta_{P}\right)\Phi\left(x\right)/\delta_{A}}}{\ell\delta_{A}\left(1-x\right)v\left(x\right)}dx\right] & \leq\frac{1}{\delta_{A}v\left(1\right)}
\end{align*}
A similar argument as before shows that the LHS of the above is decreasing
in $\mu_{A}$ and thus holds for high enough values of $\mu_{A}$
and thus establishing the result.
\end{proof}

\subsection{Proof of Proposition \ref{prop:Suppose-that-}}
\begin{proof}
We start by characterization of the steady-state level, i.e., $\mu_{P}=\mu_{P}^{*}\left(\mu_{A}\right)$
where $\mu_{P}^{*}\left(\cdot\right)$ satisfies (\ref{eq: mups}).
Given its definition in (\ref{eq: mups}), when $\mu_{A}<1/2$, then,
$\mu_{A}>\mu_{P}^{*}\left(\mu_{A}\right)$ or $\ell>1$ while when
$\mu_{A}>1/2$ , then $\mu_{A}<\mu_{P}^{*}\left(\mu_{A}\right)$ or
$\ell<1$. Moreover, since $v\left(\mu_{A}\right)=v\left(1-\mu_{A}\right)$
and $v'\left(\mu_{A}\right)=-v'\left(1-\mu_{A}\right)$, we have 
\[
\mu_{P}^{*}\left(\mu_{A}\right)+\mu_{P}^{*}\left(1-\mu_{A}\right)=1.
\]
Moreover, when $\mu_{P}^{*}\left(\mu_{A}\right)>0$ and $\mu_{A}\leq1/2$,
we must have
\begin{align*}
\frac{d}{d\mu_{A}}\mu_{P}^{*}\left(\mu_{A}\right) & =\frac{d}{d\mu_{A}}\mu_{A}\frac{v\left(\mu_{A}\right)+\left(1-\mu_{A}\right)v'\left(\mu_{A}\right)+\frac{\delta_{P}}{\delta_{A}-\delta_{P}}v\left(1\right)}{v\left(\mu_{A}\right)+\frac{\delta_{P}}{\delta_{A}-\delta_{P}}v\left(1\right)}\\
 & =\frac{\mu_{P}^{*}\left(\mu_{A}\right)}{\mu_{A}}+\frac{\left(1-\mu_{A}\right)v''\left(\mu_{A}\right)}{v\left(\mu_{A}\right)+\frac{\delta_{P}}{\delta_{A}-\delta_{P}}v\left(1\right)}-\frac{\mu_{P}^{*}\left(\mu_{A}\right)v'\left(\mu_{A}\right)}{v\left(\mu_{A}\right)+\frac{\delta_{P}}{\delta_{A}-\delta_{P}}v\left(1\right)}
\end{align*}
Since $v''\left(\mu_{A}\right)\geq0$ and $v'\left(\mu_{A}\right)<0$,
for $\mu_{A}<1/2$, the above is positive. Moreover, if $\mu_{A}>1/2$
and $\mu_{P}^{*}\left(\mu_{A}\right)<1$, then
\[
\frac{d}{d\mu_{A}}\mu_{P}^{*}\left(\mu_{A}\right)=-\frac{d}{d\mu_{A}}\mu_{P}^{*}\left(1-\mu_{A}\right)=\frac{d}{d\left(1-\mu_{A}\right)}\mu_{P}^{*}\left(1-\mu_{A}\right)>0
\]
Therefore, when $\mu_{P}^{*}\left(\mu_{A}\right)\in\left(0,1\right)$,
$\mu_{P}^{*}$ is strictly increasing in $\mu_{A}$.

Let $\lambda^{*}\left(\mu_{A}\right)$ be defined by
\[
v\left(1\right)\frac{\lambda^{*}\left(\mu_{A}\right)}{\lambda^{*}\left(\mu_{A}\right)+\delta_{A}}=v\left(\mu_{A}\right).
\]
This is the incentive constraint of the agent when
\[
\frac{G_{P,1}\left(t\right)}{\mu_{P}}=\frac{G_{P,0}\left(t\right)}{1-\mu_{P}}=e^{-\lambda^{*}\left(\mu_{A}\right)t}
\]
In line with the rest of the analysis, we assume that $\mu_{P}=\mu_{P}^{*}\left(\mu_{A}\right)\in\left(0,1\right)$.
The Lagrange multiplier $\Lambda\left(t\right)$ is defined by
\[
\Lambda\left(t\right)=\frac{1-\ell}{\ell}\frac{e^{\left(\delta_{A}-\delta_{P}\right)t}}{v'\left(\mu_{A}\right)\left(\delta_{A}-\delta_{P}\right)}
\]
Given the definition of $\mu_{P}^{*}\left(\cdot\right)$ and the property
mentioned above, $1-\ell$ and $v'\left(\mu_{A}\right)$ have the
same sign, which implies that $\Lambda\left(t\right)>0$, and $\Lambda'\left(t\right)>0$.
Given the above definition, $\Lambda\left(t\right)$ satisfies
\begin{align*}
\left(v\left(1\right)-v\left(\mu_{A}\right)+\mu_{A}v'\left(\mu_{A}\right)\right)\Lambda'\left(t\right)-\delta_{A}v\left(1\right)\Lambda\left(t\right) & =\\
\left[\left(\delta_{A}-\delta_{P}\right)\left(\mu_{A}v'\left(\mu_{A}\right)-v\left(\mu_{A}\right)\right)-\delta_{P}v\left(1\right)\right]\frac{1-\ell}{\ell}\frac{e^{\left(\delta_{A}-\delta_{P}\right)t}}{v'\left(\mu_{A}\right)\left(\delta_{A}-\delta_{P}\right)} & =\\
\left[\mu_{A}v'\left(\mu_{A}\right)-v\left(\mu_{A}\right)-\frac{\delta_{P}}{\delta_{A}-\delta_{P}}v\left(1\right)\right]\frac{\mu_{P}^{*}-\mu_{A}}{\left(1-\mu_{P}^{*}\right)\mu_{A}}\frac{e^{\left(\delta_{A}-\delta_{P}\right)t}}{v'\left(\mu_{A}\right)} & =\\
-\left[v\left(\mu_{A}\right)+\frac{\delta_{P}}{\delta_{A}-\delta_{P}}v\left(1\right)-\mu_{A}v'\left(\mu_{A}\right)\right]\times\\
\frac{\mu_{A}\left(1-\mu_{A}\right)v'\left(\mu_{A}\right)}{\left(v\left(\mu_{A}\right)+\frac{\delta_{P}}{\delta_{A}-\delta_{P}}v\left(1\right)-\mu_{A}v'\left(\mu_{A}\right)\right)\mu_{A}\left(1-\mu_{A}\right)}\frac{e^{\left(\delta_{A}-\delta_{P}\right)t}}{v'\left(\mu_{A}\right)} & =-e^{\left(\delta_{A}-\delta_{P}\right)t}
\end{align*}
where in the above we have used the definition of $\mu_{P}^{*}$ from
(\ref{eq: mups}). Similarly, we can show that
\[
\ell\left(v\left(1\right)-v\left(\mu_{A}\right)-\left(1-\mu_{A}\right)v'\left(\mu_{A}\right)\right)\Lambda'\left(t\right)-\ell\delta_{A}v\left(1\right)\Lambda\left(t\right)+e^{\left(\delta_{A}-\delta_{P}\right)t}=0
\]
A simple examination of the above illustrates that they are the First
Order Condition of the Lagrangian $L\left(\hat{x};x\right)$ with
respect to $G_{P,0}\left(t\right)$ and $G_{P,1}\left(t\right)$.
This establishes the optimality of the steady state mentioned above.

In order to prove the claim of the Proposition, we will show the case
where $\mu_{P}>\mu_{P}^{*}\left(\mu_{A}\right)$. The other case is
the mirror of this proof. Suppose that $\mu_{A}\left(t\right)$ satisfies
the differential equation (\ref{eq: ODEmu0}) in phase 1 until it
reaches the value $\mu_{A}^{*}$ which is defined by 
\[
\ell=\frac{\mu_{A}^{*}}{1-\mu_{A}^{*}}\frac{1-\mu_{P}^{*}\left(\mu_{A}^{*}\right)}{\mu_{P}^{*}\left(\mu_{A}^{*}\right)}.
\]
Such a value of $\mu_{A}^{*}$ is uniquely determined since the above
becomes
\begin{equation}
\ell=\frac{\delta_{A}v\left(1\right)-\left(\delta_{A}-\delta_{P}\right)\left[v\left(1\right)-v\left(\mu_{A}^{*}\right)+\mu_{A}^{*}v'\left(\mu_{A}^{*}\right)\right]}{\delta_{A}v\left(1\right)-\left(\delta_{A}-\delta_{P}\right)\left[v\left(1\right)-v\left(\mu_{A}^{*}\right)-\left(1-\mu_{A}^{*}\right)v'\left(\mu_{A}^{*}\right)\right]}\label{eq: muaS}
\end{equation}
whenever $\mu_{A}^{*}\in\left(0,1\right)$. Since the numerator is
decreasing in $\mu_{A}^{*}$ and the denominator is increasing, the
above have at most one solution. When no such $\mu_{A}^{*}$ exist,
we set $\mu_{A}^{*}=1$. We let $t^{*}$ be the time by which this
value of belief is reached. Additionally, let $G_{P,\omega}$ be the
associated engagement policy given by
\begin{align*}
G_{P,1}\left(t\right) & =\begin{cases}
\mu_{P} & t\leq t^{*}\\
\mu_{P}e^{-\hat{\lambda}\left(t-t^{*}\right)} & t>t^{*}
\end{cases}\\
G_{P,0}\left(t\right) & =\begin{cases}
\ell\mu_{P}\frac{1-\mu_{A}\left(t\right)}{\mu_{A}\left(t\right)} & t\leq t^{*}\\
\ell\mu_{P}\frac{1-\mu_{A}\left(t^{*}\right)}{\mu_{A}\left(t^{*}\right)}e^{-\hat{\lambda}\left(t-t^{*}\right)} & t>t^{*}
\end{cases}
\end{align*}
Given this, the rest of the proof is very similar to that of Proposition
\ref{prop:When--is-1}. We construct the Lagrange multiplier using
the ODE
\[
e^{\left(\delta_{A}-\delta_{P}\right)t}+\Lambda'\left(t\right)\left(\underbrace{v\left(1\right)-v\left(\mu_{A}\left(t\right)\right)+\mu_{A}\left(t\right)v'\left(\mu_{A}\left(t\right)\right)}_{\Delta_{0}\left(t\right)}\right)-\delta_{A}v\left(1\right)\Lambda\left(t\right)=0,\forall t\leq t^{*}
\]
for a given value of $\Lambda\left(0+\right)\geq0$. For values of
$t\geq t^{*}$, we set
\[
\Lambda\left(t\right)=\frac{1-\ell}{\ell}\frac{e^{\left(\delta_{A}-\delta_{P}\right)t}}{v'\left(\mu_{A}^{*}\right)\left(\delta_{A}-\delta_{P}\right)}
\]
where $\Lambda\left(0+\right)$ has to satisfy
\[
\frac{1-\ell}{\ell}\frac{e^{\left(\delta_{A}-\delta_{P}\right)t^{*}}}{v'\left(\mu_{A}^{*}\right)\left(\delta_{A}-\delta_{P}\right)}=e^{\int_{0}^{t^{*}}\frac{\delta_{A}v\left(1\right)}{\Delta_{0}\left(t\right)}dt}\left[\Lambda\left(0+\right)-\int_{0}^{t^{*}}\frac{e^{-\int_{0}^{t}\frac{\delta_{A}v\left(1\right)}{\Delta_{0}\left(s\right)}ds}e^{\left(\delta_{A}-\delta_{P}\right)t}}{\Delta_{0}\left(t\right)}dt\right].
\]
Since LHS is positive, $\Lambda\left(0+\right)$ that solves the above
is also positive. Finally, the Lagrange multipliers $\lambda_{0},\lambda_{1}$
are constructed according to:
\begin{align*}
\lambda_{0}= & \Lambda\left(0+\right)\Delta_{0}\left(0\right)>0\\
\lambda_{1}= & \Lambda\left(0+\right)\Delta_{1}\left(0\right)\\
 & +\int_{0}^{t^{*}}\left[e^{-\delta_{P}t}+\ell e^{-\delta_{A}t}\left(\Delta_{1}\left(t\right)\Lambda'\left(t\right)-\delta_{A}v\left(1\right)\Lambda\left(t\right)\right)\right]dt
\end{align*}
where $\Delta_{1}\left(t\right)=v\left(1\right)-v\left(\mu_{A}\left(t\right)\right)-\left(1-\mu_{A}\left(t\right)\right)v'\left(\mu_{A}\left(t\right)\right)$.
The FOC with respect to $G_{P,\omega}\left(t\right)$ is satisfied
when $\omega=0$ or when $\omega=1,t\geq t^{*}$, given that $\Lambda\left(t\right)$
satisfies the above ODE. To show optimality, it is sufficient to show
that
\[
\int_{t}^{t^{*}}e^{-\delta_{P}s}\left[e^{\left(\delta_{A}-\delta_{P}\right)s}+\ell\left(\Delta_{1}\left(s\right)\Lambda'\left(s\right)-\delta_{A}v\left(1\right)\Lambda\left(s\right)\right)\right]\geq0
\]
We prove the stronger result that for all $t\in\left(0,t^{*}\right)$
\begin{equation}
e^{\left(\delta_{A}-\delta_{P}\right)t}+\ell\left(\Delta_{1}\left(t\right)\Lambda'\left(t\right)-\delta_{A}v\left(1\right)\Lambda\left(t\right)\right)\geq0\label{eq: ineq5}
\end{equation}
To establish this inequality, we realize that $\mu_{P}>\mu_{P}^{*}\left(\mu_{A}\right)$,
and since $t^{*}$ is the lowest time for which $\mu_{P}\left(t^{*}\right)=\mu_{P}^{*}\left(\mu_{A}\left(t^{*}\right)\right)$,
it must be that for all values of $t\in\left[0,t^{*}\right)$, $\mu_{P}\left(t\right)>\mu_{P}^{*}\left(\mu_{A}\left(t\right)\right)$.

Recall that $\Lambda\left(t\right)$ satisfies the following
\[
\Lambda\left(t\right)=\Gamma\left(t\right)\left[\Lambda\left(0+\right)-\int_{0}^{t}\frac{e^{\left(\delta_{A}-\delta_{P}\right)s}}{\Delta_{0}\left(s\right)\Gamma\left(s\right)}ds\right],\log\Gamma\left(s\right)=\int_{0}^{t}\frac{\delta_{A}v\left(1\right)}{\Delta_{0}\left(s\right)}ds
\]
Let $\Omega\left(t\right)=\Lambda\left(t\right)/\Gamma\left(t\right)$.
We can use the ODE governing $\Lambda\left(t\right)$ and rewrite
the LHS of (\ref{eq: ineq5}) as
\begin{align*}
 & \left(\ell\Delta_{1}\left(t\right)-\Delta_{0}\left(t\right)\right)\Lambda'\left(t\right)+\left(1-\ell\right)\delta_{A}v\left(1\right)\Lambda\left(t\right)=\\
 & \left(\ell\Delta_{1}\left(t\right)-\Delta_{0}\left(t\right)\right)\left(\frac{\delta_{A}v\left(1\right)}{\Delta_{0}\left(t\right)}\Omega\left(t\right)\Gamma\left(t\right)-\frac{e^{\left(\delta_{A}-\delta_{P}\right)t}}{\Delta_{0}\left(t\right)}\right)+\left(1-\ell\right)\delta_{A}v\left(1\right)\Omega\left(t\right)\Gamma\left(t\right)
\end{align*}
We can divide the above by $\Gamma\left(t\right)$ and rewrite the
resulting expression as
\[
\frac{\ell\Delta_{1}\left(t\right)}{\Delta_{0}\left(t\right)}\left[\delta_{A}v\left(1\right)\Omega\left(t\right)-\frac{e^{\left(\delta_{A}-\delta_{P}\right)t}}{\Gamma\left(t\right)}\right]+\frac{e^{\left(\delta_{A}-\delta_{P}\right)t}}{\Gamma\left(t\right)}-\delta_{A}v\left(1\right)\ell\Omega\left(t\right)
\]
In order to show that the above is positive, we will show that it
is decreasing and its value at $t^{*}$ is positive. Its derivative
with respect to $t$ is given by
\begin{align}
 & \left(\frac{\ell\Delta_{1}\left(t\right)}{\Delta_{0}\left(t\right)}\right)'\left[\delta_{A}v\left(1\right)\Omega\left(t\right)-\frac{e^{\left(\delta_{A}-\delta_{P}\right)t}}{\Gamma\left(t\right)}\right]+\nonumber \\
 & \frac{\ell\Delta_{1}\left(t\right)}{\Delta_{0}\left(t\right)}\left[-\delta_{A}v\left(1\right)\frac{e^{\left(\delta_{A}-\delta_{P}\right)t}}{\Gamma\left(t\right)\Delta_{0}\left(t\right)}+\frac{\delta_{A}v\left(1\right)e^{\left(\delta_{A}-\delta_{P}\right)t}}{\Gamma\left(t\right)\Delta_{0}\left(t\right)}-\frac{\left(\delta_{A}-\delta_{P}\right)e^{\left(\delta_{A}-\delta_{P}\right)t}}{\Gamma\left(t\right)}\right]+\nonumber \\
 & \frac{e^{\left(\delta_{A}-\delta_{P}\right)t}}{\Gamma\left(t\right)}\left(\delta_{A}-\delta_{P}-\frac{\delta_{A}v\left(1\right)}{\Delta_{0}\left(t\right)}\right)+\delta_{A}v\left(1\right)\ell\frac{e^{\left(\delta_{A}-\delta_{P}\right)t}}{\Gamma\left(t\right)\Delta_{0}\left(t\right)}\nonumber \\
= & \;\;\;\left(\frac{\ell\Delta_{1}\left(t\right)}{\Delta_{0}\left(t\right)}\right)'\left[\delta_{A}v\left(1\right)\Omega\left(t\right)-\frac{e^{\left(\delta_{A}-\delta_{P}\right)t}}{\Gamma\left(t\right)}\right]+\nonumber \\
 & \;\;\;\frac{e^{\left(\delta_{A}-\delta_{P}\right)t}}{\Gamma\left(t\right)\Delta_{0}\left(t\right)}\left(\left(\delta_{A}-\delta_{P}\right)\left(\Delta_{0}\left(t\right)-\ell\Delta_{1}\left(t\right)\right)+\left(\ell-1\right)\delta_{A}v\left(1\right)\right)\label{eq: Negative}
\end{align}
We show that each of the expressions above are negative. To do so,
note that for all values of $t<t^{*}$, we must have that $\mu_{P}^{*}\left(\mu_{A}\left(t\right)\right)<\mu_{P}\left(t\right)$,
given the definition of $t^{*}$. Therefore
\begin{align*}
\ell & =\frac{1-\mu_{P}\left(t\right)}{\mu_{P}\left(t\right)}\frac{\mu_{A}\left(t\right)}{1-\mu_{A}\left(t\right)}<\frac{1-\frac{\mu_{A}\left(t\right)v'\left(\mu_{A}\left(t\right)\right)}{v\left(\mu_{A}\left(t\right)\right)+\frac{\delta_{P}}{\delta_{A}-\delta_{P}}v\left(1\right)}}{1+\frac{\left(1-\mu_{A}\left(t\right)\right)v'\left(\mu_{A}\left(t\right)\right)}{v\left(\mu_{A}\left(t\right)\right)+\frac{\delta_{P}}{\delta_{A}-\delta_{P}}v\left(1\right)}}=\frac{\delta_{A}v\left(1\right)-\left(\delta_{A}-\delta_{P}\right)\Delta_{0}\left(t\right)}{\delta_{A}v\left(1\right)-\left(\delta_{A}-\delta_{P}\right)\Delta_{1}\left(t\right)}\\
\Rightarrow & \left(\delta_{A}-\delta_{P}\right)\left(\Delta_{0}\left(t\right)-\ell\Delta_{1}\left(t\right)\right)+\left(\ell-1\right)\delta_{A}v\left(1\right)<0
\end{align*}
which is proportional to the last expression in (\ref{eq: Negative}).
Moreover, since $\mu_{A}\left(t\right)$and $\Delta_{0}\left(t\right)$
are increasing, and $\Delta_{1}\left(t\right)$ is decreasing, we
must have that $\left(\frac{\ell\Delta_{1}\left(t\right)}{\Delta_{0}\left(t\right)}\right)'\leq0$.
Finally, 
\begin{align*}
\delta_{A}v\left(1\right)\Omega\left(t\right)-\frac{e^{\left(\delta_{A}-\delta_{P}\right)t}}{\Gamma\left(t\right)} & =\frac{\delta_{A}v\left(1\right)\Lambda\left(t\right)-e^{\left(\delta_{A}-\delta_{P}\right)t}}{\Gamma\left(t\right)}\\
 & =\frac{\Delta_{0}\left(t\right)\Lambda'\left(t\right)}{\Gamma\left(t\right)}\geq0
\end{align*}
This implies that (\ref{eq: Negative}) is negative. Thus, it is sufficient
to show that (\ref{eq: ineq5}) holds at $t=t^{*}$ and this automatically
holds given that at $t=t^{*}$, $\Lambda'\left(t^{*}\right)=\left(\delta_{A}-\delta_{P}\right)\Delta\left(t^{*}\right)$
and $\Lambda\left(t^{*}\right)=e^{\left(\delta_{A}-\delta_{P}\right)t^{*}}\frac{1-\ell}{\ell v'\left(\mu_{A}^{*}\right)\left(\delta_{A}-\delta_{P}\right)}$
which implies that at $t=t^{*}$ , (\ref{eq: ineq5}) holds with equality.
This concludes the proof.
\end{proof}
\newpage{}

\subsection{Proof of Lemma \ref{lem:There-exists-a}\label{subsec:Proof-of-Lemma}}
\begin{proof}
First, we note that we can consider $\hat{T}<\overline{T}$ and its
associated solution 
\[
\hat{x}_{\hat{T}}=\left(G_{P,0}\left(T;\hat{T}\right),G_{P,1}\left(T;\hat{T}\right)\right)
\]
. Let $T_{\omega}$ to be the infimum value such that $G_{P,\omega}\left(T_{\omega};\hat{T}\right)=0$
and if such value does not exist, we set $T_{\omega}=\hat{T}$. We
note that we must have $T_{1}=T_{0}$. Suppose that $T_{1}<\hat{T}$,
then we must have that $\mu_{A}\left(T\right)=0$ for all $T>T_{1}$.
Hence, the incentive constraint is given by 
\[
D\left(T\right)G_{P,0}\left(T\right)v\left(1\right)+\int_{T}^{\hat{T}}D'\left(s\right)G_{P,0}\left(s\right)ds\geq D\left(T\right)G_{P,0}\left(T\right)v\left(1\right)
\]
Since $G_{P,0}\left(T\right)\geq0$ and $D'\left(T\right)<0$, the
only way the above constraint can hold is if $G_{P,0}\left(T\right)=0$.
This implies that $T_{0}\leq T_{1}$. Using a similar argument, we
have that $T_{1}=T_{0}$. Note that since the marginal benefit of
$G_{P,\omega}\left(T\right)$ is always 1, we can assume that that
$d\Lambda_{\hat{T}}\left(T\right)=0$ for all $T\geq T_{1}$.

Next, let us consider $x=\left(0,\max\left\{ \hat{G}_{P,1}\left(T;\hat{T}\right)-\varepsilon,0\right\} \right)\in\Omega$
where $\varepsilon>0$. We can then write the relations in (\ref{eq: FOCL})
as 
\[
\frac{\partial L_{\hat{T}}\left(\hat{x}_{\hat{T}};x\right)-\partial L_{\hat{T}}\left(\hat{x}_{\hat{T}};\hat{x}_{\hat{T}}\right)}{\varepsilon}\leq0.
\]
By sending $\varepsilon\rightarrow0$, the above becomes
\begin{align}
 & T_{1}+\label{eq: FOC1}\\
 & \ell\int_{0}^{T_{1}}\left[v\left(1\right)-v\left(\mu_{A}\left(T;\hat{T}\right)\right)-v'\left(\mu_{A}\left(T;\hat{T}\right)\right)\left(1-\mu_{A}^{T}\left(T;\hat{T}\right)\right)\right]D\left(T\right)d\Lambda_{\hat{T}}\left(T\right)+\nonumber \\
 & \ell v\left(1\right)\int_{0}^{T_{1}}\int_{T}^{T_{1}}D'\left(s\right)dsd\Lambda_{\hat{T}}\left(T\right)-\lambda_{1,\hat{T}}\leq0\nonumber 
\end{align}
Similarly, we can consider $x=\left(0,\max\left\{ \hat{G}_{P,1}\left(T;\hat{T}\right)-\varepsilon,0\right\} \right)\in\Omega\in\Omega^{T}$
and do the same as above. This would then imply that the LHS of (\ref{eq: FOC1})
is non-negative and thus has to be 0. If we let $\Delta_{1}\left(\mu\right)=v\left(1\right)-v\left(\mu\right)-v'\left(\mu\right)\left(1-\mu\right)$,
then we can write the above as
\begin{equation}
\int_{0}^{T_{1}}\left(-\Delta_{1}\left(\mu_{A}\left(T;\hat{T}\right)\right)D\left(T\right)+v\left(1\right)\left(D\left(T\right)-D\left(T_{1}\right)\right)\right)d\Lambda_{\hat{T}}=\frac{T_{1}-\lambda_{1,\hat{T}}}{\ell}.\label{eq: Eq1}
\end{equation}
Repeating the above for $\omega=0$, we arrive at
\begin{equation}
\int_{0}^{T_{1}}\left(-\Delta_{0}\left(\mu_{A}\left(T;\hat{T}\right)\right)D\left(T\right)+v\left(1\right)\left(D\left(T\right)-D\left(T_{1}\right)\right)\right)d\Lambda_{\hat{T}}=T_{1}-\lambda_{0,\hat{T}}\label{eq: Eq2}
\end{equation}
A property of the $\Delta_{1},\Delta_{0}$ functions are
\begin{align*}
\frac{\Delta_{1}\left(\mu\right)+\Delta_{0}\left(\mu\right)}{2} & =v\left(1\right)-v\left(\mu\right)+\frac{v'\left(\mu\right)\mu-v'\left(\mu\right)\left(1-\mu\right)}{2}\\
 & =v\left(1\right)-v\left(\mu\right)+\frac{v'\left(\mu\right)\left(2\mu-1\right)}{2}
\end{align*}
Convexity of $v\left(\mu\right)$ implies that
\begin{align*}
\mu\geq1/2: & \frac{v\left(\mu\right)-v\left(1/2\right)}{\mu-1/2}\leq v'\left(\mu\right)\rightarrow v\left(\mu\right)-v'\left(\mu\right)\frac{2\mu-1}{2}\leq v\left(1/2\right)\\
\mu\leq1/2: & \frac{v\left(1/2\right)-v\left(\mu\right)}{1/2-\mu}\geq v'\left(\mu\right)\rightarrow v\left(\mu\right)-v'\left(\mu\right)\frac{2\mu-1}{2}\leq v\left(1/2\right)
\end{align*}
Which further implies that 
\[
\frac{\Delta_{1}\left(\mu\right)+\Delta_{0}\left(\mu\right)}{2}\geq v\left(1\right)-v\left(1/2\right)
\]
Now, we consider (\ref{eq: Eq1}) and (\ref{eq: Eq2}) and take equally
weighted average and use the above inequality to arrive at
\[
\int_{0}^{T_{1}}\left(-\left(v\left(1\right)-v\left(1/2\right)\right)D\left(T\right)+v\left(1\right)\left(D\left(T\right)-D\left(T_{1}\right)\right)\right)d\Lambda_{\hat{T}}\geq\frac{T_{1}-\lambda_{1,\hat{T}}+\ell\left(T_{1}-\lambda_{0,\hat{T}}\right)}{2\ell}
\]
Using integration by parts, we can write
\begin{align*}
-\int_{0}^{T_{1}}\left(v\left(1\right)-v\left(1/2\right)\right)D\left(T\right)d\Lambda_{\hat{T}}-v\left(1\right)\int_{0}^{T_{1}}\Lambda_{\hat{T}}\left(T\right)D'\left(T\right)dt & \geq\frac{T_{1}-\lambda_{1,\hat{T}}+\ell\left(T_{1}-\lambda_{0,\hat{T}}\right)}{2\ell}\Rightarrow\\
\int_{0}^{T_{1}}\left(v\left(1\right)-v\left(1/2\right)\right)D\left(T\right)d\Lambda_{\hat{T}}+v\left(1\right)\int_{0}^{T_{1}}\Lambda_{\hat{T}}\left(T\right)D'\left(T\right)dt & \leq-\frac{T_{1}-\lambda_{1,\hat{T}}+\ell\left(T_{1}-\lambda_{0,\hat{T}}\right)}{2\ell}\Rightarrow\\
\int_{0}^{T_{1}}D\left(T\right)d\Lambda_{\hat{T}}+\frac{v\left(1\right)}{v\left(1\right)-v\left(1/2\right)}\int_{0}^{T_{1}}\Lambda_{\hat{T}}\left(T\right)D'\left(T\right)dt & \leq-\frac{T_{1}-\lambda_{1,\hat{T}}+\ell\left(T_{1}-\lambda_{0,\hat{T}}\right)}{2\ell\left(v\left(1\right)-v\left(1/2\right)\right)}\Rightarrow\\
\int_{0}^{T_{1}}D\left(T\right)^{\frac{-v\left(1/2\right)}{v\left(1\right)-v\left(1/2\right)}}d\left(D\left(T\right)^{\frac{v\left(1\right)}{v\left(1\right)-v\left(1/2\right)}}\Lambda_{\hat{T}}\left(T\right)\right) & \leq-\frac{T_{1}-\lambda_{1,\hat{T}}+\ell\left(T_{1}-\lambda_{0,\hat{T}}\right)}{2\ell\left(v\left(1\right)-v\left(1/2\right)\right)}\Rightarrow
\end{align*}
\[
\int_{0}^{T_{1}}d\left(D\left(T\right)^{\frac{v\left(1\right)}{v\left(1\right)-v\left(1/2\right)}}\Lambda_{\hat{T}}\left(T\right)\right)\leq\int_{0}^{T_{1}}D\left(T\right)^{\frac{-v\left(1/2\right)}{v\left(1\right)-v\left(1/2\right)}}d\left(D\left(T\right)^{\frac{v\left(1\right)}{v\left(1\right)-v\left(1/2\right)}}\Lambda_{\hat{T}}\left(T\right)\right)
\]
where in the above we have used the fact that $D\left(T\right)^{\frac{-v\left(1/2\right)}{v\left(1\right)-v\left(1/2\right)}}\geq1$.
We then have that
\begin{align*}
D\left(T\right)^{\frac{v\left(1\right)}{v\left(1\right)-v\left(1/2\right)}}\Lambda_{\hat{T}}\left(T\right)-\Lambda_{\hat{T}}\left(0\right) & \leq-\frac{T_{1}-\lambda_{1,\hat{T}}+\ell\left(T_{1}-\lambda_{0,\hat{T}}\right)}{2\ell\left(v\left(1\right)-v\left(1/2\right)\right)}\\
0\leq D\left(T\right)^{\frac{v\left(1\right)}{v\left(1\right)-v\left(1/2\right)}}\Lambda_{\hat{T}}\left(T\right) & \leq\Lambda_{\hat{T}}\left(0+\right)-\frac{T_{1}-\lambda_{1,\hat{T}}+\ell\left(T_{1}-\lambda_{0,\hat{T}}\right)}{2\ell\left(v\left(1\right)-v\left(1/2\right)\right)}
\end{align*}
Hence, in order to prove the claim, it is sufficient to show that
the right hand side stays bounded as $\hat{T}$ converges to $\overline{T}$.

We know that $T_{1}\leq\overline{T}$ and thus it is bounded above.
Moreover, since $\Delta_{1}\left(\mu\right)\left(1-\mu\right)+\Delta_{0}\left(\mu\right)\mu=v\left(1\right)-v\left(\mu\right)$,
we can use complementary slackness to write $\partial L_{\hat{T}}\left(\hat{x}_{\hat{T}};\hat{x}_{\hat{T}}\right)=0$
as
\[
\int_{0}^{T_{1}}\hat{G}_{P}\left(T;\hat{T}\right)dT=\lambda_{1,\hat{T}}\mu_{P}+\lambda_{0,\hat{T}}\left(1-\mu_{P}\right)
\]
The RHS of the above is less than $T_{1}\leq\overline{T}$ which then
implies that $\lambda_{1,\hat{T}}$ and $\lambda_{0,\hat{T}}$ stay
bounded below $\max\left\{ \frac{\overline{T}}{\mu_{P}},\frac{\overline{T}}{1-\mu_{P}}\right\} $.
Finally, plugging in $x=\left(\hat{G}_{P,0}\left(T\right),\varepsilon\mathbf{1}\left[T=0\right]+\hat{G}_{P,1}\left(T\right)\right)$
into (\ref{eq: FOCL}) and subtract (\ref{eq:FOCeq}) we arrive at
\begin{align*}
 & \Delta_{1}\left(\mu_{A}\right)\Lambda_{\hat{T}}\left(0+\right)-\lambda_{1,\hat{T}}\leq0
\end{align*}
which implies that $\Lambda_{\hat{T}}\left(0+\right)$ is also bounded.
This concludes the proof.
\end{proof}

\end{document}